\documentclass[a4paper,fleqn,usenatbib,onecolumn]{mnras}

\usepackage[T1]{fontenc}
\usepackage{ae,aecompl}

\usepackage{graphicx}	% Including figure files
\usepackage{amsmath}	% Advanced maths commands
\usepackage{amssymb}	% Extra maths symbols
\newcommand{\beq}[0]{\begin{equation}}
\newcommand{\eeq}[0]{\end{equation}}

\title[Relativistic electron-ion reconnection]{Non-thermal particle acceleration in collisionless relativistic electron--proton reconnection}

\author[G.~R.~Werner et al]{
G.~R.~Werner,$^{1}$\thanks{E-mail: Greg.Werner@colorado.edu}
D.~A.~Uzdensky,$^{1,2}$
M.~C.~Begelman,$^{3,4}$
B.~Cerutti,$^{5}$
K.~Nalewajko$^{6}$
\\
$^{1}$Center for Integrated Plasma Studies, Physics Department, 
   390 UCB, University of Colorado, Boulder, CO 80309, USA\\
$^{2}$ Institute for Advanced Study, Princeton, NJ 08540, USA \\
$^{3}$JILA, University of Colorado and National Institute of
Standards and Technology, 440 UCB, Boulder, CO 80309-0440, USA\\
$^{4}$Department of Astrophysical and Planetary Sciences, 
391 UCB, University of Colorado, Boulder, CO 80309, USA\\
$^{5}$Univ. Grenoble Alpes, CNRS, IPAG, F-38000 Grenoble, France\\
$^{6}$Nicolaus Copernicus Astronomical Center, Polish Academy of Sciences, Bartycka 18, 00-716 Warsaw, Poland
}

\date{}

\begin{document}
\label{firstpage}
\pagerange{\pageref{firstpage}--\pageref{lastpage}}
\maketitle

\begin{abstract}
Magnetic reconnection in relativistic collisionless plasmas can accelerate particles and power high-energy emission in various astrophysical systems. Whereas most previous studies focused on relativistic reconnection in pair plasmas, less attention has been paid to electron-ion plasma reconnection, expected in black hole accretion flows and relativistic jets.  We report a comprehensive particle-in-cell numerical investigation of reconnection in an electron-ion plasma, spanning a wide range of ambient ion magnetizations $\sigma_i$, from the semirelativistic regime (ultrarelativistic electrons but nonrelativistic ions, $10^{-3} \ll \sigma_i \ll 1$) to the fully relativistic regime (both species are ultrarelativistic, $\sigma_i \gg 1$). We investigate how the reconnection rate, electron and ion plasma flows, electric and magnetic field structures, electron/ion energy partitioning, and nonthermal particle acceleration depend on~$\sigma_i$. Our key findings are:
(1) the reconnection rate is about 0.1 of the Alfv\'enic rate across all regimes;
(2) electrons can form concentrated moderately relativistic outflows even in the semirelativistic, small-$\sigma_i$ regime; 
(3) while the released magnetic energy is partitioned equally between electrons and ions in the ultrarelativistic limit, the electron energy fraction declines gradually with decreased $\sigma_i$ and asymptotes to about 0.25 in the semirelativistic regime;
(4) reconnection leads to efficient nonthermal electron acceleration with a $\sigma_i$-dependent power-law index, $p(\sigma_i)\simeq {\rm const}+0.7\sigma_i^{-1/2}$.
These findings are important for understanding black hole systems and lend support to semirelativistic reconnection models for powering nonthermal emission in blazar jets, offering a natural explanation for the  spectral indices observed in these systems. 
%
%This is a simple template for authors to write new MNRAS papers.
%The abstract should briefly describe the aims, methods, and main results of the paper.
%It should be a single paragraph not more than 250 words (200 words for Letters).
%No references should appear in the abstract.
\end{abstract}


%{\bf \color{red} DMITRI: 
%I have selected keywords that I thought might be suitable. There are 8 in the list above, and we'll need to select 6 from them. Most of these keywords coincide with Greg's original ApJ list. I think all journals may actually be using the same universal list now. The full list is available here: \\
%% http://oxfordjournals.org/our_journals/mnrasl/for_authors/mnraskey.pdf
%http://oxfordjournals.org/our\_journals/mnrasl/for\_authors/mnraskey.pdf
%}

%%%%%%%%%%%%%%%%%%%%%%%%%%%%%%%%%%%%%%%%%%%%%%%%%%

%%%%%%%%%%%%%%%%% BODY OF PAPER %%%%%%%%%%%%%%%%

%***********************************************************************************

%***********************************************************************************

\section{Introduction}
\label{sec-intro}

Magnetic reconnection is a basic plasma process often leading to the violent release of magnetic energy through rapid rearrangement of magnetic field geometry \citep[e.g.,][]{Biskamp:2005,Zweibel_Yamada-2009,Yamada_etal-2010}.
Relativistic magnetic reconnection can occur when,
as in many high-energy astrophysical environments, the magnetic energy density exceeds the energy density of the ambient plasma (including the rest-mass energy density).
This situation is characterized by a large magnetization parameter $\sigma_{\rm hot} \equiv B^2/4\pi w$, where 
$B$ is the ambient reconnecting magnetic field strength and
$w \equiv u+p$ is the relativistic enthalpy density, the sum of the relativistic energy density $u$ (including rest mass) and the plasma pressure~$p$.  
When $\sigma_{\rm hot}\gg 1$, the (relativistic) Alfv\'en velocity $V_A = c\, \sigma_{\rm hot}^{1/2} (1+\sigma_{\rm hot})^{-1/2}$  approaches the speed of light~$c$, and magnetic reconnection proceeds in the relativistic regime \citep{Blackman_Field-1994, Lyutikov_Uzdensky-2003, Lyubarsky-2005}, leading to ultrarelativistic bulk flows, plasma heating to relativistic temperatures, nonthermal particle acceleration to ultrarelativistic energies, etc. 

There are many examples of high-$\sigma$ ($\sigma_{\rm hot} \gg 1$) and medium-$\sigma$ ($\sigma_{\rm hot} \sim 1$) 
astrophysical environments where relativistic magnetic reconnection is believed to take place, including pulsar magnetospheres, pulsar winds, and pulsar wind nebulae (PWNe)
\citep{Coroniti:1990,Lyubarsky:2001,Zenitani_Hoshino-2001,Kirk_Skjaeraasen-2003,Kirk-2004,Contopoulos-2007a,Contopoulos-2007b,Sironi_Spitkovsky-2011,Uzdensky_etal-2011,Cerutti_etal-2012a,Cerutti_etal-2012b,Cerutti_etal-2013,Cerutti_etal-2014a,Cerutti_etal-2014b,Uzdensky:2014,Philippov_etal-2014a,Philippov_etal-2014b,Cerutti_etal-2015}; gamma-ray bursts (GRBs) \citep{Drenkhahn:2002,Giannios:2006,McKinney:2012}; magnetospheres of magnetars \citep{Lyutikov:2003,Lyutikov:2006,Uzdensky:2011}; coronae of accreting black holes (BHs) in both galactic X-ray binaries (XRBs) and active galactic nuclei (AGNs) \citep{Galeev_etal-1979,DiMatteo-1998,Hoshino_Lyubarsky-2012,Uzdensky_Goodman-2008,Goodman_Uzdensky-2008}, as well as in relativistic jets powered by these systems, including blazars \citep{Giannios:2009,Giannios:2010,Nalewajko:2011,Nalewajko:2012, Giannios-2013, Sironi_etal-2015, Petropoulou_etal-2016}.
Many of these systems are observed to emit nonthermal broad-band radiation (characterized by a power-law energy spectrum).
Whether quiescent or in the form of bright, impulsive X-ray and $\gamma$-ray flares with rapid time variability, such nonthermal radiation is thought to result from energetic electrons (and positrons, if present) emitting via synchrotron and/or inverse Compton (IC) mechanisms; the energy spectrum of emitting particles is therefore inferred to be nonthermal as well (a power law, sometimes extending over orders of magnitude in energy).
This highlights the importance of understanding the physical mechanisms of non-thermal particle acceleration operating in collisionless energy-conversion processes, of which magnetic reconnection is a leading candidate [see, e.g., \citep{Hoshino_Lyubarsky-2012} for a review], taking place in relativistic astrophysical plasmas. 

Nonthermal particle acceleration in plasma processes has been studied within fluid (e.g., magnetohydrodynamic, MHD) frameworks by tracking distributions of passive test particles 
\citep{Nodes:2003,Onofri:2006,Drake:2009,Ding:2010,Gordovskyy:2010b,Kowal:2011}. 
However, such studies, though valuable in many respects, cannot account for the effects of nonthermally accelerated particles acting back on the electromagnetic fields and fluid motions.  These self-consistent effects are likely to be important, especially when the particle energy distributions $f(\varepsilon) \sim \varepsilon^{-p}$ have power-law indices $p<2$ \citep[as in ultrarelativistic electron-positron reconnection:][]{Sironi_Spitkovsky-2014,Guo:2015,Werner_etal-2016}, indicating that most of the particle kinetic energy resides in particles with energies much higher than average.
Because of this, a self-consistent study of nonthermal particle acceleration inherently requires a kinetic approach that evolves the entire particle distribution.  
A powerful computational tool often employed in numerical studies of kinetic plasma processes is particle-in-cell (PIC) simulation. In particular, there have been a large number of PIC studies of relativistic collisionless reconnection in electron-positron pair plasmas in the high-$\sigma$ limit in the past few years 
\citep{Zenitani_Hoshino-2001, Zenitani_Hoshino-2005, Zenitani_Hoshino-2007, Zenitani_Hoshino-2008, Jaroschek_etal-2004a, Lyubarsky_Liverts-2008, Jaroschek_Hoshino-2009, Liu_etal-2011, Bessho_Bhattacharjee-2012, Kagan_etal-2013, Cerutti_etal-2012b, Cerutti_etal-2013, Cerutti_etal-2014a, Cerutti_etal-2014b, Sironi_Spitkovsky-2011, Sironi_Spitkovsky-2014, Guo:2014, Guo:2015, Liu_etal-2015,Nalewajko:2015, Sironi_etal-2015, Sironi_etal-2016, Werner_etal-2016}.  
These studies have concentrated on pair plasmas for two main reasons. 
First, some of the most important astrophysical applications of relativistic reconnection involve pair plasmas, most notably pulsar magnetospheres, pulsar winds, and~PWN. The second reason is the relative simplicity of pair plasma reconnection due to the absence of the scale separation, present in conventional electron-ion plasmas, between the important electron and ion kinetic scales such as the Larmor radii, $\rho_{e}$ and~$\rho_i$, and the collisionless skin depths, $d_e$ and~$d_i$. 
Computational advances have allowed the more recent of these PIC studies \citep[e.g.,][]{Sironi_Spitkovsky-2014, Guo:2014, Werner_etal-2016} to add substantially to our understanding of relativistic pair-plasma reconnection by clearly demonstrating efficient nonthermal particle acceleration and mapping out its quantitative characteristics, such as the power-law index and high-energy cutoff, as functions of the system parameters. 

However, {\it electron-ion} ({\it ei}) plasmas predominate in many relativistic high-energy astrophysical environments in which magnetic reconnection is believed to be important (throughout this paper, the ions are protons).  For example, {\it ei}-plasma reconnection has been invoked as the main mechanism responsible for electron heating and subsequent high-energy radiation in many of the systems mentioned above, e.g., in accretion disc coronae (ADC) in black hole systems, in gamma-ray flares in relativistic blazar jets (and AGN jets in general), including ultrarapid TeV flares, and for powering prompt emission in~GRBs.  Despite its importance, and in stark contrast to nonrelativistic reconnection, relativistic reconnection in {\it ei} plasmas has so far received relatively little attention; PIC studies of {\it ei} reconnection in the relativistic regime have started only recently \citep{Melzani:2014a,Melzani:2014b,Werner_etal-2013,WernerDPP:2015,Guo:2016}. 

When considering ultrarelativistic reconnection in electron-ion plasmas, it is important to note that the key fundamental plasma time- and length-scales, such as the plasma and cyclotron frequencies, the collisionless skin depth, and the Larmor radius, become independent of the particle rest mass and instead depend only on the average particle energy in this ultrarelativistic limit.  Therefore, if both electrons and ions become ultrarelativistic as a result of magnetic energy release in a reconnection process, then any differences in the dynamical behaviour of the two species disappear (in the absence of radiative losses).  Correspondingly, one {\it a priori} expects ultrarelativistic {\it ei}-plasma reconnection to become essentially similar to ultrarelativistic pair-plasma reconnection, as was indeed confirmed in a recent study by \cite{Guo:2016}. 

Importantly, however, in many real astrophysical plasmas, one finds a more complex and interesting situation---the {\it semirelativistic} regime---where typical electrons are ultrarelativistic (with the average Lorentz factor $\bar{\gamma}_e = 1 + \bar{\varepsilon}_e / m_e c^2 \gg 1$) 
but ions are subrelativistic ($\bar{\varepsilon}_i \ll m_i c^2$); 
here, $\bar{\varepsilon}_{e,i}$ is the average kinetic energy of electrons/ions, excluding the rest-mass energy.
Whether typical ions in the reconnection outflow are expected to be relativistic is governed by the upstream `cold' magnetization $\sigma_i \equiv B_0^2/(4\pi n_{bi,0} m_i c^2)$, where $B_0$ is the reconnecting magnetic field and $n_{bi,0}$ is the ion density upstream of the reconnection region; this quantity is basically twice the ratio of the available magnetic energy density to ion rest-mass energy density.
Magnetic reconnection in the semirelativistic regime, characterized by $\sigma_i \lesssim 1$, may be of great importance in, e.g., accreting black hole coronae and AGN (including blazar) jets.  Because of the very large separation between the electron and proton rest masses, $\mu = m_p/m_e = 1836$, the range of applicability of this regime is actually quite broad:  it essentially covers three orders of magnitude in~$\sigma_i$ and hence in the expected average dissipated energy per particle, and correspondingly several orders of magnitude in the energy of emitted photons (since $\varepsilon_{\rm ph}\sim \gamma^2$).  It is thus critical to understand this new, so far unexplored, reconnection regime, as well as the transition from this semirelativistic regime to the pair-plasma-like regime of ultrarelativistic ions as $\sigma_i$ increases from $\sigma_i < 1$ to $\sigma_i \gg 1$.  
We note that, in contrast to our present study, \cite{Guo:2016} focused solely on
the ultrarelativistic regime, $\sigma_i \geq 10$, 
%with real and reduced mass ratios, 
while \cite{Melzani:2014a,Melzani:2014b} studied greatly reduced mass ratios ($\mu\leq 25$) with $\sigma_i \approx 1$ and mildly relativistic electrons, as well as with $\sigma_i = 0.04$ and barely relativistic electrons.

Here we report the results of our systematic investigation
of collisionless {\it ei}-plasma magnetic reconnection, using the real mass ratio $\mu=1836$, from
the semirelativistic through the ultrarelativistic regime---covering the entire range from $\sigma_i = 0.03$  to 
$\sigma_i = 10^4$---using the relativistic PIC code {\sc Vorpal}/{\sc VSim} \citep{Nieter:2004}.
We focus on several important quantitative characteristics of the reconnection process for large system sizes $L_x$---such as the reconnection rate, energy partitioning between electrons and ions, Hall effect signatures, power-law slopes and cutoff energies of nonthermal particle populations---and study how they scale with the upstream ion magnetization~$\sigma_i$. 

This paper is organized as follows. 
In~\S\ref{sec-sims} we describe our numerical simulations, and then 
present the results in \S\ref{sec-results-dynamics}--\ref{sec-results-kinetics}.  In~\S\ref{sec-results-dynamics} we discuss the basic fluid dynamics and electrodynamics of reconnection---reconnection rate, plasma flows, and Hall effect signatures. We then describe our findings regarding the energetics and kinetic aspects of the reconnection process in~\S\ref{sec-results-kinetics}, namely, the energy partitioning between electrons and ions, and nonthermal particle acceleration. 
After that, in~\S\ref{sec-astro} we discuss some of the most intriguing astrophysical applications of our results, in particular, as a possible natural explanation for ultrarapid TeV flares in AGN/blazar jets [with connections to the minijet model of \cite{Giannios:2009}], as well as the non-flaring emission at lower energies. Finally, in~\S\ref{sec-conclusions} we summarize our conclusions. 

%----------------------------------------------------------------------------------

\section{Simulations}
\label{sec-sims}

%----------------------------------------------------------------------------

Our simulations---all 2D---of relativistic electron-proton plasma reconnection presented here are performed with the explicit electromagnetic particle-in-cell {\sc Vorpal}/{\sc VSim} code \citep{Nieter:2004}, using a fully electromagnetic field solver \citep{Yee:1966} and a relativistic Boris particle-push \citep{Boris:1970} with divergence-preserving current deposition \citep{Villasenor:1992,Esirkepov:2001}.  
The simulations are initialized with an electron-proton plasma, using the real mass ratio $\mu=m_p/m_e=1836$, in a double-periodic relativistic Harris equilibrium, which is described below in detail. 
Our investigation focuses on the effects of varying ion magnetization $\sigma_i$ (defined below); besides describing the setup below, we also relate the initial simulation parameters to $\sigma_i$.

The fundamental aspects of the initial configuration are the reversing magnetic field and ambient (background) plasma.  To allow the use of easily-implemented periodic boundary conditions, it is convenient to simulate two field reversals (the so-called double-periodic configuration).  As is commonly done, we add a relativistic Harris current layer \citep{Hoh-1966, Kirk_Skjaeraasen-2003}---a non-uniform, drifting plasma component---at each field reversal to provide current and pressure to balance the reversing magnetic field.

The simulations are conducted in a 2D rectangular computational box with a regular Cartesian grid and periodic boundary conditions in both directions.   
The initial double-reversing magnetic field configuration (cf. Figs.~\ref{fig:AzTimeEvolutionSigp1} and \ref{fig:AzTimeEvolutionSig10}, top left panel), before perturbation (described below), is:
\begin{eqnarray}
  B_x(y) &=& \left\{ \begin{array}{l@{\quad}l}
   \phantom{-}
   B_0 \tanh\left( \displaystyle \frac{y-y_{\rm cs,2}}{\delta} \right)
    & %|y-y_{\rm cs,2}| < |y-y_{\rm cs,1}| 
      \textrm{ (upper layer)}
      \vspace{0.05in}
      \\
   -B_0 \tanh\left(\displaystyle  \frac{y-y_{\rm cs,1}}{\delta} \right)
    & %|y-y_{\rm cs,2}| > |y-y_{\rm cs,1}| 
      \textrm{ (lower layer)}\\
    \end{array} \right.
\end{eqnarray}
where $B_0$ is the asymptotic upstream reconnecting magnetic field strength, $y_{cs,1}$ and $y_{cs,2}$ are the mid-planes of the lower and upper current sheets where $B_x(y)=0$, and $\delta$ is the initial layer half-thickness, determined to be consistent with the other Harris sheet equilibrium parameters. 

Thus, our coordinate system places $x$ in the direction of the reconnecting magnetic field (the plasma outflow direction), and $y$ in the direction of reconnected magnetic field (the inflow direction, perpendicular to the current layers), with the ignorable $z$-direction parallel to the initial Harris current. 
For simplicity, we limit this study to the case without a guide magnetic field component ($B_z=0$, initially). 
Also, unlike most previous double-periodic simulations, our present simulations use a tall box, $L_y = 2L_x$ (where $L_x$ and $L_y$ are the box sizes in the $x$- and $y$-directions, respectively) to reduce interactions between the two layers.

In addition to the two concentrated Harris layers, which are described in more detail below, we initially fill the entire box with a low-density uniform and stationary background plasma, described by non-drifting relativistic Maxwell-J\"uttner electron and ion distributions with equal initial electron and ion densities, $n_{be,0}=n_{bi,0}$, and temperatures, $T_{be}=T_{bi} = \theta_{be} m_e c^2 = \theta_{bi} m_i c^2$. 
We then characterize the fundamental reconnection setup through the relationships between the upstream reconnecting magnetic field~$B_0$ and the parameters of this ambient background plasma.
In particular, we use the initial electron and ion `cold magnetizations' $\sigma_e=B_0^2/(4\pi n_{be,0} m_e c^2)$ and $\sigma_i=B_0^2/(4\pi n_{bi,0} m_i c^2) = \sigma_e/\mu$ to relate the magnetic field energy density $B_0^2/8\pi$ to the background electron and ion densities, $n_{be,0}=n_{bi,0}$. 
In addition, we relate the background plasma temperature to the upstream magnetizations: for all simulations presented here we have chosen $\theta_{be}=\sigma_e/200$, $\theta_{bi}=\theta_{be}/\mu = \sigma_i/200$. 
These background density and temperature choices yield an initial upstream plasma $\beta$ parameter of
$\beta_{\rm up} = \beta_{be} + \beta_{bi} = 8 \pi (n_{be,0} T_{be} + n_{bi,0} T_{bi})/B_0^2 = 4  \theta_{bi}/\sigma_i = 1/50 \ll 1$, so the upstream region is magnetically dominated.

Because $\sigma_e m_e c^2 = \sigma_i m_i c^2$ is roughly the available magnetic energy per particle, the `cold' magnetizations set the basic scale for average particle energy gain during reconnection.
It is convenient to define a corresponding length-scale $\rho_c=\sigma_e \rho_{e0} = \sigma_i \rho_{i0}$, where $\rho_{s0} \equiv m_s c^2 / (eB_0)$ are nominal relativistic electron and ion Larmor radii; thus defined, $\rho_c$ is the Larmor radius of an ultrarelativistic electron with energy $\sigma_e m_e c^2$ (or, if $\sigma_i \gg 1$, of an ion with energy $\sigma_i m_i c^2$).  The corresponding time-scale is $\omega_c^{-1}=\rho_c/c = B_0/(4\pi e n_{be,0} c)$.
Because the Larmor radius $\rho_c$ and Larmor period of typical electrons (after being energized by
reconnection) scale with $\sigma_i$, simulation grid cell size and timestep are set basically proportional to~$\sigma_i$, with small adjustments as $\sigma_i$ decreases (because the scale separation between electrons and ions becomes larger).
We note that $\rho_c$ can also be interpreted as the half-thickness of the current sheet that balances (via Ampere's Law) the jump in $B_x$ equal to~$2B_0$, assuming the sheet has density $n_{be,0}$ and drift velocity $c\hat{\bf z}$.
That is, a uniform electron current density of $J_e = e n_{be,0} c$ in a layer of total thickness $2\rho_c =2 \sigma_e \rho_{e0} = 2 B_0/(4\pi e n_{be,0})$ would support a magnetic field discontinuity $\Delta B = (4\pi/c) J_e (2\rho_c) = 2B_0$.

\begin{figure}
\includegraphics[width=8.cm,trim=0mm 0 0mm 0]{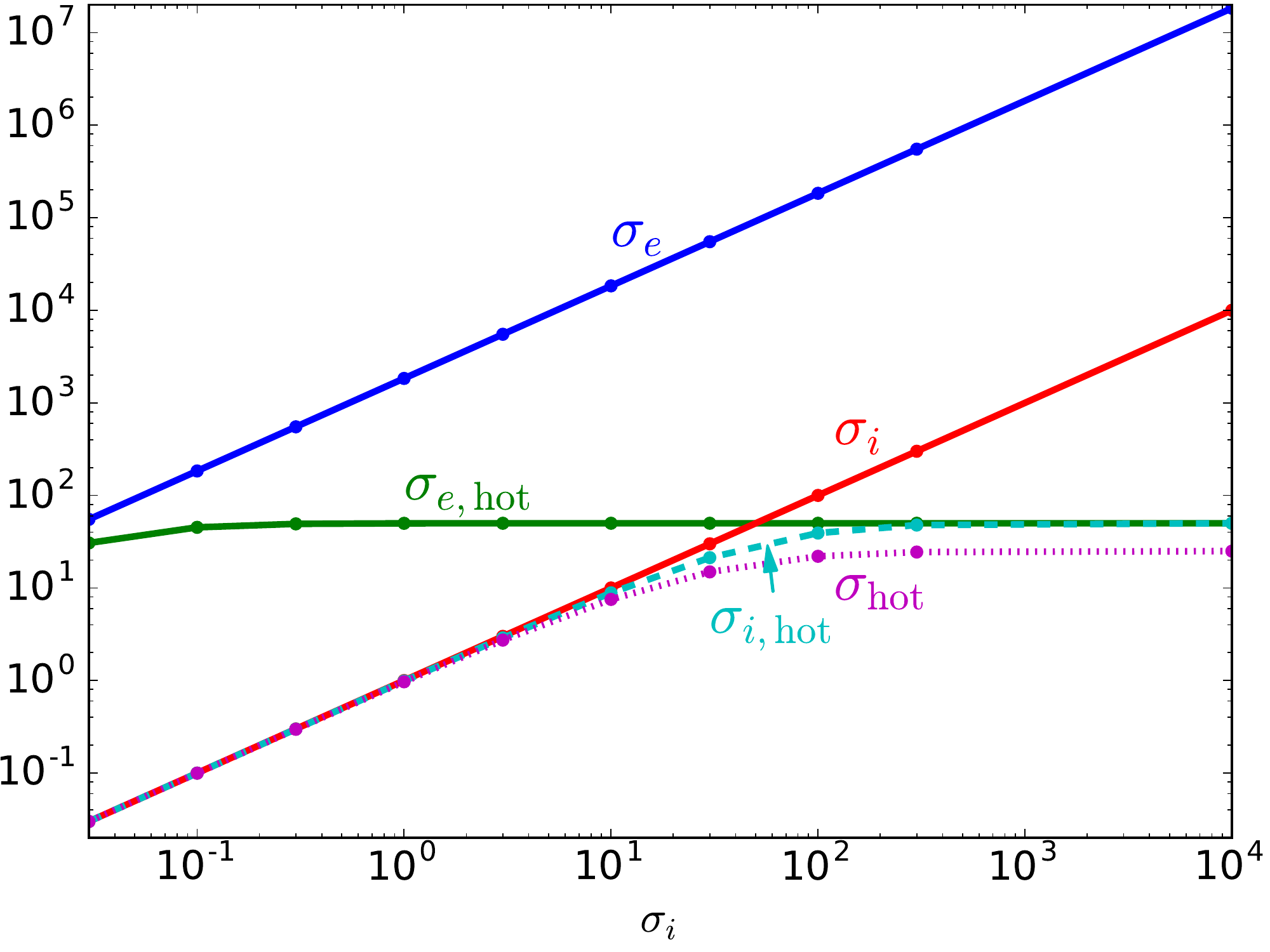}
\caption{The dependence of cold and hot magnetization parameters on
$\sigma_i$ (the `cold' $\sigma$ with respect to background ions).
Circles mark the values of $\sigma_i$ for simulations described in this paper.
\label{fig:sigmas}
}
\end{figure}

Linking the initial upstream temperature to $\sigma_i$ has both advantages and disadvantages. 
Importantly, it helps to ensure that our simulations are always properly resolved, namely, that the Debye length of the background plasma, $\lambda_{D,b} \equiv [T_{be}/4\pi (n_{be,0}+n_{bi,0}) e^2]^{1/2}=\rho_c/20$, does not become small compared to the grid cell size (which also scales with $\rho_c$---see below).
However, choosing $\theta_{be}/\sigma_e=\theta_{bi}/\sigma_i=1/200$ to avoid extreme resolution requirements has the drawback of limiting the maximum~$\sigma_{\rm hot}$.
The `hot' $\sigma$ with respect to electrons is defined as $\sigma_{e,\rm hot}=B_0^2/4\pi w_{be}$ where $w_{be}$ is the relativistic enthalpy density of background electrons; 
for ultrarelativistic electrons, $w_{be} \approx 4\theta_{be} n_{be,0} m_e c^2$, and so $\sigma_{e,\rm hot}\approx \sigma_{e} / 4 \theta_be = 50$ in our simulations.
Thus $\sigma_{e,\rm hot}$ describes the ratio between the available magnetic energy and (roughly) the relativistic thermal energy (plus the negligible electron rest-mass energy) in the upstream region; if $\sigma_{e,\rm hot}$ is not much larger than one, then typical electrons will not gain significant energy compared to the average (thermal) energy they already have.
For subrelativistic ions, however, the rest mass dominates the relativistic enthalpy, and $\sigma_{i,\rm hot}\approx \sigma_{i}$.
The total $\sigma_{\rm hot}$ includes the enthalpies of both electrons and ions; for $\sigma_i \lesssim 10$, we have $\sigma_{\rm hot} \approx \sigma_{i,\rm hot} \approx \sigma_i$.
Figure~\ref{fig:sigmas} shows how all these magnetization parameters vary with $\sigma_i$ (given our choice of initial background plasma temperature).

With the upstream parameters settled, we now turn to the description of the two initial relativistic Harris current sheets that are introduced to balance the magnetic field reversals \citep{Hoh-1966, Kirk_Skjaeraasen-2003}.  
We choose the drifting electrons and ions to contribute equally to the initial layer pressure and to drift with equal but opposite speeds (in the simulation, or `lab' frame): 
for all simulations, the drift speed is $\beta_{\rm drift} c = 0.3c$ and the peak drift-to-background plasma density ratios are $n_{de,0}/n_{be,0} = n_{di,0}/n_{bi,0} = 5$ (all densities are specified in the lab frame).
The remaining Harris-layer parameters (layer half-thickness $\delta$ and co-moving temperatures $T_{de}=T_{di}=\theta_{de} m_e c^2=\theta_{di} m_i c^2$ of the relativistic Maxwell-J\"uttner distributions) are determined by Ampere's law 
[$B_0 / \delta = (4\pi/c) e (n_{de,0}+n_{di,0}) \beta_{\rm drift} c$], 
and pressure balance [$(n_{de,0}\theta_{de} m_e +n_{di,0} \theta_{di} m_i)c^2/\gamma_{\rm drift} = B_0^2/8\pi$].  
This yields $\theta_{di} = (\gamma_{\rm drift}/4)(n_{be,0}/n_{de,0})\, \sigma_i \approx 0.05\, \sigma_i$ and $\delta=2(\theta_{di}/\gamma_{\rm drift} \beta_{\rm drift}) \rho_{i0}=\rho_c/3$.
To be consistent with the magnetic profile, the drift plasma density varies with $y$ as $n_{de}(y) = n_{di}(y) = n_{de,0} \cosh^{-2} [(y-y_{\rm cs,\ell})/\delta]$ for layers $\ell=1,2$.

To accelerate the onset of reconnection, we add a slight (1\%) initial magnetic perturbation in the $z$-component of the vector potential (magnetic flux function), so that the total initial vector potential is (e.g., for the lower layer)
\begin{eqnarray}
  A_{z} &=& 
  \left[
    1 + 0.01 \cos \frac{2 \pi x}{L_x} \cos^2 \frac{2\pi (y-y_{cs,1})}{L_y} 
    \right] 
  B_0 \delta 
  \left[\ln \cosh \frac{y_{cs,2}-y_{cs,1}}{2\delta}
       -\ln \cosh \frac{y-y_{cs,1}}{\delta}
       \right].
\end{eqnarray}

This study focuses mainly on the dynamics, energetics, and kinetics (i.e., nonthermal particle acceleration) of electron-ion magnetic reconnection as one transitions from the semirelativistic regime (nonrelativistic ions but ultrarelativistic electrons: $\sigma_i < 1$, $\sigma_e \gg 1$) to the fully ultrarelativistic regime (both ions and electrons being ultrarelativistic, $\sigma_e \gg \sigma_i \gg 1$). To accomplish this goal, we conduct an extensive parameter-space study with respect to the upstream ion magnetization~$\sigma_i$, specifically studying simulations with $\sigma_i = 0.03, 0.1, 0.3, 1.0, 3.0, 10, 30, 100, 300, 10^4$, for system size $L_x=120\rho_c$.

As $\sigma_i$ varies, the key initial energy-scales and length-scales---namely, $T_{be}=T_{bi}$, $T_{de}=T_{di}$, $\rho_c$, and~$\delta$---change in proportion to $\sigma_i$, as described above.  We note that the background electrons are always somewhat relativistic and usually very relativistic; even for our lowest $\sigma_i$ of $0.03$, we have $\theta_{be} \approx 0.3$, yielding an average kinetic energy nearly equal to the electron rest mass for upstream electrons---and these electrons become significantly more relativistic as they enter the reconnection layer.  Ions, on the other hand, range from sub- to ultra-relativistic as $\sigma_i$ increases.

Our main $\sigma_i$-comparison study is conducted using the same fiducial system size $L_x=120 \rho_c$, which is chosen to be as large as possible while remaining computationally feasible.  
Thus, unless specifically stated otherwise, all results shown in this work are for simulations with $L_x=120\rho_c$.
This size is sufficiently large that the number and total energy of initially-drifting particles is small compared to background particles, and dynamical quantities, such as reconnection rate, are nearly independent of~$L_x$. 
In addition, however, in order to examine the extent to which these fiducial sizes reach the asymptotic `large-system regime,' we explore the effects of varying $L_x$ for two values of $\sigma_i=0.1,1$, over a range $20 \leq L_x/\rho_c \leq 160$.
The simulations run until the magnetic energy essentially stops decreasing; for $L_x=120\rho_c$, reconnection ends after roughly 
$4 L_x/c$ for large $\sigma_i$ and after $12 L_x/c$ for $\sigma_i=0.1$.

The discretization parameters of our simulations are as follows. The spatial resolution is represented by the ratio of grid cell size $\Delta x = \Delta y$ to the fundamental microphysical scale~$\rho_c$, and 
the timestep is set just below the usual (2D) Courant-Friedrichs-Lewy maximum stable step, $\Delta t = 0.99(\Delta x/c)/\sqrt{2}$.
For the system size $L_x=120\rho_c$ the resolution and number of initial particles per cell (including all species) are set to be 
$\Delta x/\rho_c = 1/16$ (256 particles per cell) for $\sigma_i \leq 0.3$,
$\Delta x/\rho_c = 1/20$ (168 particles per cell) for $\sigma_i = 1,3$,
and
$\Delta x/\rho_c = 1/24$ (128 particles per cell) for $\sigma_i \geq 10$.%
\footnote{For the two largest simulations ($\sigma_i=0.1$ and~1), $L_x=160\rho_c$, computational feasibility led us to use $\Delta x/\rho_c = 16$ (128 particles per cell); because of their size and fewer particles, they experienced energy non-conservation greater than 1\% (1.5\% and 1.1\%).
For smaller simulations we used the same resolution as for $L_x=120 \rho_c$, but increased the number of particles per cell purely to improve the statistics of binning particles in energy distributions so that we could attribute, e.g., a high-energy cutoff (of an energy spectrum), to system size and not to a probability becoming small compared to the inverse of the total number of particles in the simulation.}

%-------------------------------------------------------------------------------------------------

In this study we are able to adopt the real ion-to-electron rest-mass ratio, $\mu=m_i/m_e = 1836$, because relativistic effects decrease the scale separation between electrons and ions.  
Essentially, the electron rest mass becomes irrelevant for highly relativistic electrons;
instead, the effective relativistic mass $\gamma_e m_e$, or
equivalently the electron energy $\gamma_e m_e c^2$, 
determines the characteristic electron length-scales, such as the Larmor radius
$\rho_e \approx \gamma_e m_e c^2/eB_0$. 
As we show in this study, both electrons and ions on average gain comparable amounts of energy as a result of reconnection, of order $\sigma_e m_e c^2 = \sigma_i m_i c^2$.  As long as $\sigma_i \gg 1/\mu$ and hence $\sigma_e \gg 1$ (which is satisfied in our study), so that typical electrons are ultrarelativistic, the characteristic electron Larmor radius is 
$\rho_e \sim \sigma_e m_e c^2/eB_0 = \sigma_i m_i c^2/eB_0 =\rho_c$. 
At the same time, the characteristic ion Larmor radius scales as 
$\rho_i \sim \sqrt{\sigma_i (2 + \sigma_i)}\, m_i c^2/eB_0 = \sqrt{(2 + \sigma_i)/\sigma_i}\, \rho_c$.  
Thus the ion/electron scale separation, characterized by the ratio of the Larmor radii, 
$\rho_i/\rho_e \sim  \sqrt{(2+\sigma_i)/\sigma_i}$, is independent of the rest-mass ratio~$\mu$ 
and remains well below its nonrelativistic value of~$\sqrt{\mu}$ for $\sigma_i \mu \gg 1$. 
In particular, the scale separation scales as $(2/\sigma_i)^{1/2}$ in the semirelativistic regime $\sigma_i \ll 1$, diminishing to $\rho_i/\rho_e \rightarrow 1$ in the ultrarelativistic regime $\sigma_i \gg 1$. 
Thus, the traditional challenge created by the large disparity between electron and ion scales (in nonrelativistic reconnection) is greatly reduced, allowing us to use the mass ratio $\mu=1836$.

%----------------------------------------------------------------------------

%----------------------------------------------------------------------------------

\section{Results---dynamics of reconnection}  
\label{sec-results-dynamics}
%\label{subsec-dynamics}

\subsection{Basic time evolution}

\begin{figure}
\includegraphics*[width=16.cm,trim=0mm 0 0mm 0]{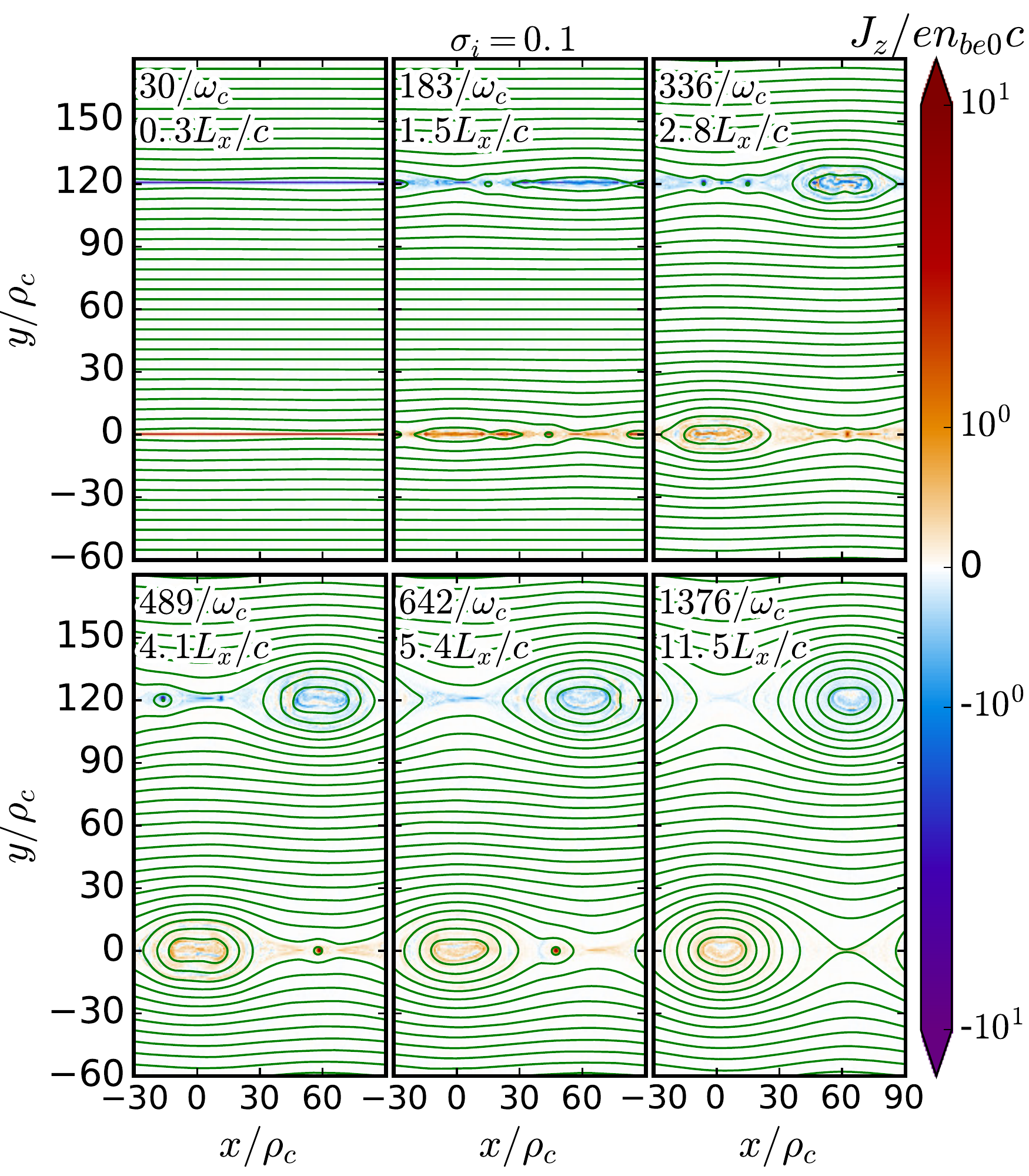}
\caption{(For $\sigma_i=0.1$) In-plane magnetic field lines (green lines) and out-of-plane current density $J_z$ (red/blue shading) at several different times, expressed in $1/\omega_c$ and also in terms of
light-crossing times $L_x/c$;
The magnetic field starts with a small perturbation (upper left) of reversing magnetic field and evolves through reconnection to a nearly steady state (lower right).
\label{fig:AzTimeEvolutionSigp1}
}
\end{figure}

\begin{figure}
\includegraphics*[width=16.cm,trim=0mm 0 0mm 0]{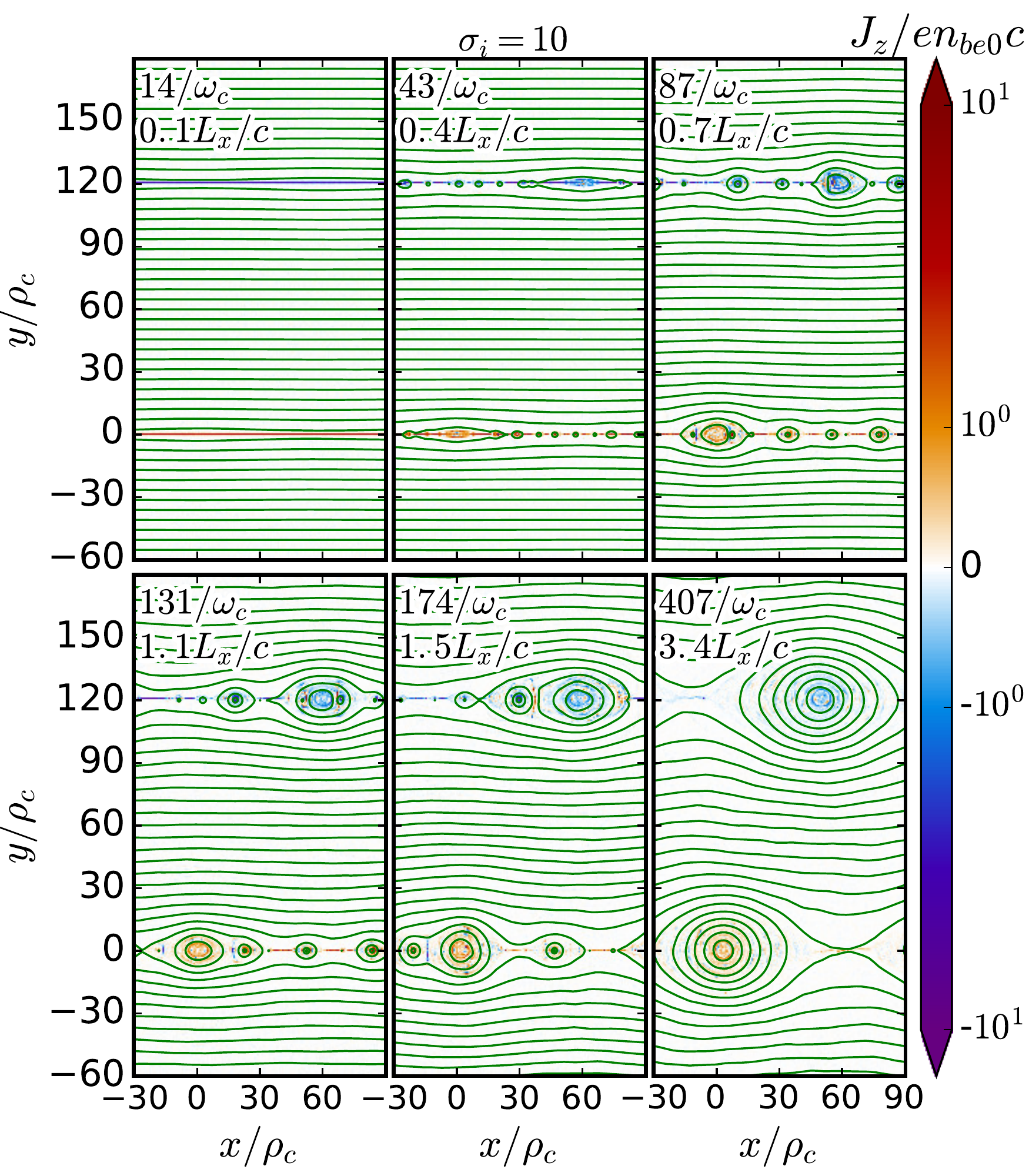}
\caption{For higher $\sigma_i=10$ (compare to Fig.~\ref{fig:AzTimeEvolutionSigp1}), the current layer is thinner (relative to $\rho_c$) and more plasmoids form during reconnection.  Anti-reconnection layers can be seen forming between merging plasmoids (e.g., lower middle plot, upper current layer,
$x/\rho_c \approx 37$)
\label{fig:AzTimeEvolutionSig10}
}
\end{figure}

The typical evolution of reconnection is described by sequences of snapshots in Figs.~\ref{fig:AzTimeEvolutionSigp1} and~\ref{fig:AzTimeEvolutionSig10}. 
These plots show magnetic flux contours (i.e., in-plane magnetic field lines) superimposed on a colour map showing the out-of-plane total electric current density~$J_z(x,y)$, at several simulation snapshots for two simulations, $\sigma_i=0.1$ and $\sigma_i=10$, both performed with our fiducial size $L_x=120\,\rho_c$.  
Since the system is closed, the reconnection process is not stationary, and reconnection eventually ceases as the system asymptotically approaches the relaxed end state.  
The initially long and thin Harris current layers ($t=0$, top left panel) that support the (slightly perturbed) reversing magnetic field $B_x$ quickly become unstable to the tearing instability, creating a chain of several first-generation, primary magnetic islands (plasmoids), clearly visible in the top centre panel (for $\sigma_i=10$ there are about seven primary plasmoids in the lower layer at $t=43/\omega_c$). Subsequently, the plasmoids undergo a complex and stochastic hierarchical evolution,
as illustrated in Figs.~\ref{fig:AzTimeEvolutionSigp1} and~\ref{fig:AzTimeEvolutionSig10}.
Secondary current sheets form between the plasmoids; the plasmoids grow, fed by reconnection in the secondary current sheets and plasmoid mergers/coalescence; meanwhile, new plasmoids form as a result of tearing instability in secondary current sheets.
This process proceeds for a few global light crossing times, until the free magnetic energy for reconnection is exhausted; eventually the number of plasmoids decreases as they merge with each other, and the system reaches the final tearing-stable equilibrium state with just one large magnetic island (plasmoid) and one significant X-point in each of the two initial layers (lower right panels of Figs.~\ref{fig:AzTimeEvolutionSigp1} and \ref{fig:AzTimeEvolutionSig10})---these final major X-points are relaxed, with separatrices crossing at nearly 90 degree angles.

\begin{figure}
\includegraphics*[width=16.cm,trim=0mm 1mm 0mm 0]{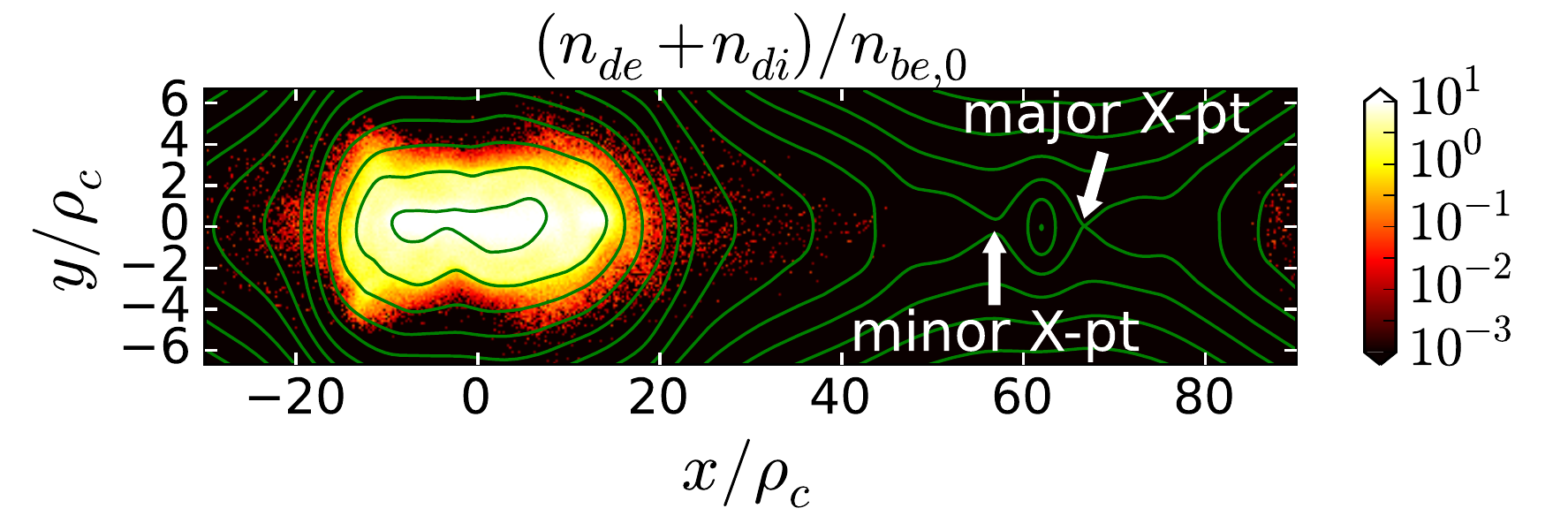}\\
\caption{
The combined number density of initially-drifting electrons and ions (normalized to the background electron density) in a narrow region around the lower layer,
for $\sigma_i=0.1$ at $t\omega_c=337$.  By this time, the initially-drifting particles have been almost entirely swept up into the major plasmoid around $x=0$, and are nearly absent from the region containing active X-points.
\label{fig:ndrift}
}
\end{figure}

In plasmoid-dominated reconnection, such as we see in this study, there typically exist, except at the end of reconnection, more than one X- and O-point in each layer.
It is useful to define the major X-point and O-point in each layer as the extrema of the flux function; however, the major X-point is not necessarily the only or even the most active X-point in terms of the reconnection electric field $E_z$ at any given time (although it is usually one of the most active X-points).
For example, if we zoom-in on the lower layer, as in Fig.~\ref{fig:ndrift}, the major O-point is immediately apparent (near $x=0$), and there is another prominent O-point near $x=60\rho_c$ flanked  by two prominent X-points; here (and in subsequent figures) we have drawn the magnetic field line (separatrix) that goes directly through the major X-point (to the right of the minor O-point, at $x\approx 67 \rho_c$).

\begin{figure}
(a)%
\includegraphics*[width=16.cm,trim=0mm 17.7mm 0mm 0]{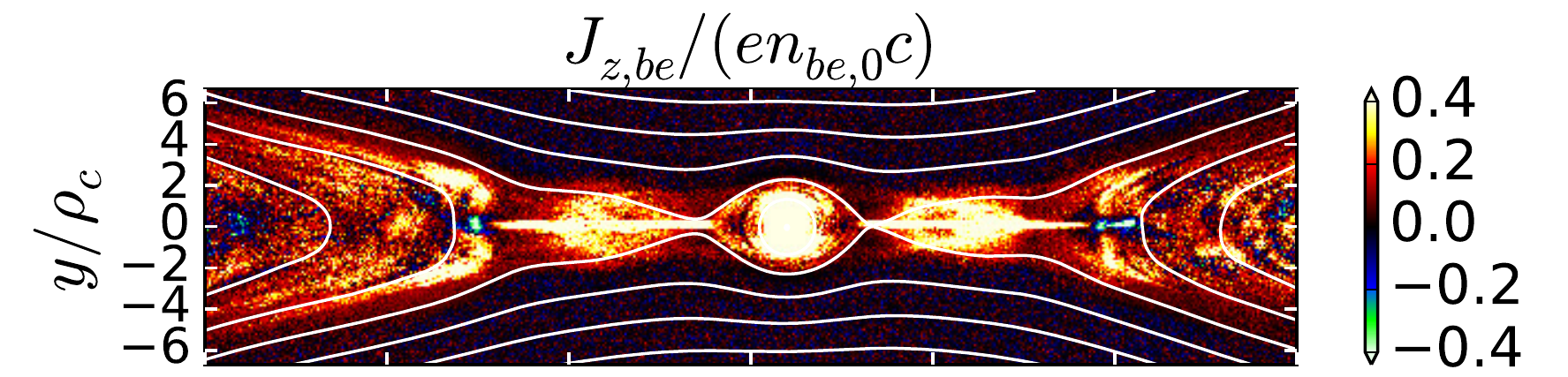}\\
(b)%
\includegraphics*[width=16.cm,trim=0mm 17.7mm 0mm 0]{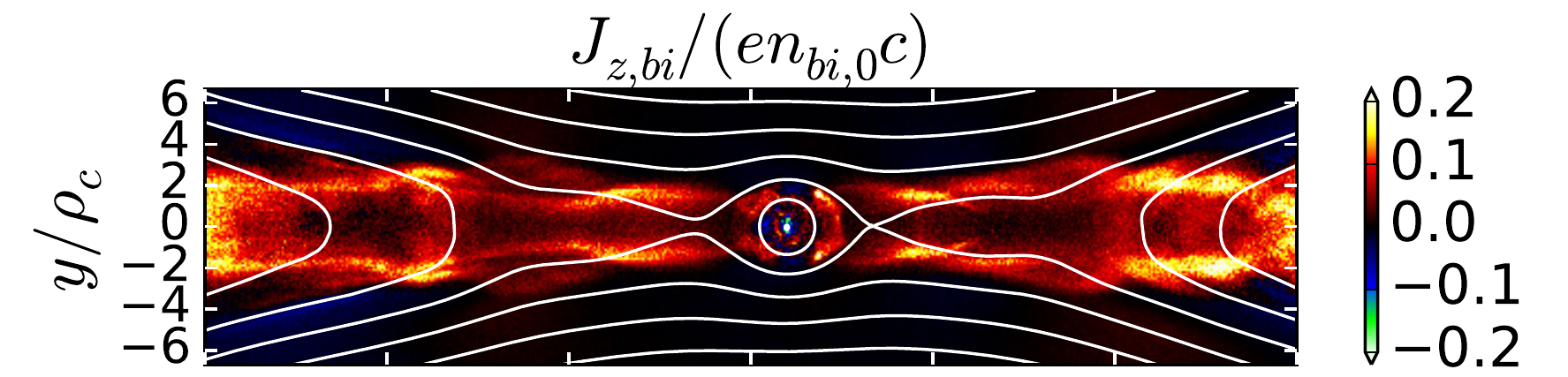}\\
(c)%
\includegraphics*[width=16.cm,trim=0mm 17.4mm 0mm 0]{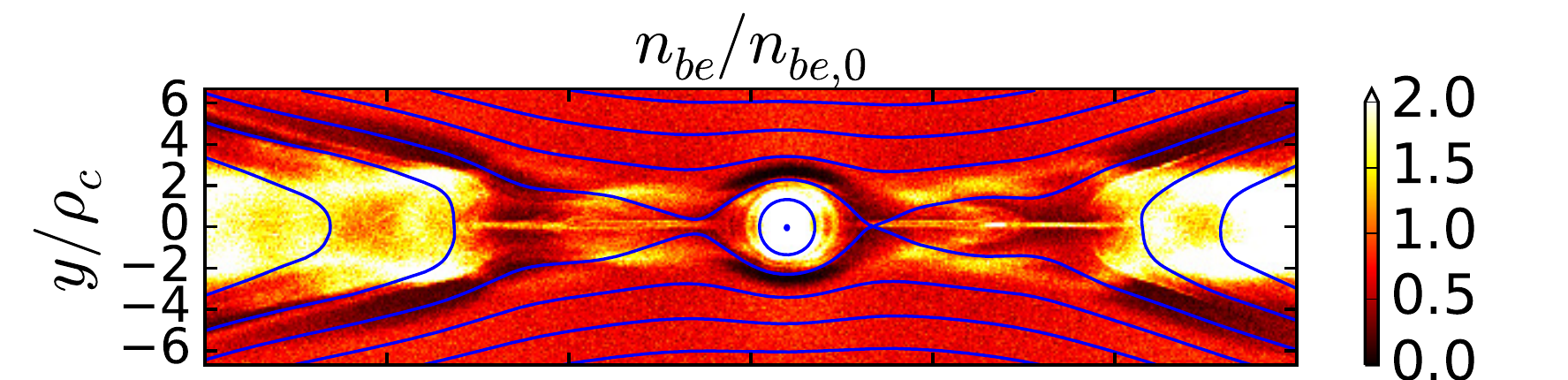}\\
(d)%
\includegraphics*[width=16.cm,trim=0mm 0 0mm 0]{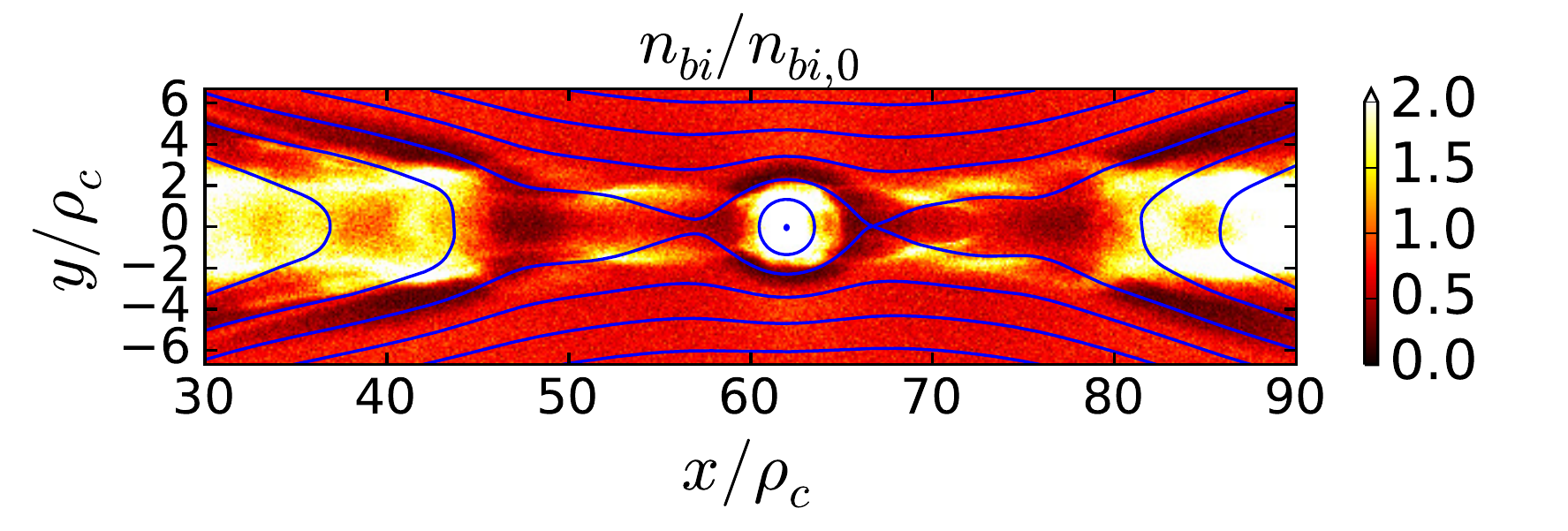}\\
\caption{For $\sigma_i=0.1$ and $L_x=120\rho_c$, the above plots show the current ($J_z$) and number density for background electrons and ions, in a narrow region ($-6 < y/\rho_c < 6$) around the right half of the lower layer at $t\omega_c=337$ ($t \approx 2.8 L_x/c$); at this time all the particles in the original drifting layer are in the largest plasmoids (not shown here). 
\label{fig:nAndJz}
}
\end{figure}

After the onset of reconnection the drifting particles in the initial Harris layers are quickly swept out of the initial X-points and trapped around the major O-points, as shown in Fig.~\ref{fig:ndrift} at $t=337\,\omega_c^{-1} \approx 2.8\, L_x/c$ for $\sigma_i=0.1$, where one can see that even the prominent minor O-point contains no initially-drifting particles. 
Meanwhile, background particles that flow into the layer are accelerated (near X-points, where $B_x=B_y=0$) in the $\pm \hat{\bf z}$-direction by the reconnection electric field $E_z$ to maintain the current sheets supporting the~$B_x$ reversal.  Figure~\ref{fig:nAndJz} zooms-in on the background electron and ion currents and densities in the region near the major (and also a prominent minor) X-point in the lower layer at $t\omega_c=377$ for $\sigma_i=0.1$ (cf. Fig.~\ref{fig:AzTimeEvolutionSigp1}, upper right).   
With the initially-drifting particles already wrapped up in the major plasmoid, the background particles are entirely responsible for subsequent secondary current sheets and plasmoids, such as in Fig.~\ref{fig:nAndJz}.
In time, the small plasmoid at the centre of Fig.~\ref{fig:nAndJz} (and any other newly created plasmoids) will grow and merge with the major plasmoid, leaving a single X-point and O-point in the layer as reconnection winds down.

\begin{figure}
\centering
\includegraphics*[width=8.cm,trim=0mm 0 0mm 0]{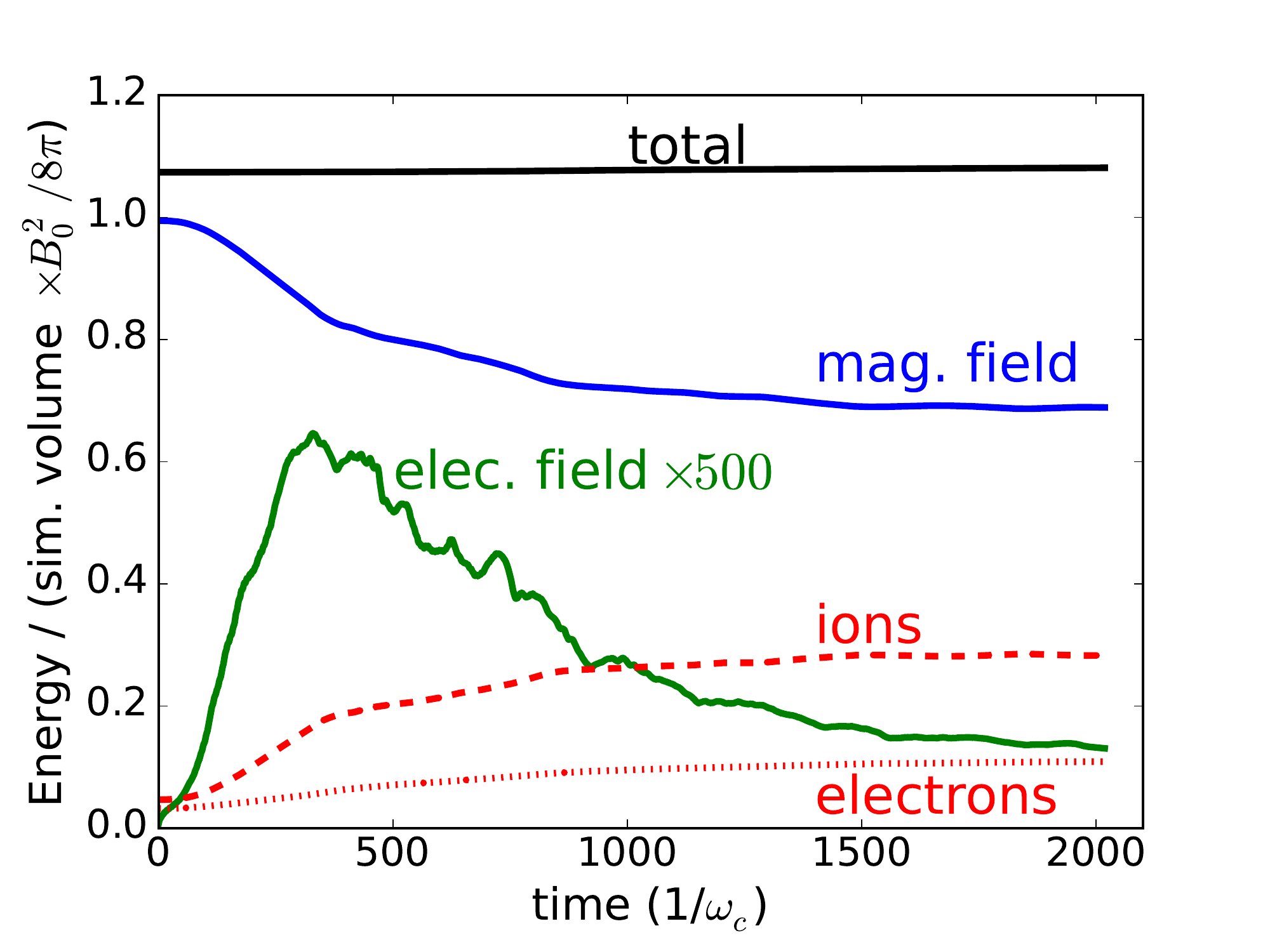}
\caption{
The energy versus time for field and particle components (with the electric field energy magnified by $500\times$),
for $\sigma_i=0.1$ and $L_x=120 \sigma_{i} \rho_{i0}$.  
\label{fig:energyVsTime}
}
\end{figure}

The evolution of the basic energetics of the system is illustrated in Fig.~\ref{fig:energyVsTime}, which shows different forms of energy---magnetic, electric, and particle kinetic---as functions of time. Since the system is closed, the total energy in the box is conserved,% 
\footnote{The PIC algorithm does not conserve energy exactly, but does conserve it approximately; 
all the simulations presented here conserved energy to better than 1\% over the course of reconnection,
except for the two $L_x=160\rho_c$ simulations, which experienced energy non-conservation of~1.5\% ($\sigma_i=0.1$) and~1.0\% ($\sigma_i=1$).
}
as the magnetic energy is converted to particle kinetic energy. 
The electric energy indicates the active reconnection phase, decaying as the reconnection process slows down.

\begin{figure*}
\includegraphics*[width=8.cm,trim=0mm 0 0mm 0]{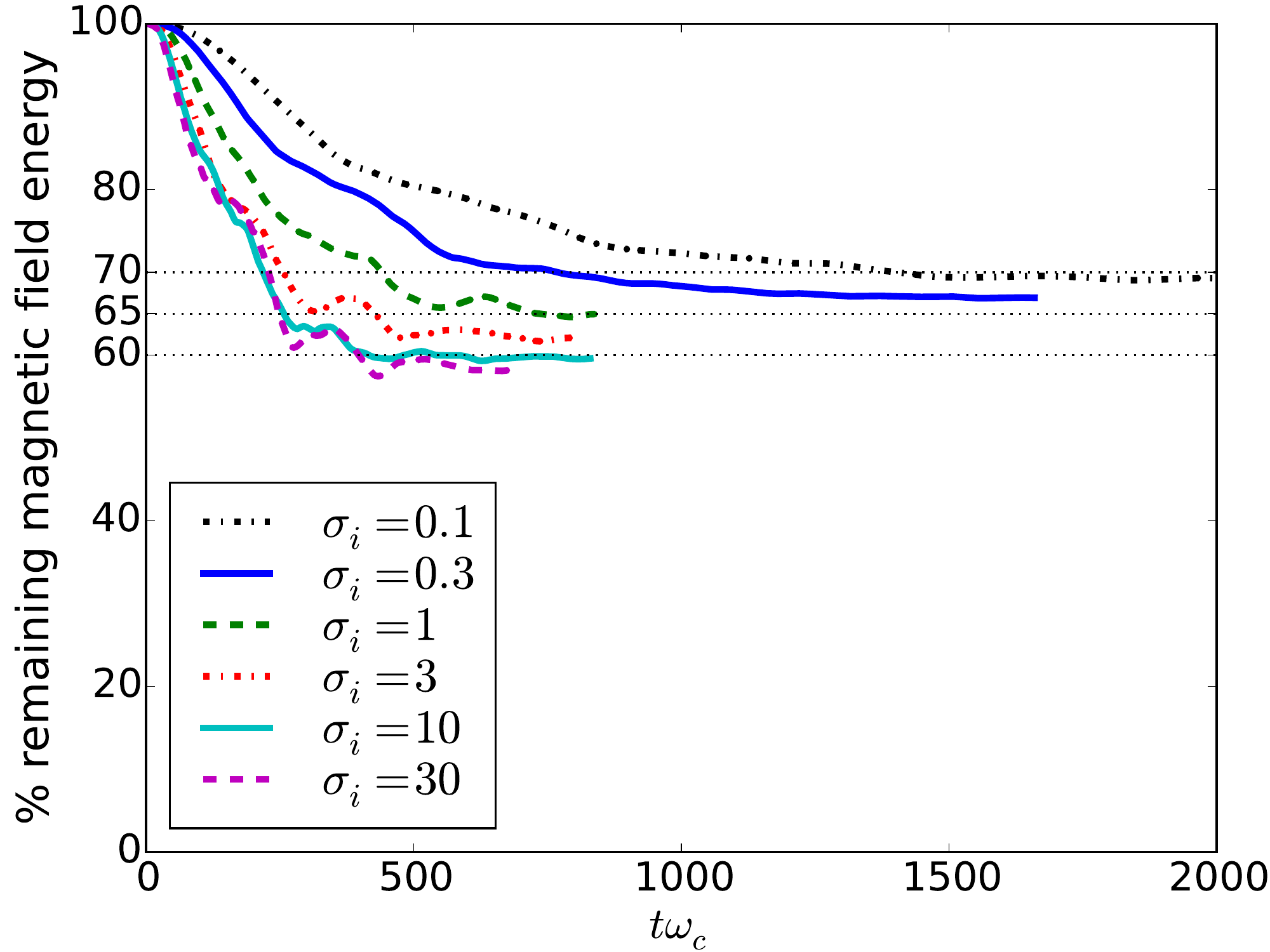}%
\includegraphics*[width=8.cm,trim=0mm 0 0mm 0]{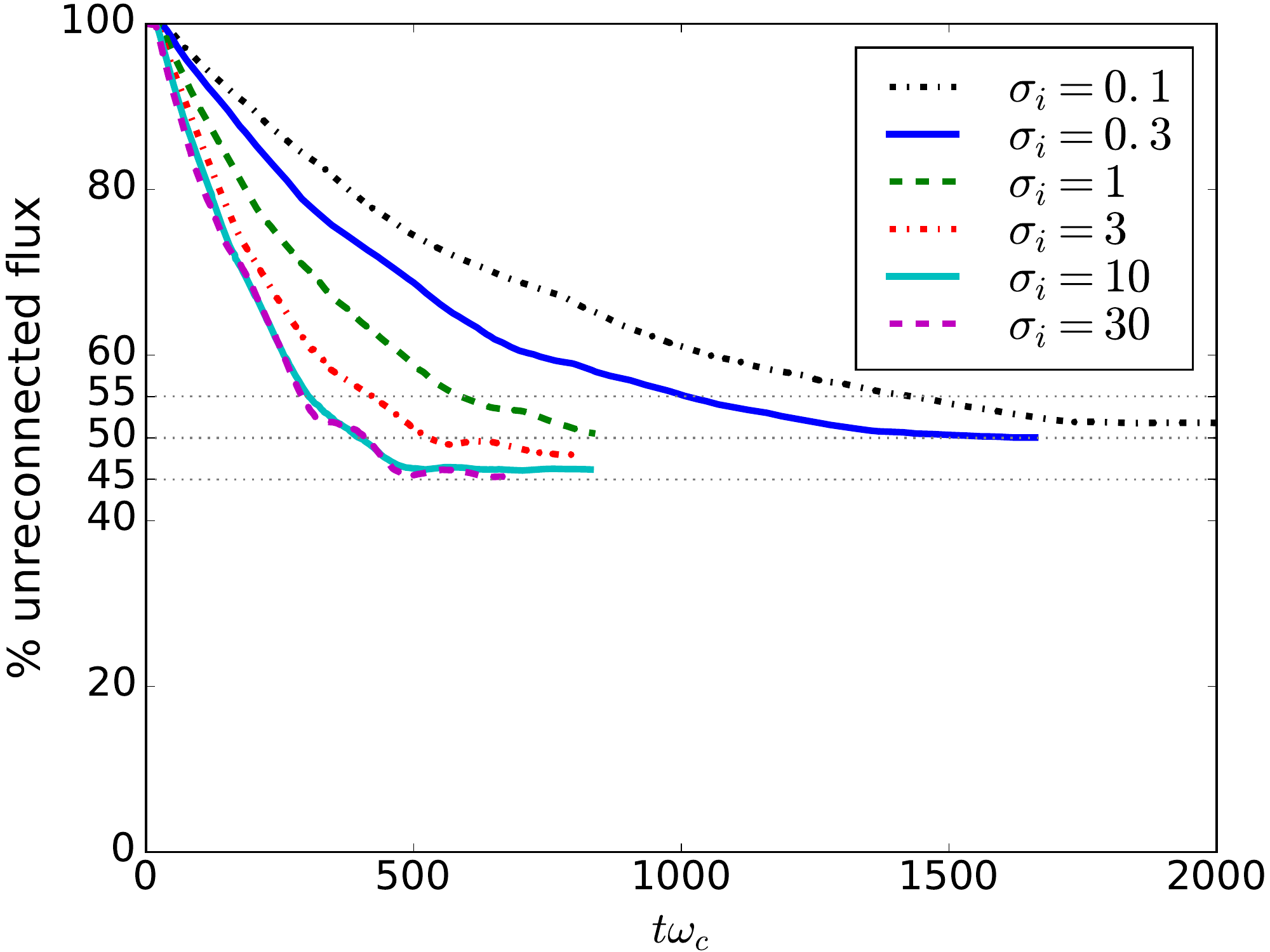}
\caption{
Time evolution of the percentage of remaining magnetic energy (left) and unreconnected flux $\Psi(t)/\Psi(0)$ (right),
for simulations with a range of $\sigma_i$.
\label{fig:reconFluxVsSigma}
}
\end{figure*}

\begin{figure*}
\includegraphics*[width=8.cm,trim=0mm 0 0mm 0]{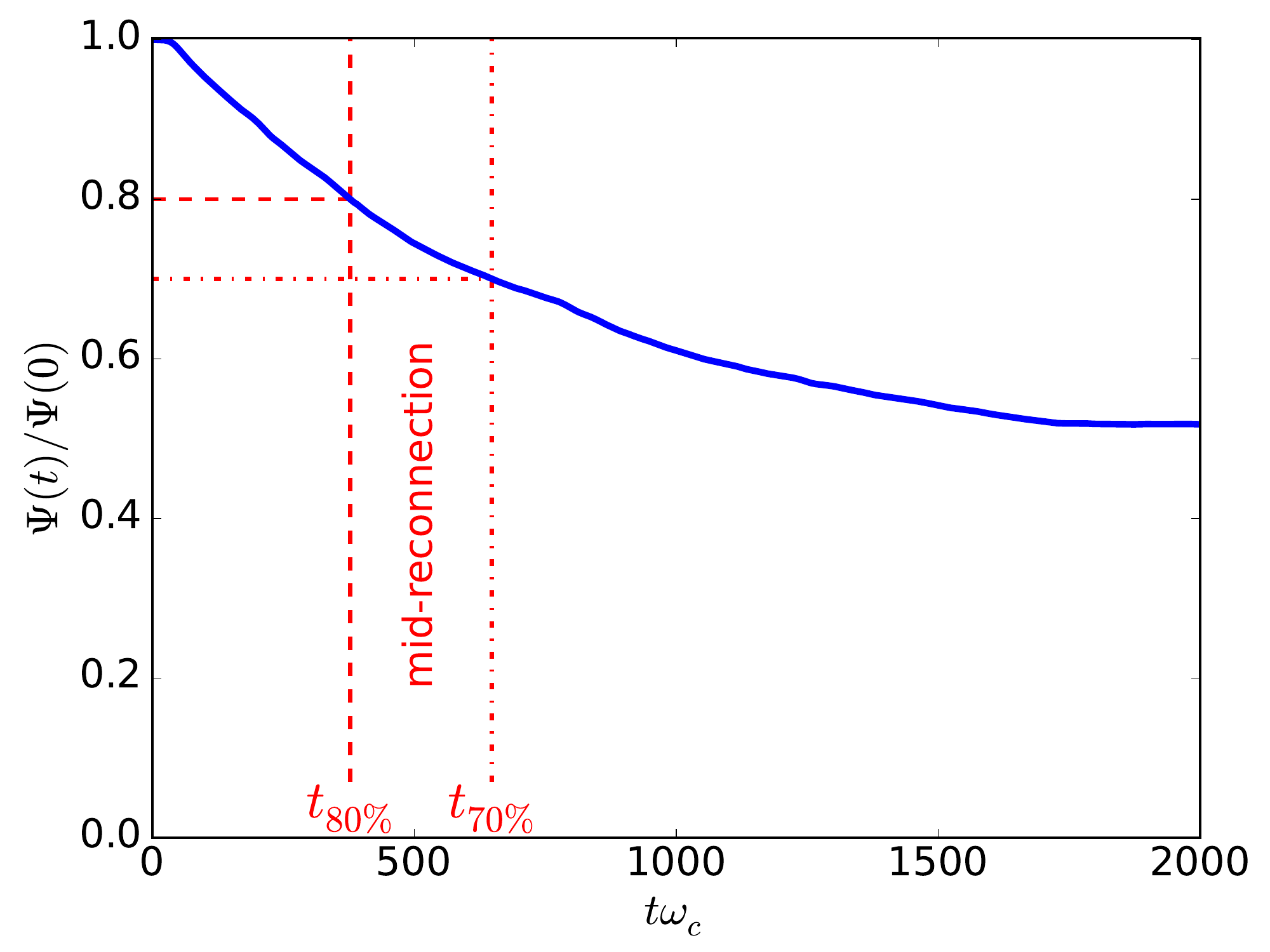}
\includegraphics*[width=8.cm,trim=0mm 0 0mm 0]{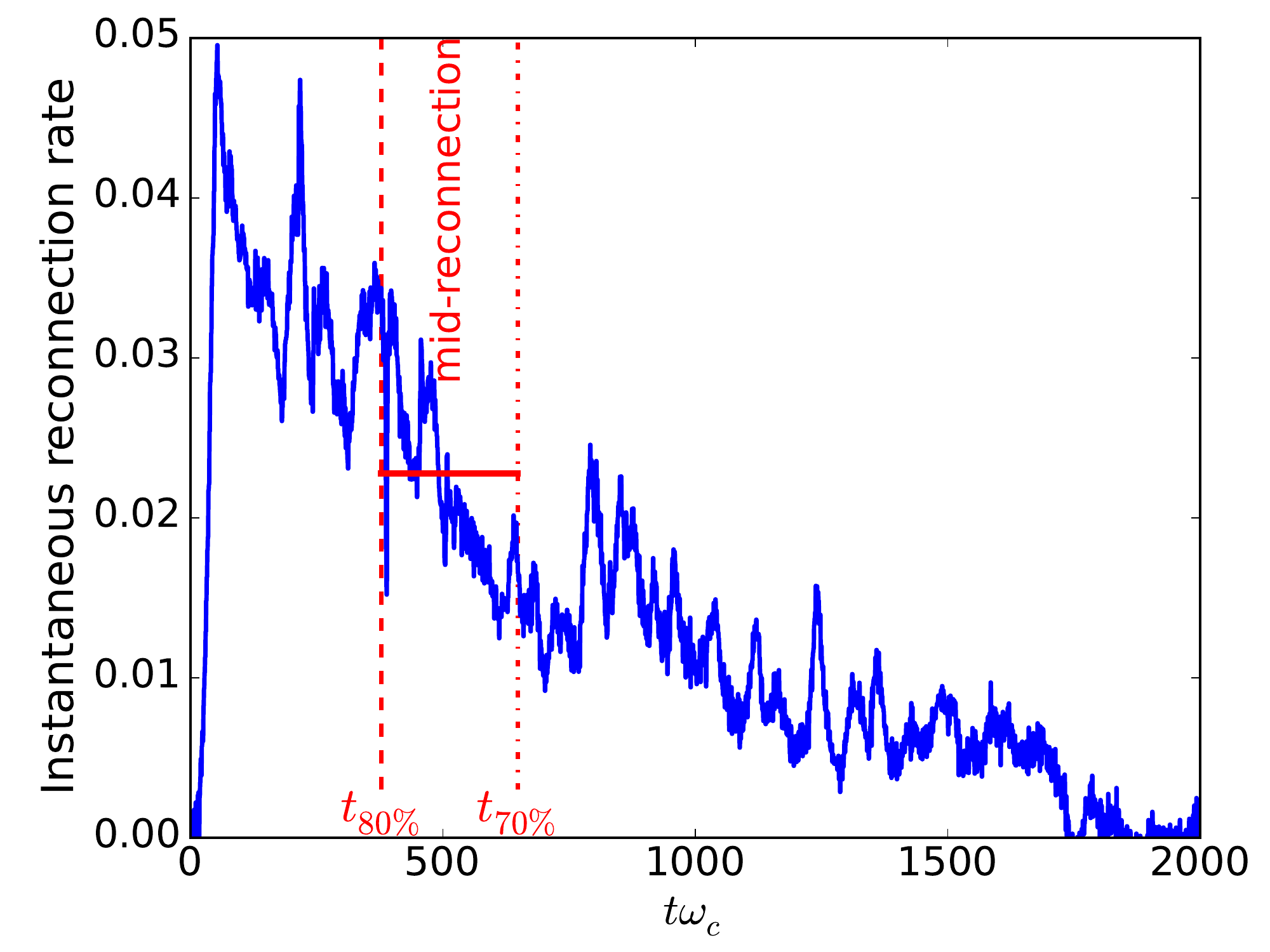}
\caption{
Time evolution of the normalized unreconnected flux $\Psi(t)$ (left panel) for a simulation with $\sigma_i=0.1$; the right panel shows the corresponding normalized reconnection rate $E_z(t)/B_0$, with a horizontal red line indicating the average value $\beta_{\rm rec}$ over the mid-reconnection period. 
\label{fig:reconRate}
}
\end{figure*}

The reconnection process 
reduces the unreconnected magnetic flux (associated mostly with~$B_x$)
between the major X-points in the two current layers, while increasing the reconnected flux (mostly $B_y$) between the major X- and O-points within each layer.
Mathematically, we define the full unreconnected flux $\Psi(t)$ (which is twice the unreconnected flux per current layer) 
as the total integrated magnetic flux between the two major X-points in the upper and lower layers (or, equivalently, the difference in $z$-component of the vector electromagnetic potential at the two major X-points).  Figure~\ref{fig:reconFluxVsSigma} (right-hand panel) shows how $\Psi(t)$ decays during reconnection, for several different~$\sigma_i$.   

Since our closed system does not allow steady-state reconnection forever, 
we analyse reconnection dynamics during the period of `mid-reconnection'---after initial transients but long before the end of reconnection.  
Specifically, we define mid-reconnection (shown between dashed lines in
Fig.~\ref{fig:reconRate}) as the time interval $[t_{80\%},t_{70\%}]$ such that $\Psi(t_{80\%})=0.8\, \Psi(0)$ and $\Psi(t_{70\%})=0.7\,  \Psi(0)$, i.e., starting when the unreconnected flux drops to 80\% of its initial value and ending when it reaches 70\%.  Because of the aspect ratio $L_y/L_x=2$,  the final system state retains a band of unreconnected flux between the layers---around 50\% of the original unreconnected flux (Fig.~\ref{fig:reconFluxVsSigma}, right)---and therefore mid-reconnection is approximately the middle fifth of active reconnection (as measured by reconnected flux).
The final state also retains about 60--70\% of the initial magnetic energy still in magnetic form (Fig.~\ref{fig:reconFluxVsSigma}, left).

The reason why we define mid-reconnection in terms of the initial flux (rather than, e.g., half-way between initial and final unreconnected flux) is that it is difficult determine the precise final value of unreconnected flux.
At the end of each simulation, a slow, relatively small-amplitude `breathing' mode alternately reconnects and anti-reconnects magnetic flux between the remaining X-point and O-point in each layer, resulting in a mild oscillation of unreconnected flux and magnetic energy.  
Determining precise final values would thus require running simulations for several extra light-crossing times to determine the centre of the oscillation; to save computation time, we therefore use well-defined initial values to determine the mid-reconnection time interval.

We emphasize that the fiducial system size $L_x=120\rho_c$ is large enough that most of the particles forming the initially-drifting Harris sheet have been trapped around the major O-points by the beginning of mid-reconnection.  
This can be seen in Fig.~\ref{fig:ndrift}, a snapshot of the initially-drifting particle density at the beginning of the mid-reconnection interval for $\sigma_i=0.1$, namely $t=337 \omega_c^{-1}$.
Therefore, we believe that, during and after the period of mid-reconnection, X-point and plasmoid dynamics (except when involving the major plasmoid) are controlled primarily by the upstream (background) environment, and are much less influenced by the initial Harris sheets.

%----------------------------------------------------------------------------------

\subsection{Rate of reconnection}
\label{subsec-recn_rate}

One of the main quantities of interest in any reconnection study is the reconnection rate, a measure of how rapidly the magnetic flux is transferred from the upstream (unreconnected) region to the downstream (reconnected) region and, correspondingly, how rapidly the available magnetic energy is converted into plasma energy.  
The instantaneous reconnection rate is essentially the time derivative of the unreconnected flux $\Psi(t)$ between the two layers (times $-1/2$ because reconnection proceeds in both layers and we want the rate to be defined as a positive quantity). 
The corresponding dimensionless instantaneous reconnection rate is then $- (1/2B_0 c)\, d\Psi/dt $.
However, $d\Psi/dt$ exhibits both a complex chaotic short-time behaviour due to multiple X-points and plasmoid dynamics, and a secular long-time behaviour due the exhaustion of the unreconnected magnetic flux and magnetic free energy (see Fig.~\ref{fig:reconRate}, right panel).  
Therefore, constructing a robust and simple measure of reconnection rate requires some care; in this paper, we characterize `the' dimensionless reconnection rate for a given simulation as the normalized time-average $\langle -d\Psi(t)/dt \rangle /(2cB_0)$ during the mid-reconnection interval:
$\beta_{\rm rec} = (1/2B_0 c) [\Psi(t_{80\%})-\Psi(t_{70\%})]/(t_{70\%}-t_{80\%})$.
This reconnection rate corresponds to a reconnection inflow velocity of $v_{\rm rec} = \beta_{\rm rec} c$ and a reconnection electric field $E_{\rm rec} = \beta_{\rm rec} B_0$; with Yee finite-difference electromagnetics \citep{Yee:1966}, $d\Psi/dt$ is exactly equivalent to the sum of out-of-plane electric fields at the two major X-points (by Faraday's Law), and so $E_{\rm rec}$ is the average electric field $E_z$ at the major X-points during mid-reconnection.

\begin{figure}
\begin{tabular}{ll}
\includegraphics*[width=8.cm,trim=0mm 0 0mm 0]{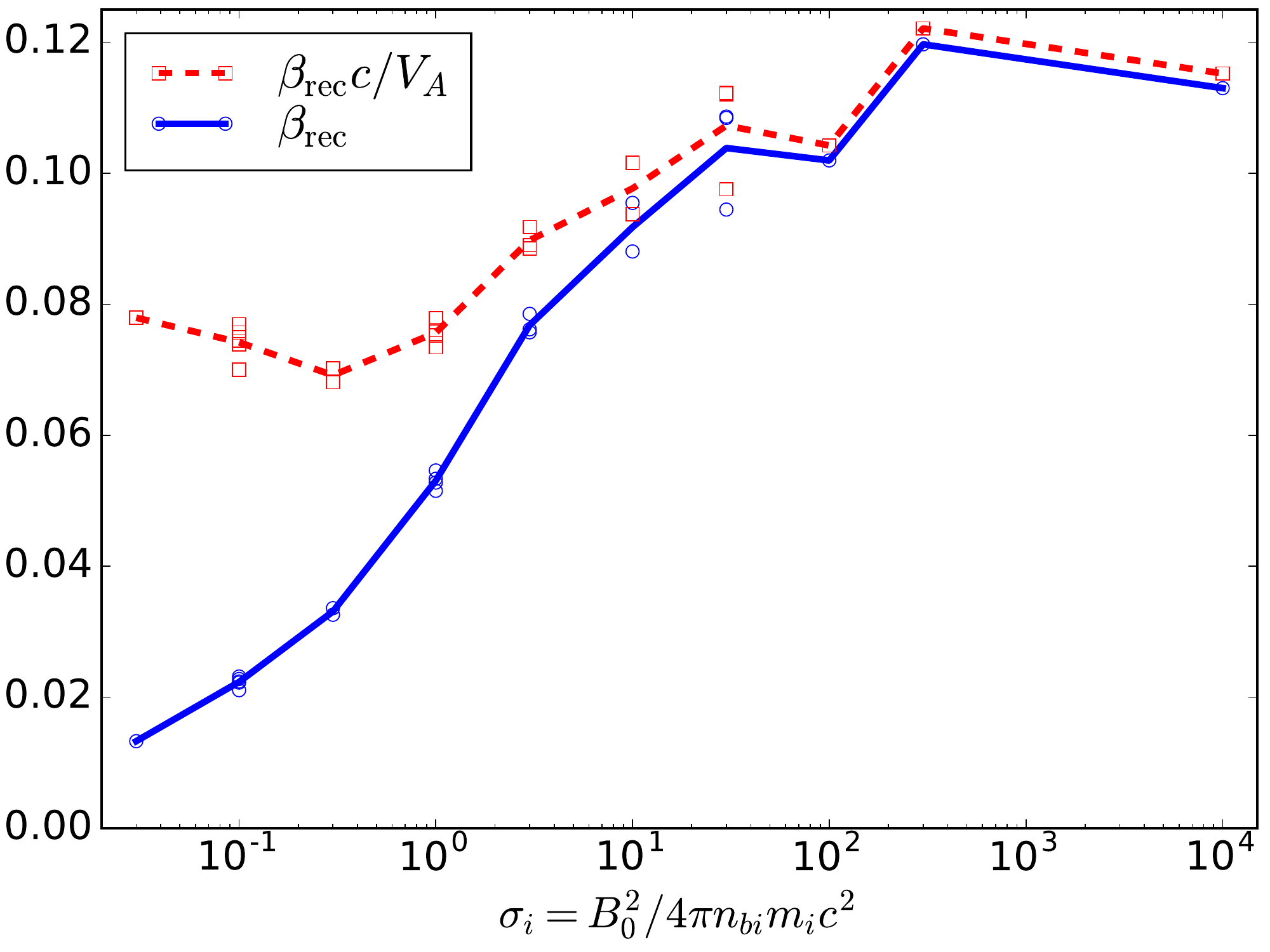}%
\end{tabular}
\caption{
The dimensionless reconnection rate versus~$\sigma_i$ (blue, solid line), normalized by the relativistic Alfv\'{e}n velocity (red, dashed).
Each data point represents a separate simulation.
\label{fig:betar}
}
\end{figure}

With this characterization of the dimensionless reconnection rate $\beta_{\rm rec}$, we can see how $\beta_{\rm rec}$ depends on various input parameters of the reconnecting system---most importantly, on $\sigma_i$, as shown in Fig.~\ref{fig:betar}. 
We find that $\beta_{\rm rec}$ grows with $\sigma_i$ for $\sigma_i \lesssim 10$, while approaching a finite asymptotic value~$\sim 0.1$ in the ultrarelativistic-ion limit $\sigma_i \gg 1$. The latter is broadly consistent with the results of previous studies of ultrarelativistic reconnection in both pair and electron-ion plasmas \citep{Sironi_Spitkovsky-2014,Guo:2014,Guo:2015,Melzani:2014a}.
However, our results also show that the variation of $\beta_{\rm rec}$ in the semirelativistic (small $\sigma_i$) regime can be attributed almost entirely to the scaling of the upstream relativistic Alfv\'en speed 
$V_A = c \sigma_{\rm hot}^{1/2} /(1+ \sigma_{\rm hot})^{1/2}
\approx c \sigma_{i}^{1/2} /(1+ \sigma_{i})^{1/2}$ with $\sigma_i$ \citep[cf.][]{Melzani:2014a}.
That is, if one normalizes the reconnection inflow velocity $v_{\rm rec} =c \beta_{\rm rec}$ by the Alfv\'en speed rather than by the speed of light, and correspondingly normalizes the reconnection electric field by $V_A B_0/c$, then the result becomes essentially independent of~$\sigma_i$.
This properly normalized average reconnection rate, $\beta_{{\rm rec},A} \equiv c \beta_{\rm rec} /V_A$, varies only modestly across the physical regimes of interest, namely, within the range from about 0.07 to~0.12 (i.e., within about 50\%), over a broad range of $\sigma_i$ spanning more than 3 orders of magnitude (see Fig.~\ref{fig:betar}).

\begin{figure}
\includegraphics*[width=8.cm,trim=0mm 0 0mm 0]{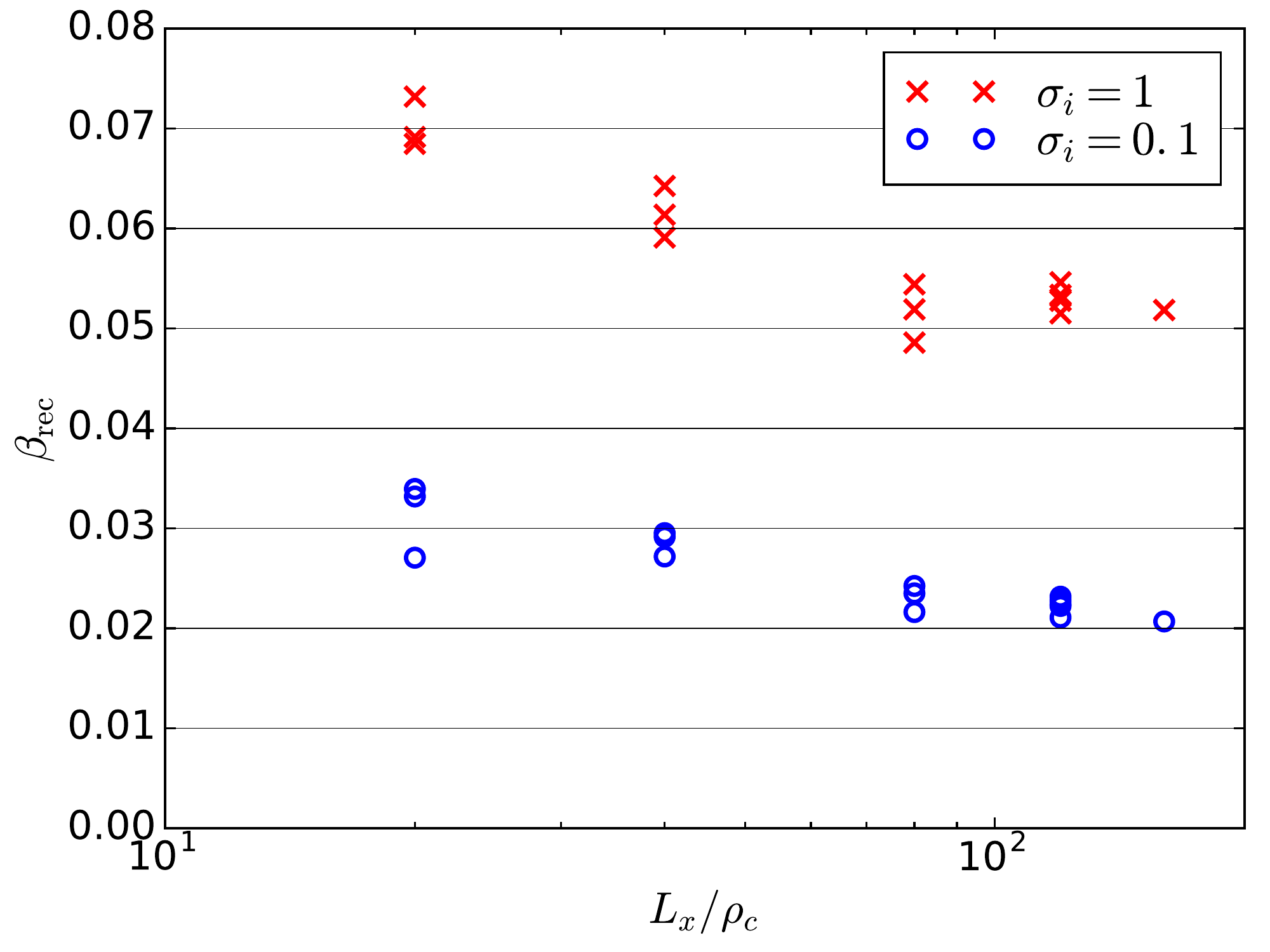}%
\caption{
The dimensionless reconnection rate $\beta_{\rm rec}$ versus system size $L_x$,
for $\sigma_i=0.1$ and~1.  Each data point represents a separate
simulation.
\label{fig:betarVsL}
}
\end{figure}

We also find that the reconnection rate has a weak dependence on the system size $L_x$ for small $L_x$, but becomes  independent of $L_x$ for $L_x \gtrsim 80\rho_c$, as demonstrated by Fig.~\ref{fig:betarVsL}.
The slightly higher reconnection rate for small systems may be due to the residual effects of the initial Harris current sheets.  
Figure~\ref{fig:betarVsL} also shows the statistical variation of the reconnection rate in different simulations with the same $L_x$ and $\sigma_i$ which can be attributed to the stochastic behaviour of plasmoids.  
Larger systems yield more consistent reconnection rates by virtue of averaging over longer times (over more X-points and plasmoids);  
for $L_x=120 \rho_c$ measurements of $\beta_{\rm rec}$ have an estimated statistical uncertainty on the order of~10\%.

%----------------------------------------------------------------------------------

\subsection{Plasma flows}
\label{sec:PlasmaFlows}

\begin{figure}
\raggedbottom
\begin{minipage}[c]{0.25in}(a)\end{minipage}%
\begin{minipage}[c]{16cm}
\includegraphics*[width=16.cm,trim=0mm 17.3mm 0mm 0]{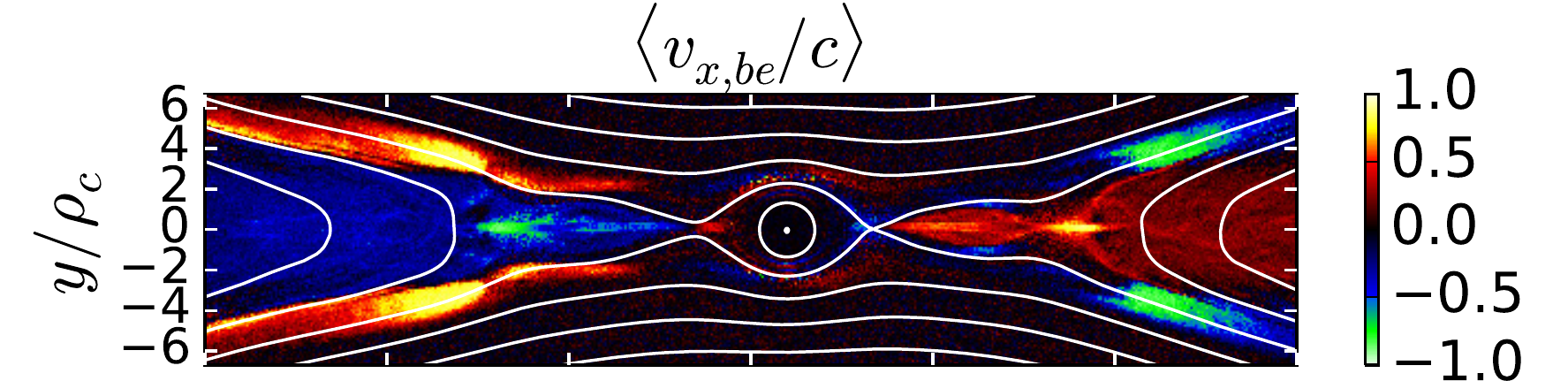}
\end{minipage}\\
\begin{minipage}[c]{0.25in}(b)\end{minipage}%
\begin{minipage}[c]{16cm}
\includegraphics*[width=16.cm,trim=0mm 17.7mm 0mm 0]{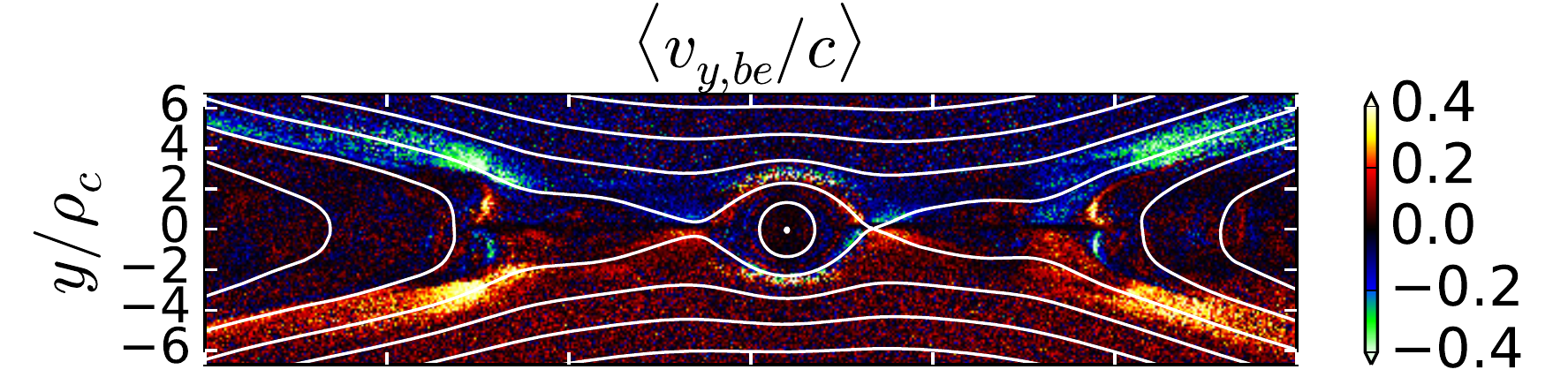}
\end{minipage}\\
\begin{minipage}[c]{0.25in}(c)\end{minipage}%
\begin{minipage}[c]{16cm}
\includegraphics*[width=16.cm,trim=0mm 2.5mm 0mm 0]{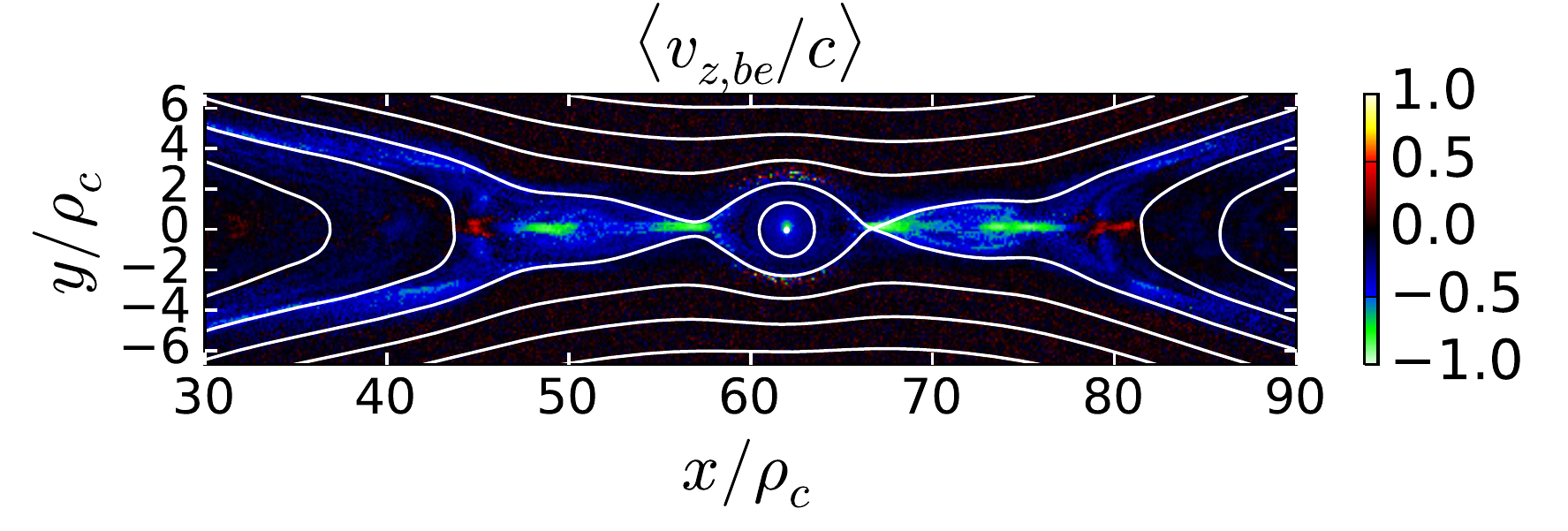}
\end{minipage}\\
\caption{Bulk (Eckart) flow three-velocities of background electrons for $\sigma_i=0.1$  and $t\omega_c=337$.
\label{fig:electronFlows}
}
\end{figure}

\begin{figure}
\raggedbottom
\begin{minipage}[c]{0.25in}(a)\end{minipage}%
\begin{minipage}[c]{16cm}
\includegraphics*[width=16.cm,trim=0mm 18mm 0mm 0]{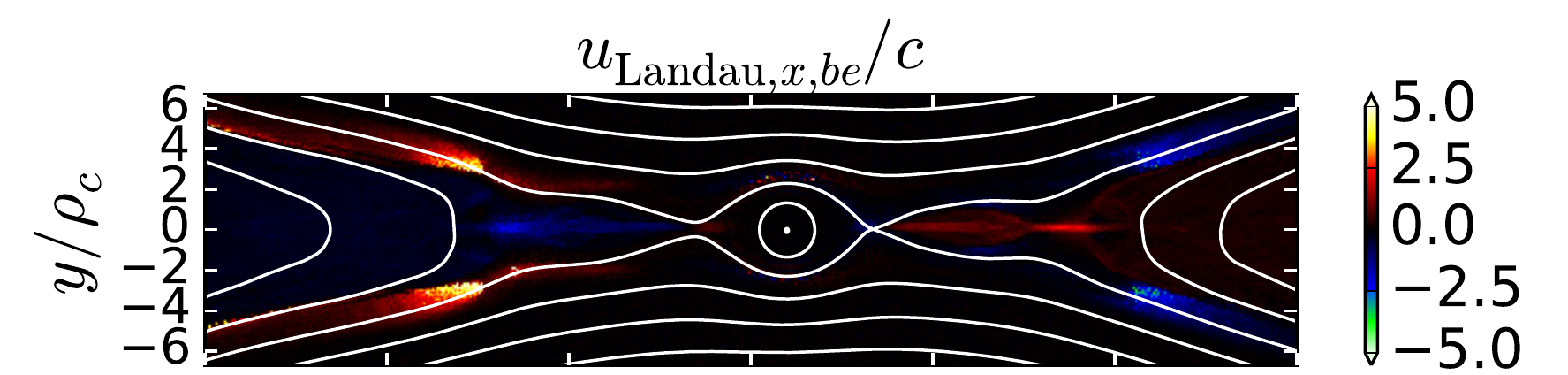}
\end{minipage}\\
\begin{minipage}[c]{0.25in}(b)\end{minipage}%
\begin{minipage}[c]{16cm}
\includegraphics*[width=16.cm,trim=0mm 17.2mm 0mm 1mm]{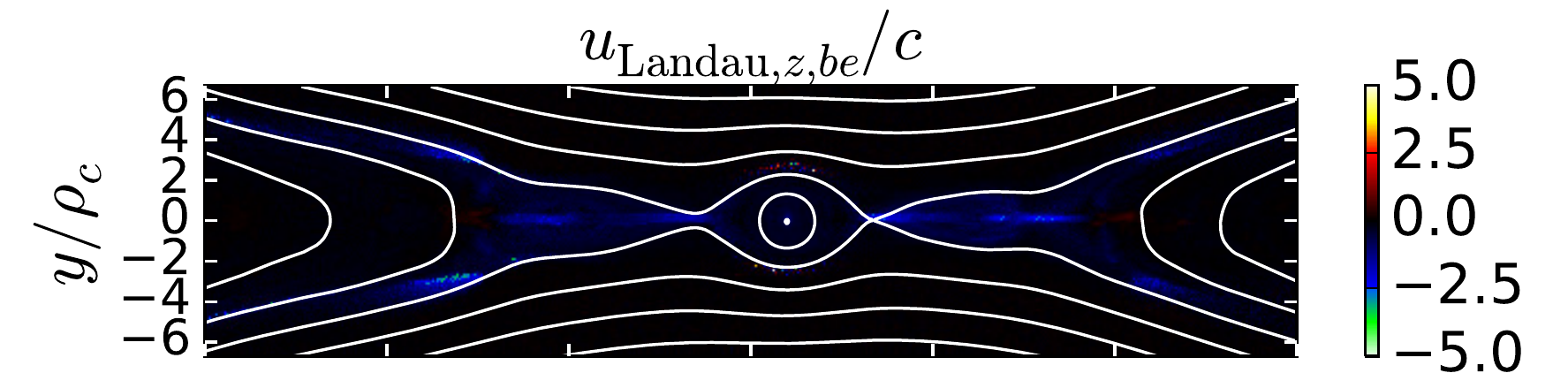}
\end{minipage}\\
\begin{minipage}[c]{0.25in}(c)\end{minipage}%
\begin{minipage}[c]{16cm}
\includegraphics*[width=16.cm,trim=0mm 17.3mm 0mm 0]{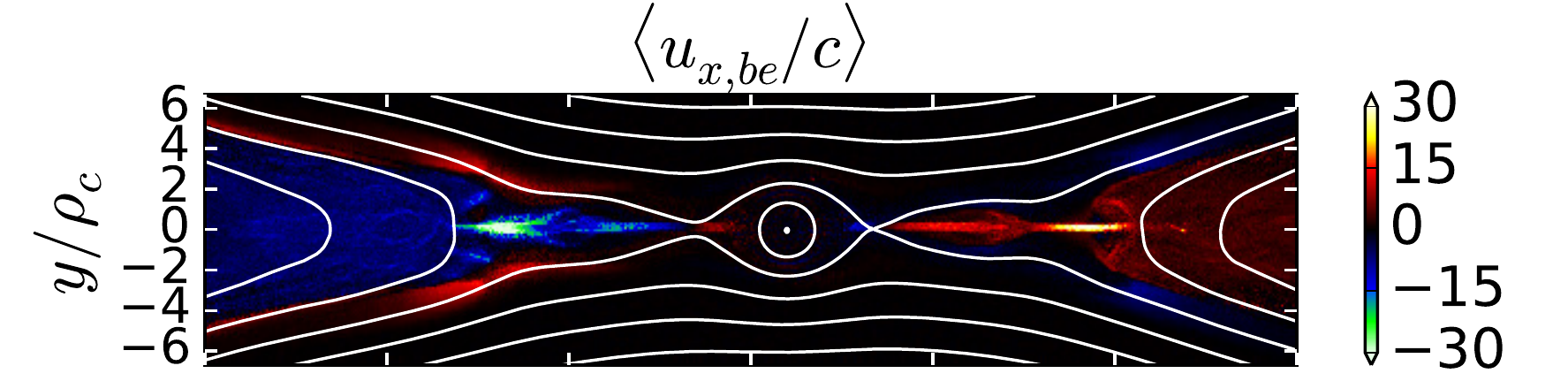}
\end{minipage}\\
\begin{minipage}[c]{0.25in}(d)\end{minipage}%
\begin{minipage}[c]{16cm}
\includegraphics*[width=16.cm,trim=0mm 2.5mm 0mm 0]{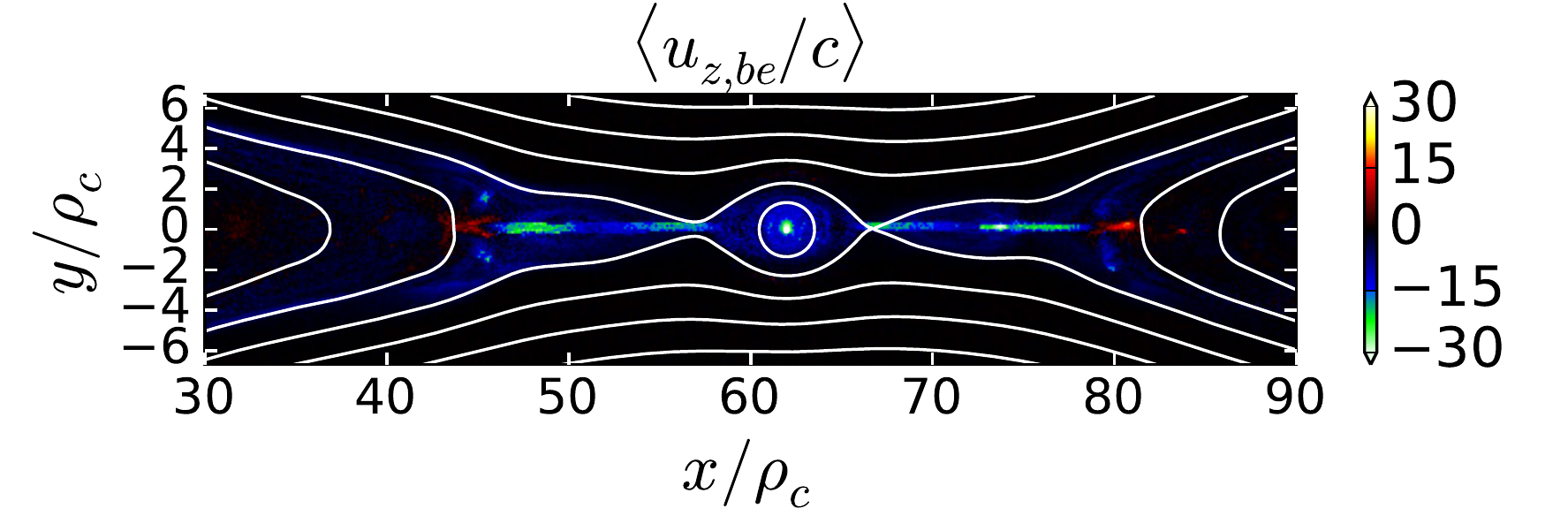}
\end{minipage}\\
\caption{Bulk Landau and average four-velocities ($x$-and $z$-components) of background electrons for $\sigma_i=0.1$  and $t\omega_c=337$ ($y$-velocities are not very relativistic). 
\label{fig:electronRelFlows}
}
\end{figure}

\begin{figure}
\begin{minipage}[c]{0.25in}(a)\end{minipage}%
\begin{minipage}[c]{16cm}
\includegraphics*[width=16.cm,trim=0mm 17.7mm 0mm 0]{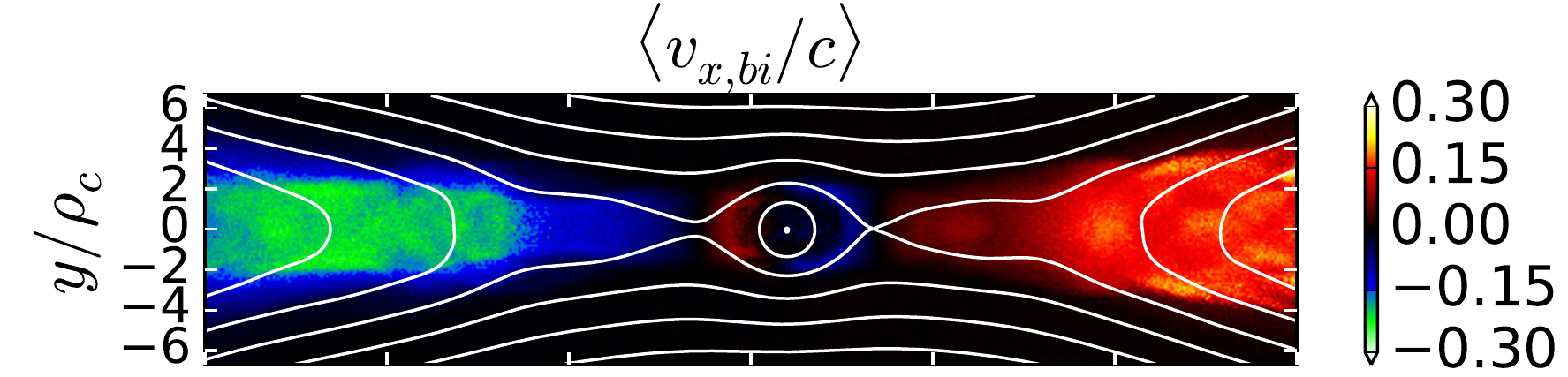}
\end{minipage}\\
\begin{minipage}[c]{0.25in}(b)\end{minipage}%
\begin{minipage}[c]{16cm}
\includegraphics*[width=16.cm,trim=0mm 17.7mm 0mm 0]{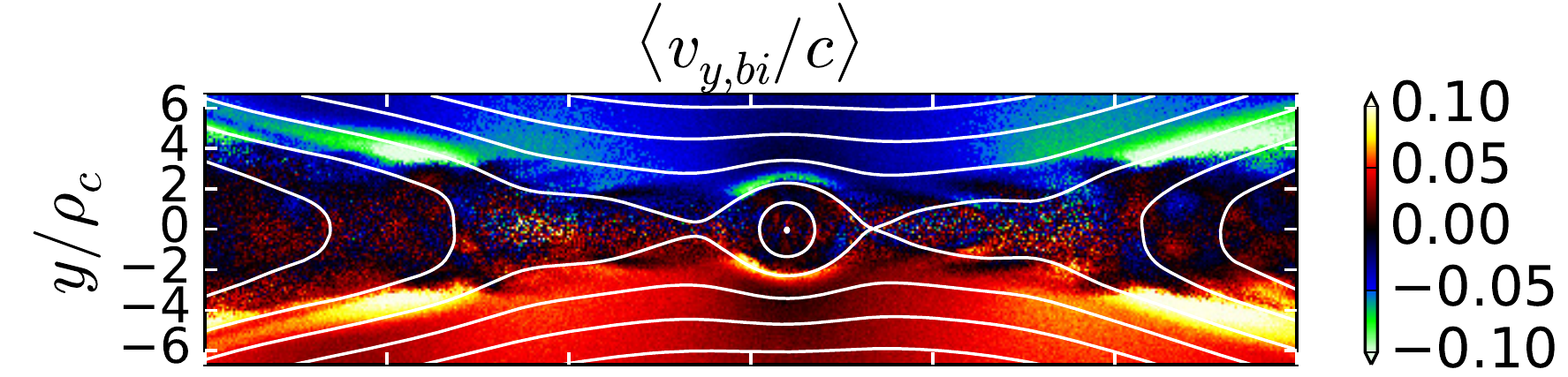}
\end{minipage} \\
\begin{minipage}[c]{0.25in}(c)\end{minipage}%
\begin{minipage}[c]{16cm}
\includegraphics*[width=16.cm,trim=0mm 0 0mm 0]{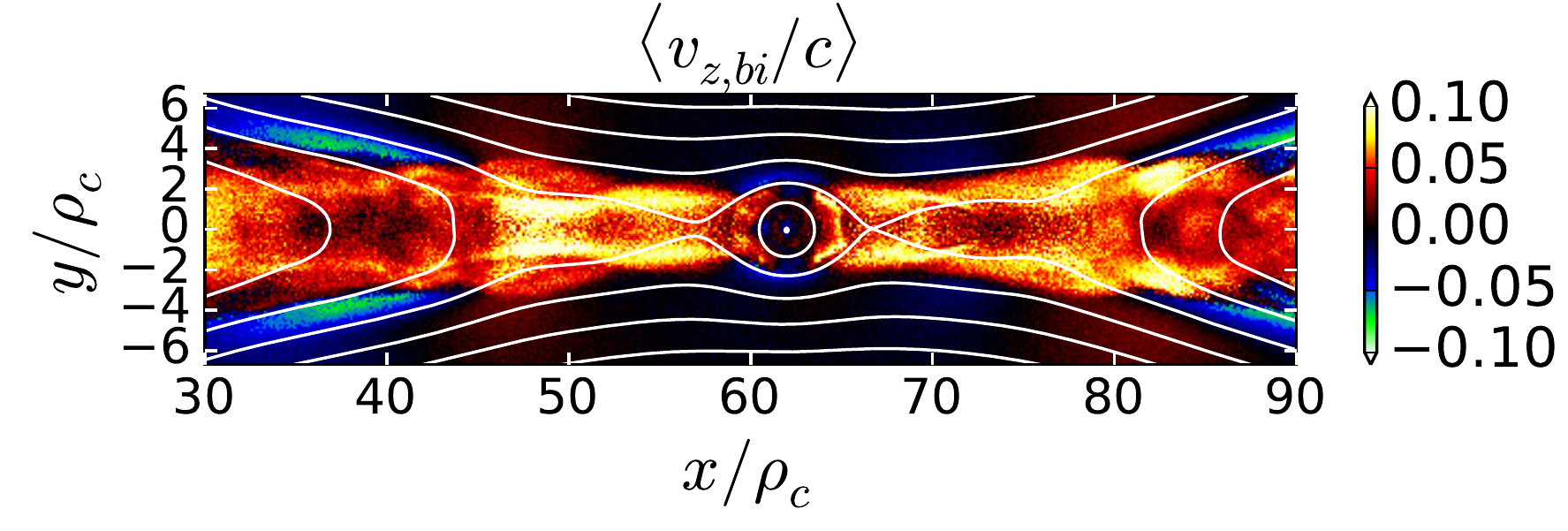}
\end{minipage}
\caption{Bulk flow velocities of background ions, for $\sigma_i=0.1$ and $t\omega_c=337$.
\label{fig:ionFlows}
}
\end{figure}

To connect our kinetic simulations with fluid/MHD descriptions of reconnection, we calculated fluid flows locally within each simulation cell (except for high-$\sigma_i$ runs, where we used higher grid resolution and fewer particles per cell, and consequently measured flows in $2\!\times \! 2$-cell volumes).  We analysed background-particle flows, in a narrow region encompassing $30 < x/\rho_c < 90$ and $-6 < y/\rho_c < 6$, and containing the active X-points; 
as previously noted, the initially-drifting particles are almost entirely absent from this region after the initial stages of reconnection.  
While the nonrelativistic ion flows can be described simply by the average velocity, the electron flows are relativistic and require more care.
We characterized the electron flows in three different ways: the average three-velocity $\langle {\bf v}\rangle$ or Eckart flow velocity; the Landau flow velocity; and the average four-velocity $\langle {\bf u}\rangle = \langle \gamma {\bf v} \rangle$.
The Eckart flow reflects the boost to the local rest frame in which the net particle current vanishes; it is thus directly related to the electric current.  In contrast, the Landau flow velocity boosts to the frame in which the net momentum vanishes.  Finally, the average four-velocity, which does not generally satisfy
$\langle {\bf u} \rangle = \langle {\bf v} \rangle / \sqrt{1-\langle v \rangle^2/c^2}$,
yields a completely different result, which meaningfully characterizes the energy and motion of high-energy particles, but cannot be said to represent a bulk fluid flow in the usual sense.
The average four-velocity can be calculated straightforwardly, as can the
average three-velocity, which provides the boost to the Eckart (zero net current) frame.  The Landau frame, however, is more difficult to calculate, involving the entire $4\times 4$ stress-energy tensor (see Appendix~\ref{sec:fluidFrames}); in our simulations, we find that the Landau flow closely resembles the Eckart flow, and both these quantities differ greatly from the average four-velocity $\langle {\bf u} \rangle$.
Figure~\ref{fig:electronFlows}, \ref{fig:electronRelFlows}, and \ref{fig:ionFlows} show snapshots of different components of the electron and ion fluid flows during mid-reconnection, namely at~$t\omega_c=337$ for the semirelativistic case $\sigma_i=0.1$. 

Figures \ref{fig:electronFlows} and \ref{fig:ionFlows} demonstrate a general inflow (in the $\pm y$-directions) into X-points and outflow (in the $\pm x$-directions) towards O-points as flux reconnects, as well as flows in the $z$-direction consistent with the magnetic field discontinuity, in opposite directions for electrons and ions.
The electron and ion flows are (for $\sigma_i \lesssim 1$) qualitatively quite similar to plasma flows found in nonrelativistic reconnection.

The three-velocity electron flows in the reconnection plane, $\langle v_{x,be} \rangle$ and $\langle v_{y,be} \rangle$, shown in Figs.~\ref{fig:electronFlows}(a) and (b), reflect the general inflow (in $y$) and outflow (in $x$), but furthermore show, as in nonrelativistic reconnection, diagonal inflow (nearly) along magnetic separatrices towards X-points.
As the magnetic field weakens just upstream of the X-points, where reconnection happens, field lines become farther apart; because electrons are still attached to field lines until very near the X-points, the electron density decreases with the spread of field lines, but is replenished by the diagonal inflow parallel to the field lines \citep{UzdenskyKulsrud:2006}.
The out-of-plane ($\langle v_{z,be} \rangle$) electron flow [Fig.~\ref{fig:electronFlows}(c)] is strongest in the current layer close to the X-points, where it generates the current $J_z$ corresponding to the reversing $B_x$ across the layer.  
Interestingly, one can see, comparing Figs.~\ref{fig:electronFlows}(a) and (c) within the electron current layer for $43 < x/\rho_c < 50$, that the electron outflow in the $-x$-direction can wrap around reconnected magnetic field $B_y$ to produce a small patch of electron flow in the $+z$
-direction [opposite the predominating current in the layer, cf. \cite{Biskamp-1994}].

Figure \ref{fig:electronRelFlows} displays electron four-velocities for $\sigma_i=0.1$, with panel (a) showing the $x$-component of the 
Landau flow ${\bf u}_{\rm Landau}/c$.
This mildly relativistic outflow is notable because it is somewhat relativistic, even though $\sigma_{\rm hot} \ll 1$; however, it is also notable that, in agreement with \citet{Lyubarsky-2005}, the bulk fluid flow is not highly relativistic, even though $\sigma_e = \mu \sigma_i \gg 1$ and, as shown in Fig.~\ref{fig:electronRelFlows}(c), the average four-velocity is highly relativistic.
Panels (b) and (d) show that the flows in the $z$-direction are weaker, but of similar magnitude to those in the $x$-direction.
We have examined the average four-velocity in outflows for simulations with larger $\sigma_i$ and found $\langle u_x \rangle \sim \sigma_e$; this is not surprising given a mildly directional flow involving particles with energies scaling as $\sim\sigma_e$.  
At the same time, the Landau (and Eckart) flow velocities continue to be mildly relativistic, despite very large $\langle {\bf u} \rangle\sim \sigma_e$ and $\sigma_{\rm hot} \gg 1$; for example, $u_{{\rm Landau},x}/c$ reaches only $\sim 6$ for $\sigma_i=10^4$ and $\sigma_{\rm hot}\approx 25$.
Because the electrons---themselves highly relativistic as indicated by $\langle {\bf u} \rangle$---do not travel
in precisely the same direction, the fluid electron flow is only mildly relativistic.

\begin{figure}
\includegraphics*[width=16.cm,trim=0mm 0 0mm 0]{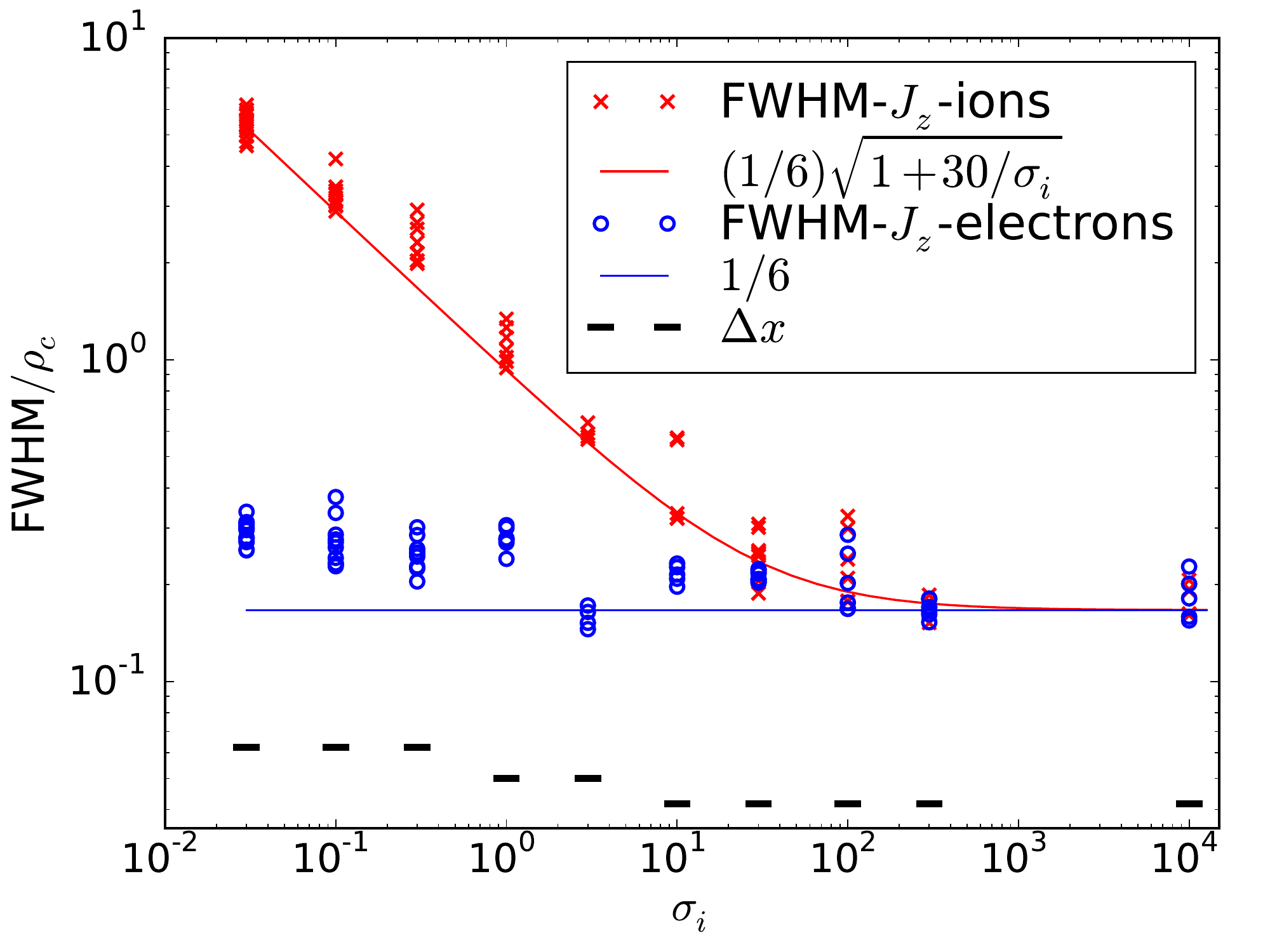}\\
\caption{The electron and ion current-layer thicknesses [e.g., the FWHM of $J_z(y)$], normalized to~$\rho_c$, at the major X-point at several times spaced throughout the mid-reconnection interval, along with simple, empirical fits.  The simulation cell size $\Delta x$ (normalized to $\rho_c$) is shown (by the black bars) to be significantly smaller than the layer thicknesses.
\label{fig:layerThicknesses}
}
\end{figure}

Figure \ref{fig:ionFlows} shows (subrelativistic) ion fluid velocities.
Typical ion inflow speeds (upstream of the $X$-points) are similar to electrons', while the ion outflow speeds are smaller than electrons'; the ion outflow channel is broader than the electron outflow. 
Unlike electrons, ions are less magnetized near X-points, and Fig.~\ref{fig:ionFlows}(b) correspondingly shows ions crossing field lines to go from upstream to downstream regions, short-circuiting the X-point.
Just as in nonrelativistic reconnection, the differences between electron and ion flow patterns lead to the Hall effect  quadrupolar out-of-plane magnetic field, which we discuss in the next section.
Like the out-of-plane electron flow, the out-of-plane ion flow [Fig.~\ref{fig:ionFlows}(c)] creates a current that supports the reversing magnetic field; while the ion current density is roughly 10 times less than the electron current density, the ion current layer is about 10 times thicker for $\sigma_i=0.1$ (see~Fig.~\ref{fig:layerThicknesses}), so both electrons and ions contribute comparable amounts to the total integrated current. In general, we find that the dependence of the ion layer thickness on $\sigma_i$ is reasonably well fitted by $\delta_i \simeq (\rho_c/6) \, (1+30/\sigma_i)^{1/2}$, as shown in Fig.~\ref{fig:layerThicknesses}, i.e., scaling approximately with the expected average ion Larmor radius.

Examining the current layers (after the initially-drifting particles have ceased to contribute), as in Fig.~\ref{fig:nAndJz}, shows that the electron current layer is (for $\sigma_i=0.1$) much thinner (in the $y$-direction) than the ion current layer, just as in nonrelativistic reconnection; this is not surprising because ions have larger Larmor radii and skin depth than electrons.
The difference in current-layer thickness is shown versus~$\sigma_i$ in Fig.~\ref{fig:layerThicknesses}, which plots the full width at half-maximum (FWHM) of $J_{ze}(y)$ and $J_{zi}(y)$ at (the $x$-position of) the major X-point at several times during the mid-reconnection interval.  Although the major X-point and the current-layer thickness fluctuate with time as indicated by the spread of points in Fig.~\ref{fig:layerThicknesses}, some general trends are clear.
The electron current layer thickness is roughly constant around $\rho_c/6$, although it does increase a bit (by $\sim$50\%) at small $\sigma_i$.
The ion current-layer thickness, on the other hand, increases significantly with decreasing $\sigma_i$, roughly as $(\rho_c/6)\sqrt{1+30/\sigma_i}$.
The disparity in current-layer thicknesses is strong for small $\sigma_i$, and disappears at larger $\sigma_i$ as ions become highly relativistic. Figure~\ref{fig:layerThicknesses} also shows that the cell size is sufficiently small to resolve the layer thickness.

%----------------------------------------------------------------------------------

\subsection{Two-fluid/Hall effect signatures}
\label{subsec-Hall}

An important feature in electron-ion plasma reconnection is the Hall effect, which has received much attention in nonrelativistic reconnection studies \citep{Sonnerup-1979,Terasawa-1983,Shay_etal-1998,Birn_etal-2001,Rogers_etal-2001,Wang_etal-2000,Ren_etal-2005,UzdenskyKulsrud:2006}, and has also been observed in relativistic electron-ion reconnection in \cite{Melzani:2014a}.
The Hall effect arises from the different behaviours of electrons and ions at intermediate length-scales between electron and ion kinetic scales. The hallmark signatures of the Hall effect in reconnection are the emergence of a quadrupolar pattern of out-of-plane magnetic field ($B_z$) concentrated along the separatrices near an X-point, and the corresponding development of a dipolar in-plane potential electric field.
We expect to see these signatures for small $\sigma_i$, where ion and electron length-scales are quite disparate; on the other hand, the Hall effect signatures must disappear as $\sigma_i$ increases into the ultrarelativistic regime (as in pair reconnection, where the equal masses of positive and negative charges preclude any Hall effect).  
In the following we will show results confirming this expectation and investigate the decline of the Hall effect signatures as $\sigma_i$ increases.

\begin{figure}
\centering
\includegraphics*[width=16.cm,trim=0mm 0 0mm 0]{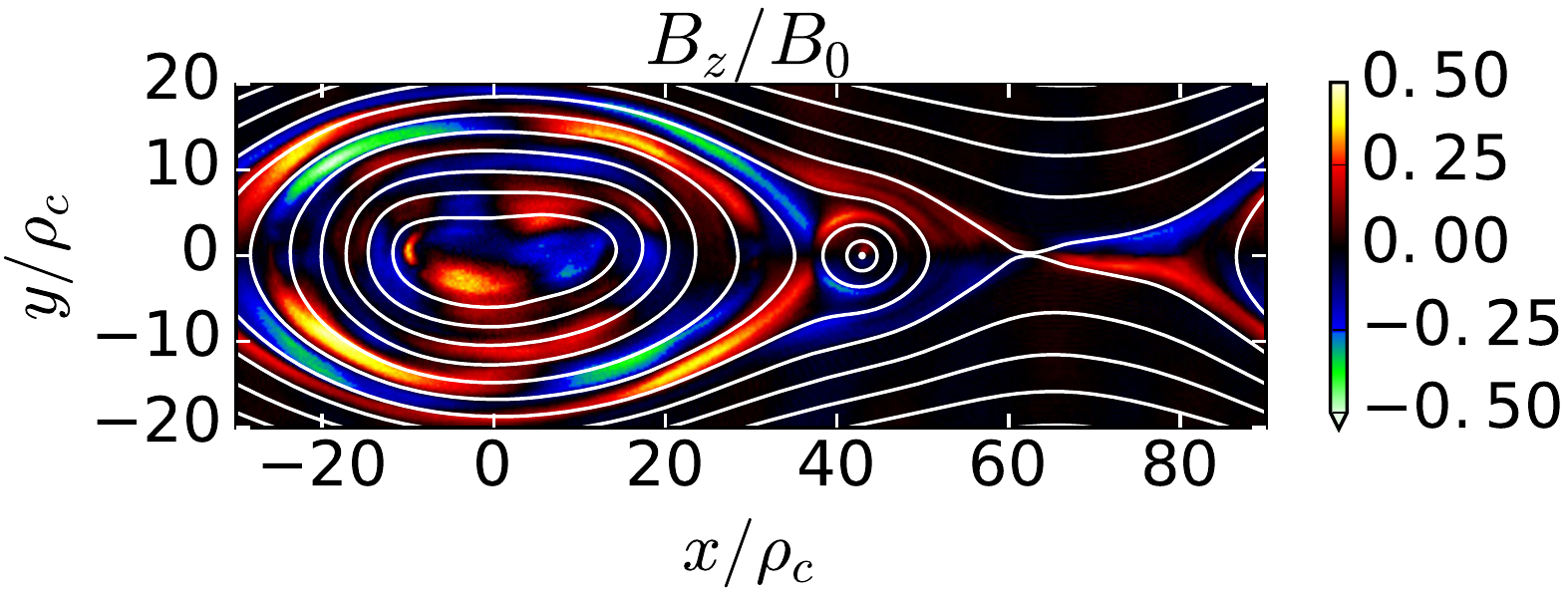}\\
\includegraphics*[width=16.cm,trim=0mm 0 0mm 0]{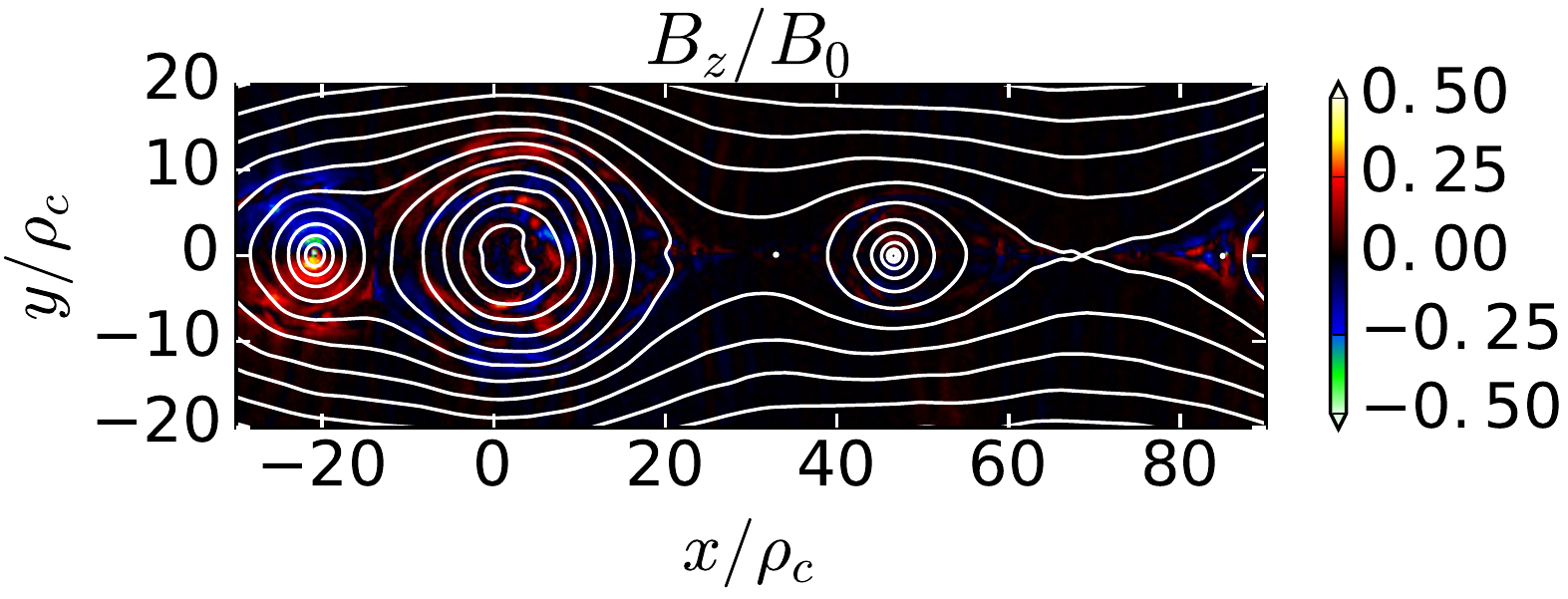}%
\llap{\raisebox{4.3cm}{
\includegraphics*[width=4.5cm,trim=0mm 0 0mm 0]{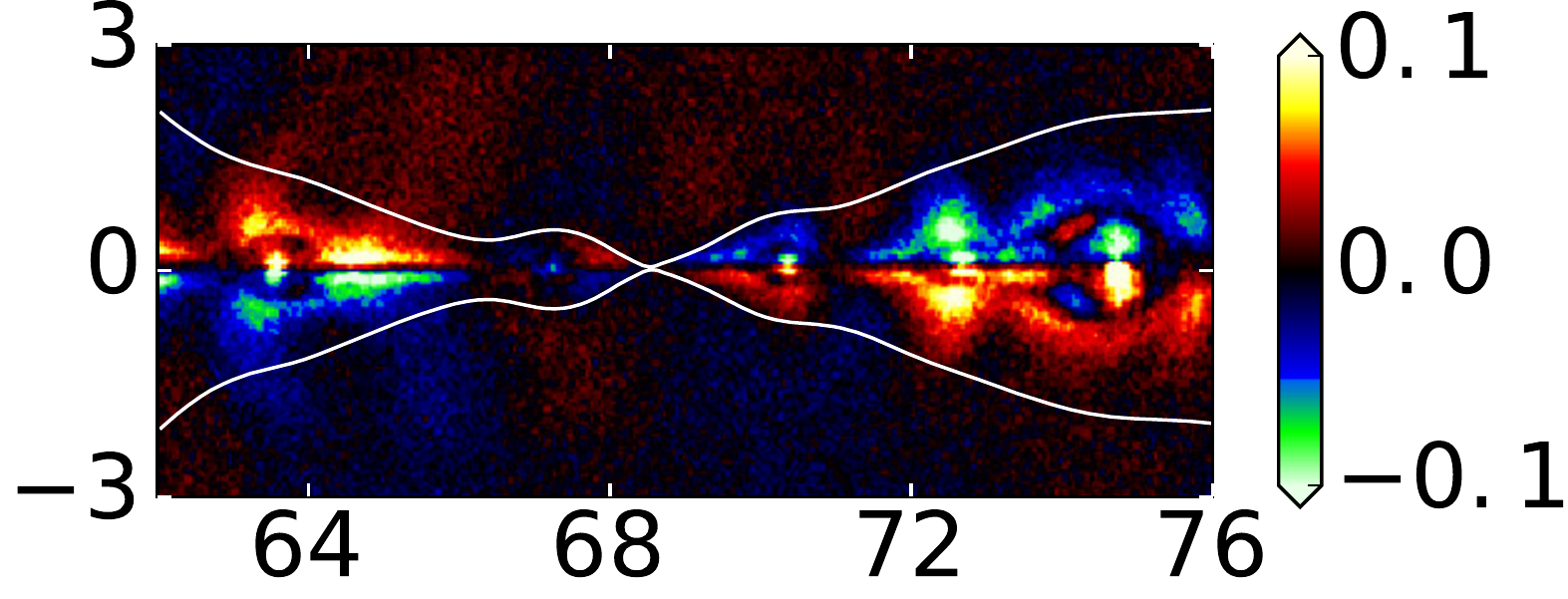}}\hspace{2.6cm}}
\caption{The out-of-plane magnetic field $B_z/B_0$ for $\sigma_i=0.1$ (top) at $t\omega_c=673$ shows the Hall quadrupolar field around the major X-point at $x/\rho_c\approx 63$; (bottom) the Hall effect is much weaker for $\sigma_i=10$ ($t\omega_c=175$).
In-plane magnetic field lines are shown in white.
\label{fig:Bz}
}
\end{figure}

As shown in \S\ref{sec:PlasmaFlows}, the scale separation between ions and electrons (when $\sigma_i$ is small) leads to different ion and electron flows:
electrons flow inward along the magnetic separatrix nearly to the X-point before crossing the separatrix to the outflow, while ions (with their larger Larmor radii) can cross the separatrix farther from the X-point.  The decoupled flows create a net electric current in the $xy$ reconnection plane, which generates a quadrupolar $B_z$ magnetic field around each X-point
(e.g., Fig.~\ref{fig:Bz}, top): $B_z > 0$ to the upper left and lower right of an X-point, and $B_z < 0$ to the lower left and upper right---this pattern is independent of whether the main current layer has $J_z>0$ or $J_z<0$, as expected 
\citep[e.g.,][]{Sonnerup-1979, Terasawa-1983}.  This $B_z$ appears prominently around the major $X$-point for $\sigma_i=0.1$ in Fig.~\ref{fig:Bz} (top), but it is much weaker for $\sigma_i=10$ (Fig.~\ref{fig:Bz}, bottom).
Similar to the work of \cite{Melzani:2014a}, which recorded $B_z$ around X-points at the level of 1--10\% of $B_0$, we observe $B_z$ around X-points at strengths $B_z \approx 0.2 B_0$ for $\sigma_i=0.1$ down to $B_z \approx 0.05 B_0$ for $\sigma_i=10$, etc.
We further observe, however, that $B_z$ is typically stronger around O-points than X-points, approaching levels around one-half $B_0$.

\begin{figure}
\includegraphics[width=8.cm,trim=0mm 0 0mm 0]{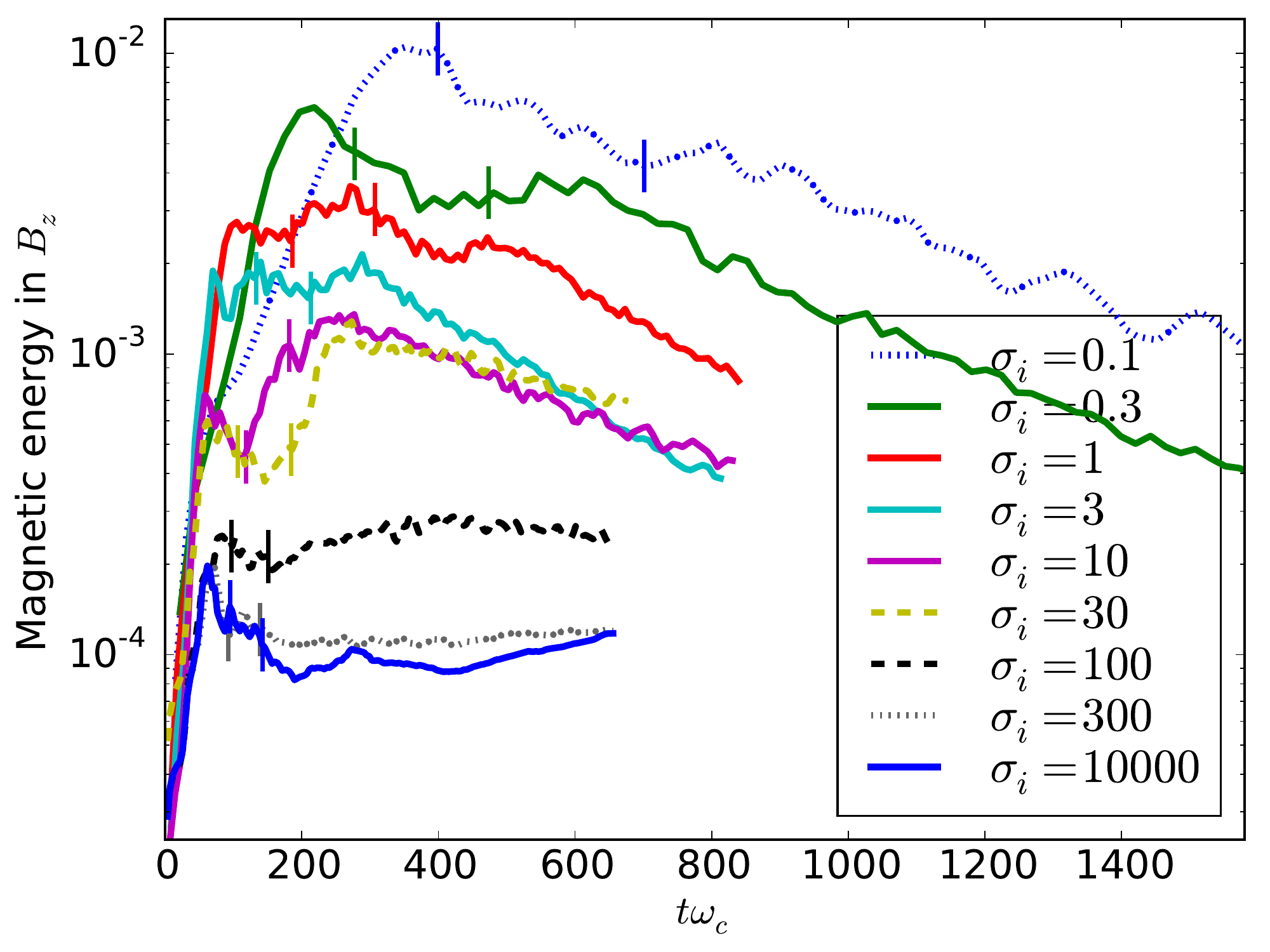}
\caption{The magnetic energy fraction stored in the out-of-plane magnetic field $B_z$ (normalized to $B_0^2/8\pi$ times the simulation volume) versus time for several $\sigma_i$ (with vertical bars to delineate the mid-reconnection interval for each case): as $\sigma_i$ increases, the scale separation between electrons and ions diminishes, and less energy goes into $B_z$.
\label{fig:BzFrac}
}
\end{figure}

With our systematic exploration across a range of $\sigma_i$, we have been able to show the effect of $\sigma_i$ on the out-of-plane Hall magnetic field~$B_z$.
A rough global measure of the strength of the Hall effect (in antiparallel reconnection, with no initial guide field) is simply the magnetic field energy associated with $B_z$ (normalized to $B_0$), i.e.,  $\int B_z^2 dV / \int B_0^2 dV$, where the integrals are carried out over the entire simulation domain.
Figure~\ref{fig:BzFrac} shows the magnetic energy fraction in $B_z$ versus time for a range of $\sigma_i$, for simulations with the same box size relative to $\rho_c$ (i.e., $L_x=120\rho_c$).
The fraction of energy in $B_z$ has a complicated relationship with the quadrupole field surrounding $X$-points, due to persistence of the $B_z$ magnetic flux in magnetic islands;
 however, it is clear that this measure is distinctly different for low and high $\sigma_i$, peaking at much higher values for low $\sigma_i$ before decaying.

\begin{figure}
\includegraphics*[width=8.cm,trim=0mm 0 0mm 0]{%
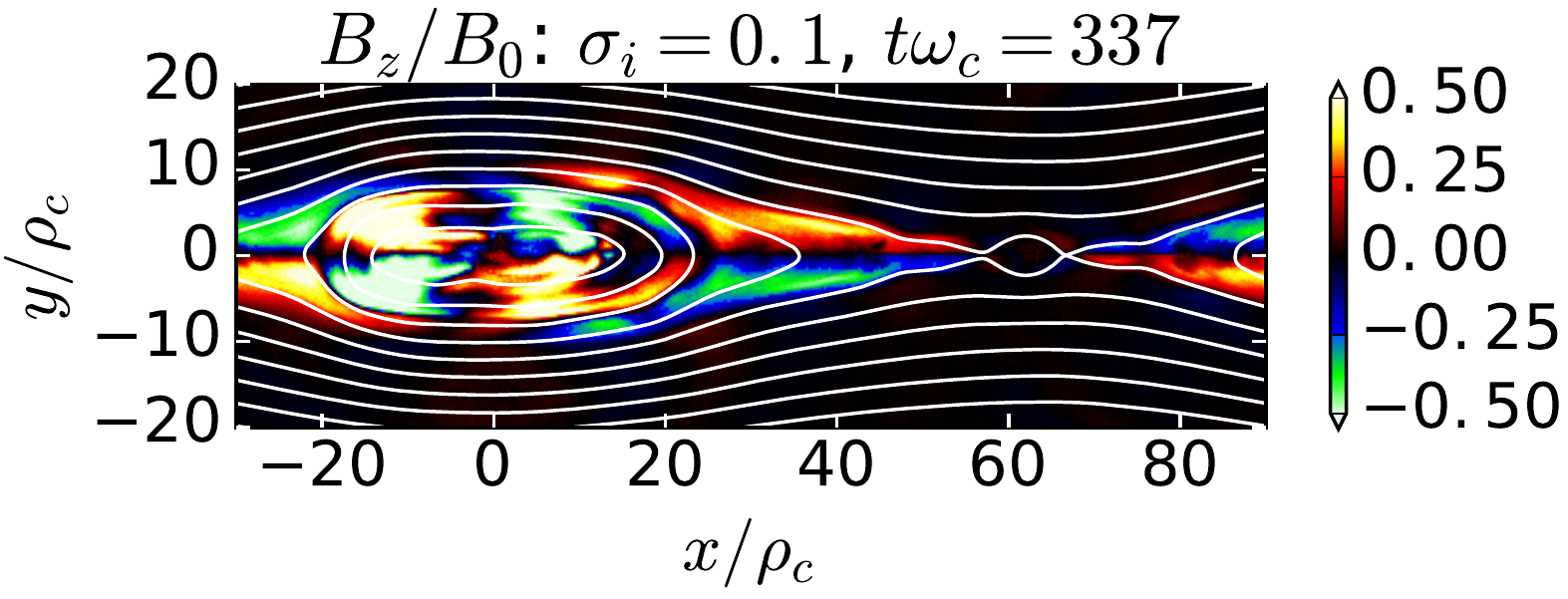}
\includegraphics*[width=8.cm,trim=0mm 0 0mm 0]{%
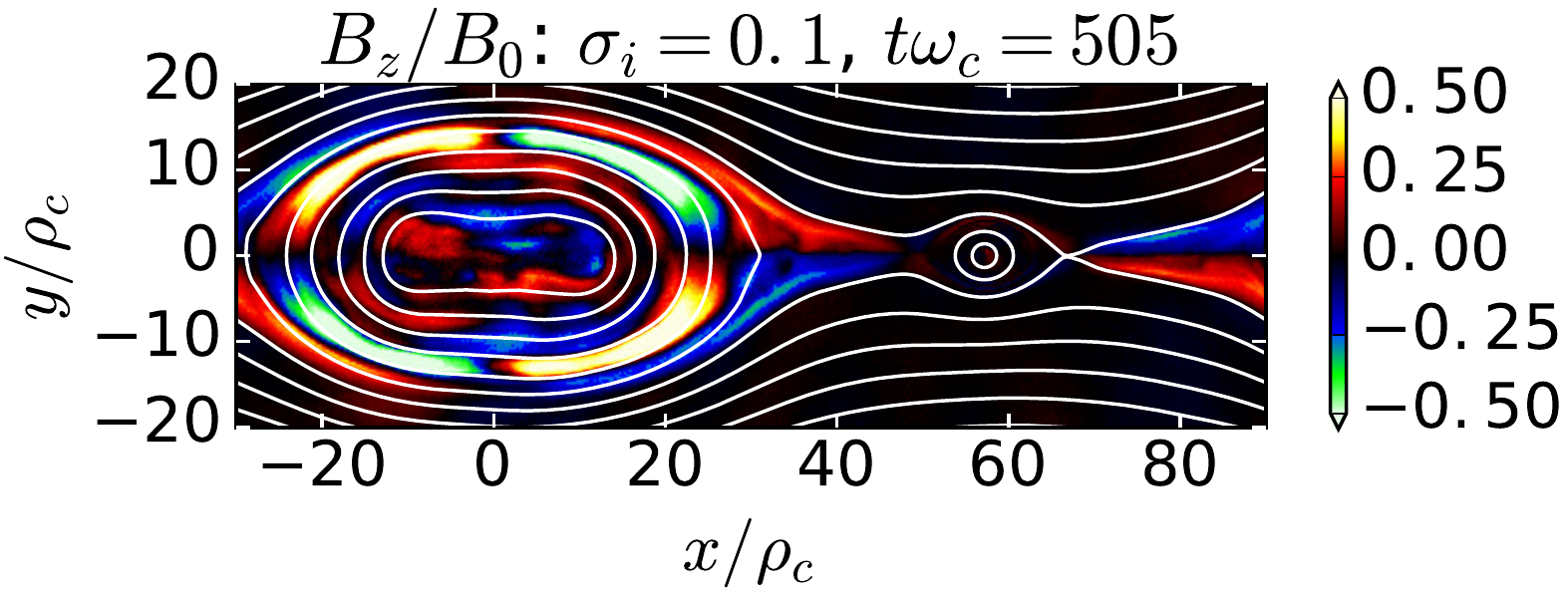}\\
\includegraphics*[width=8.cm,trim=0mm 0 0mm 0]{%
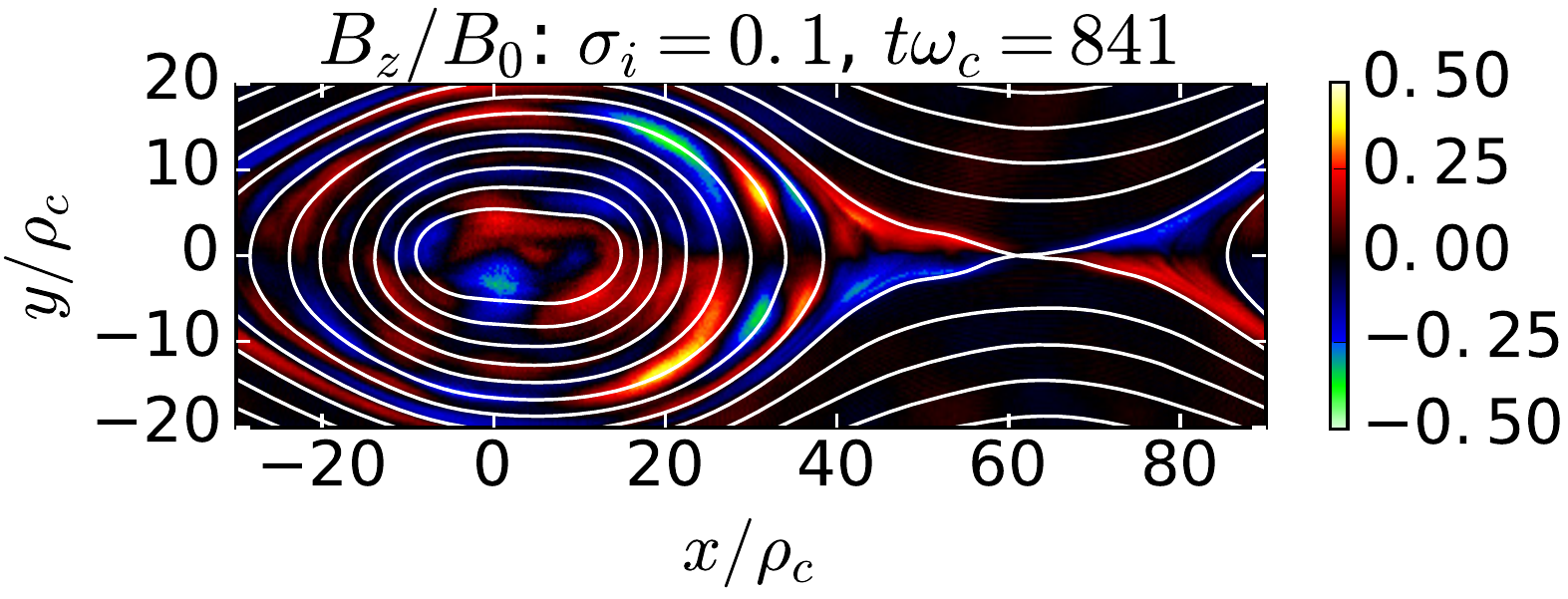}
\includegraphics*[width=8.cm,trim=0mm 0 0mm 0]{%
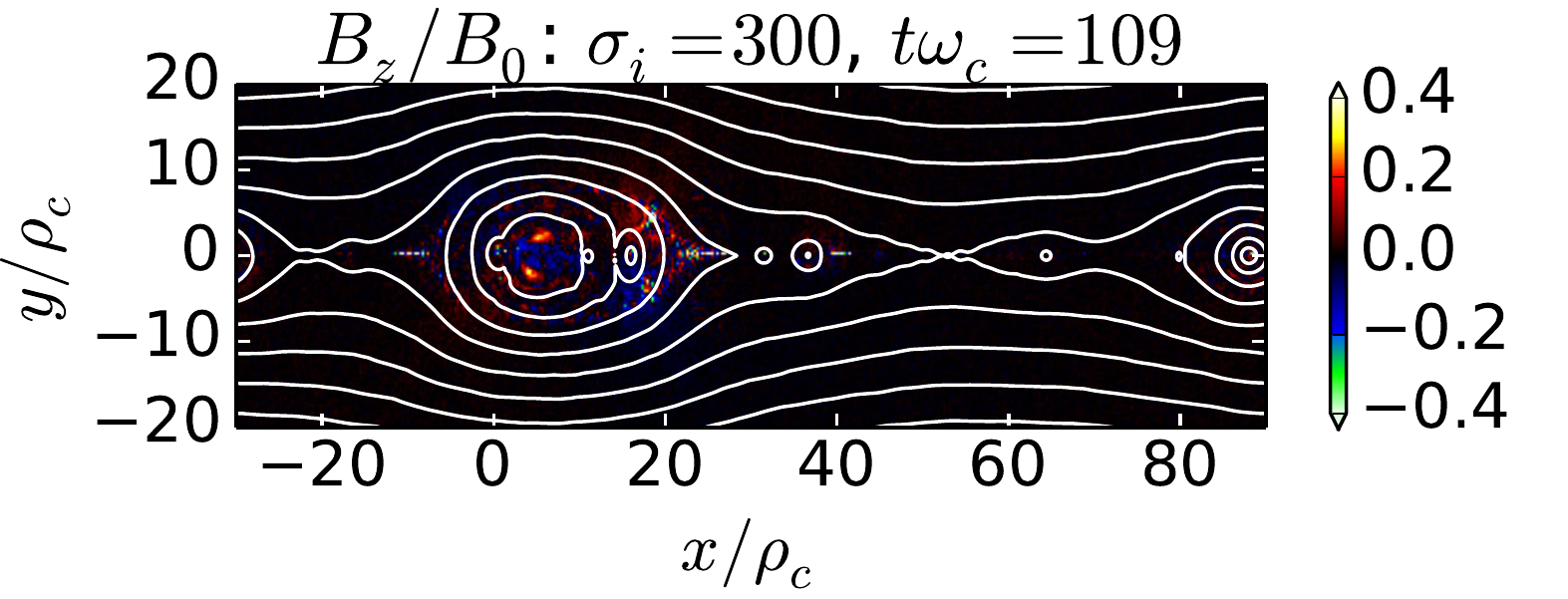}
\llap{\raisebox{0.5cm}{
\includegraphics*[width=3cm,trim=0mm 0 0mm 0]{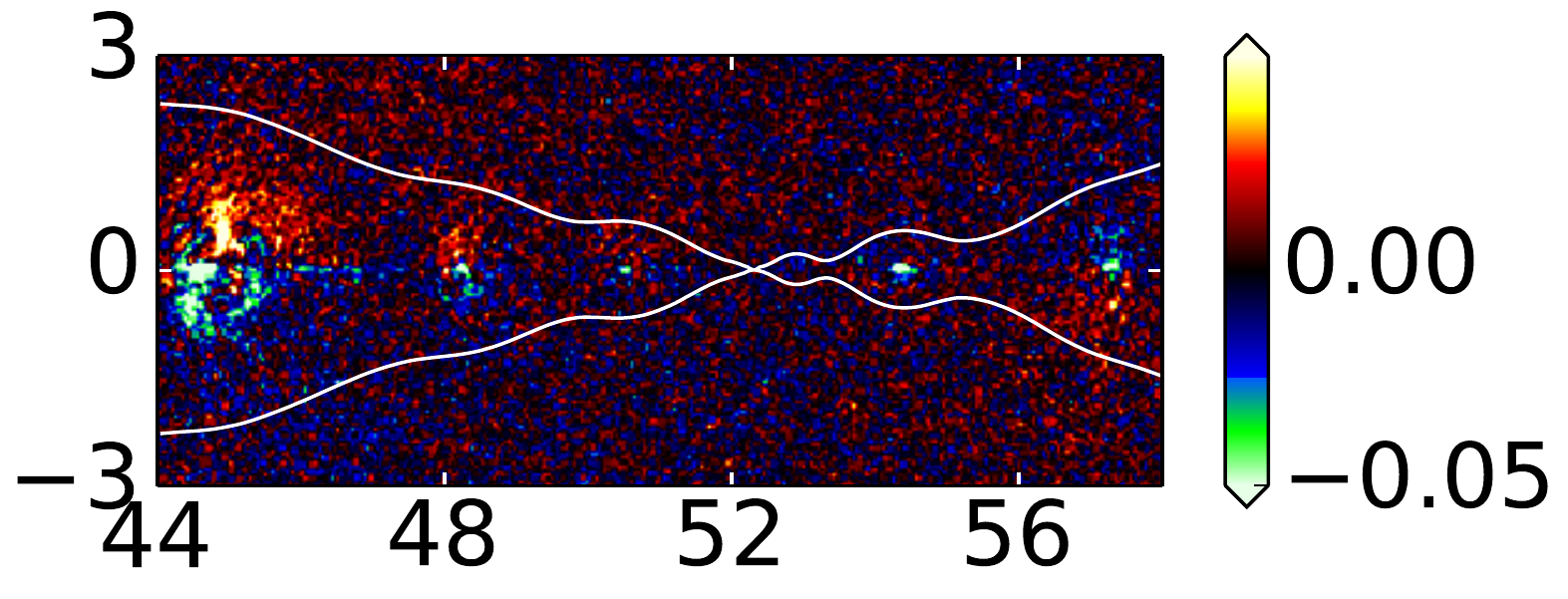}}\hspace{1.4cm}}
\caption{
The out-of-plane magnetic field $B_z/B_0$ for $\sigma_i=0.1$ at three different
times shows how the Hall quadrupolar field gets wrapped up 
in magnetic islands;
for comparison (lower right), the case of $\sigma_i=300$, where the Hall
field is almost entirely absent, but a spontaneous $B_z$ still develops
within magnetic islands.
\label{fig:BzEvolution}
}
\end{figure}

Interestingly, the $B_z$ field generated by the differing electron and ion flows near $X$-points does not simply disappear away from the $X$-points, but seems to be carried by the plasma outflow into and around magnetic islands.  
There it sets up a complex pattern of alternating $B_z$-field that survives for a long time within the major island.
Figure~\ref{fig:BzEvolution} shows $B_z$ at several different times for a simulation with $\sigma_i=0.1$ and~$L_x=120\rho_c$.  
As the main plasmoid grows, $B_z$ appears strong around the edges, while $B_z$ closer to the core decays.
We believe this pattern, in a large plasmoid, may contain a sort of record of the plasmoid's history of accumulating outflows and merging with other plasmoids.  Thus, the plasmoids are not simply dull, boring rings of magnetic field with plasma, but rather are interesting objects by themselves, with a nontrivial and complex internal structure; however, the more thorough investigation they deserve is beyond the scope of this work, and is therefore left for a future study.

\begin{figure}
\includegraphics*[width=8.cm,trim=0mm 0 0mm 0]{%
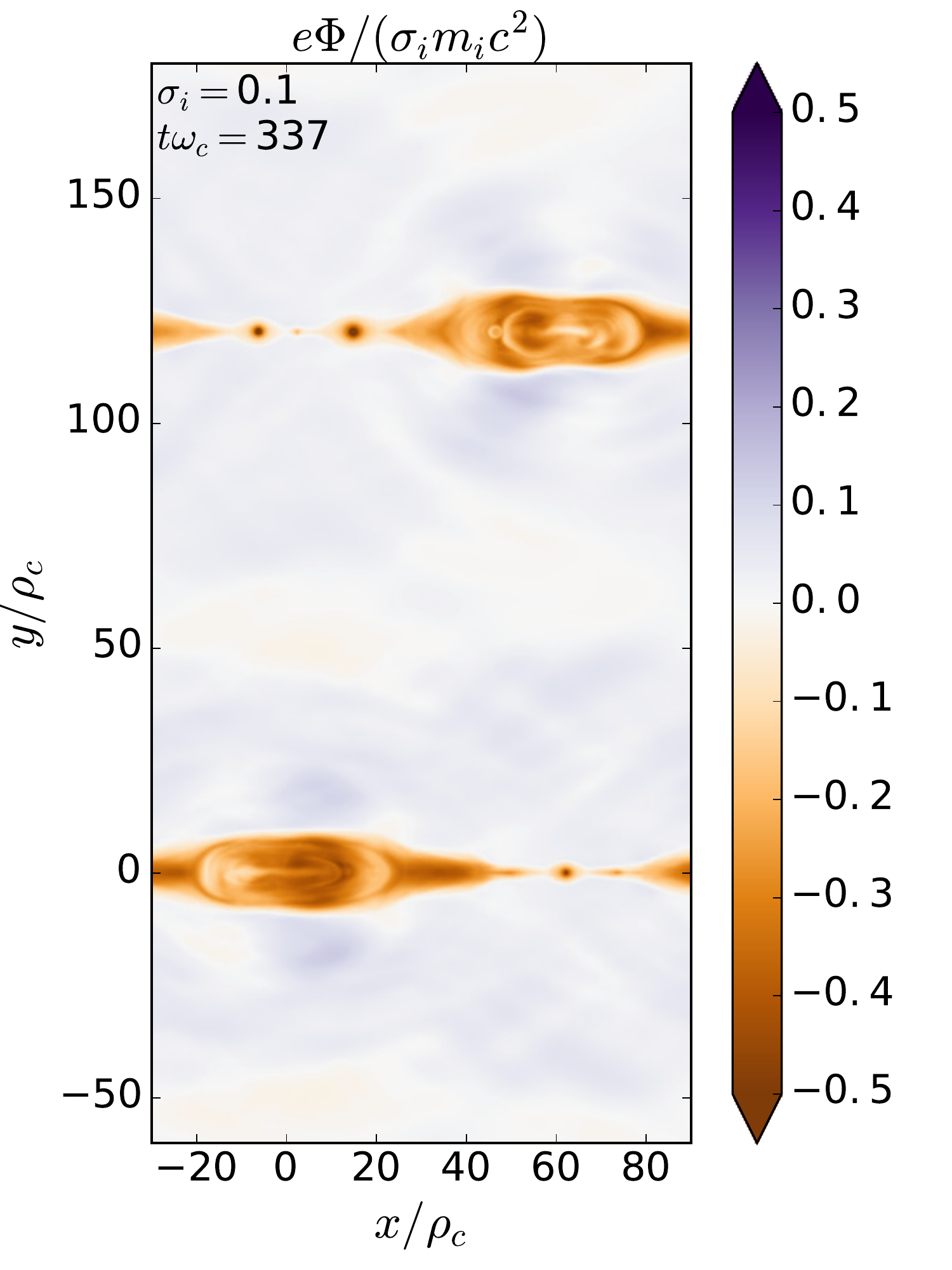}
\caption{
A snapshot of the electrostatic potential (satisfying $-\nabla^2 \Phi = 4\pi \rho$) at a time during mid-reconnection for $\sigma_i=0.1$.
\label{fig:PhiVsXY}
}
\end{figure}

\begin{figure}
\includegraphics*[width=8.cm,trim=0mm 2.5mm 0mm 2.5mm]{%
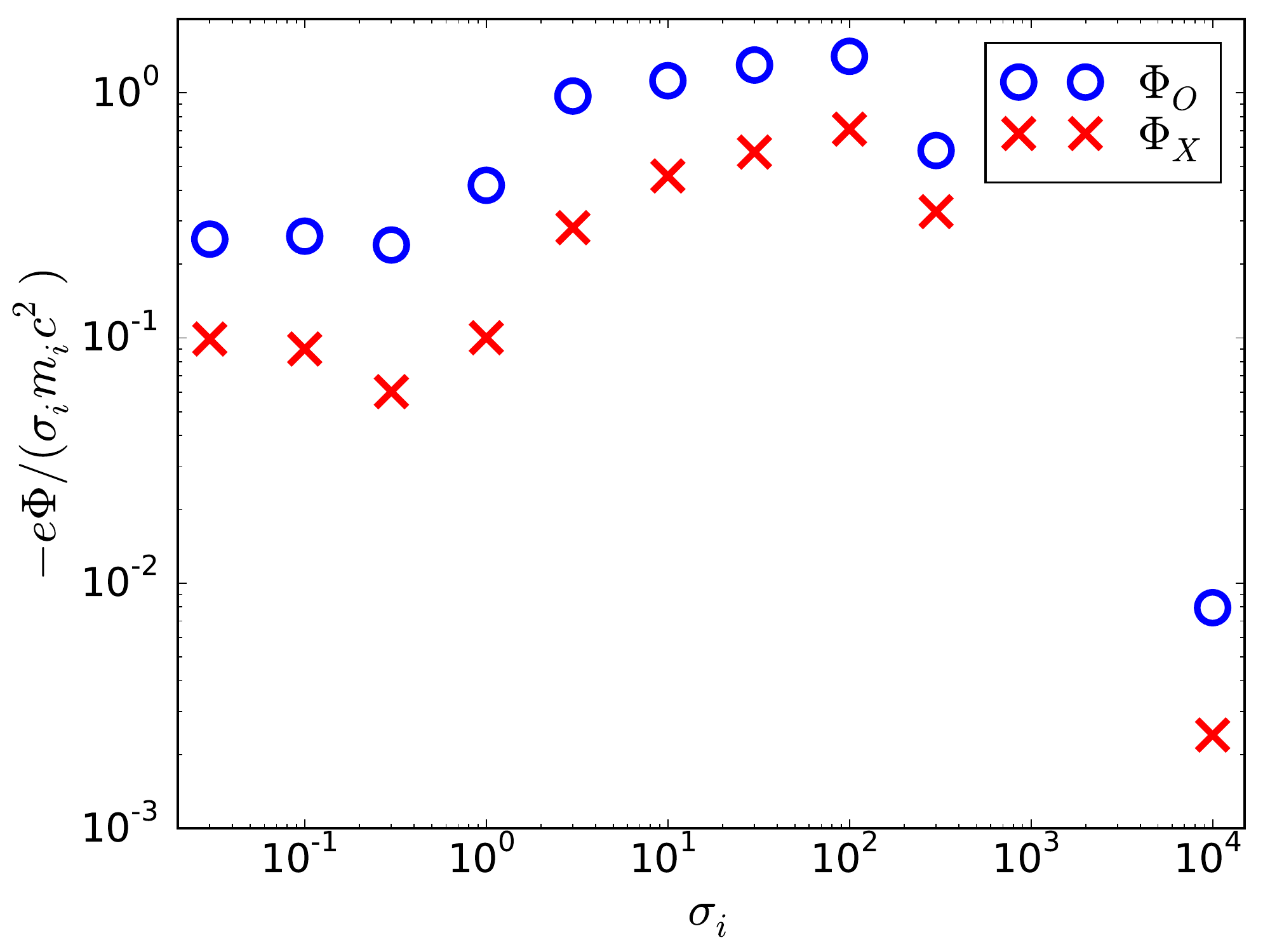}
\caption{
The electrostatic potential difference between upstream and the major X- and O-points (averaged over mid-reconnection) versus~$\sigma_i$.
\label{fig:xoPotentials}
}
\end{figure}

Another important signature of the Hall effect in reconnection is the emergence of an in-plane electrostatic electric field, $E_x$ and~$E_y$. 
The decoupling of electrons and ions leads to charge separation, which can be characterized by the electrostatic potential $\Phi$ 
related to the net charge density~$\rho$ via $-\nabla^2 \Phi = 4\pi \rho$.
As seen in Fig.~\ref{fig:PhiVsXY}, the potential remains near zero in most of the upstream region, decreasing towards X-points, and then further decreasing towards O-points; the deepest potential wells are located near plasmoid centres (these wells have a net negative charge, attracting ions).
A qualitatively similar in-plane potential has been observed in laboratory reconnection experiments \citep{Yoo_etal-2013}.
The electrostatic field can therefore act to accelerate ions crossing from the upstream inflow to the outflows into plasmoids.
We can quantify this electrostatic potential by measuring the potential
difference between the upstream region and the major X- and O-points
(Fig.~\ref{fig:xoPotentials});
at low $\sigma_i \ll 1$, these potential differences (normalized to 
$\sigma_i m_i c^2$) are fairly constant; they increase with $\sigma_i$
until $\sigma_i\sim 100$ when they decrease as the electron and ion
Larmor radii become asymptotically equivalent and the Hall effect disappears.

%----------------------------------------------------------------------------------

\section{Results---Energetics and Nonthermal Particle Acceleration}
\label{sec-results-kinetics}

%----------------------------------------------------------------------------------

\subsection{Energy partition}
\label{subsec-energy_partition}

\begin{figure}
\includegraphics*[width=8.cm,trim=0mm 0 0mm 0]{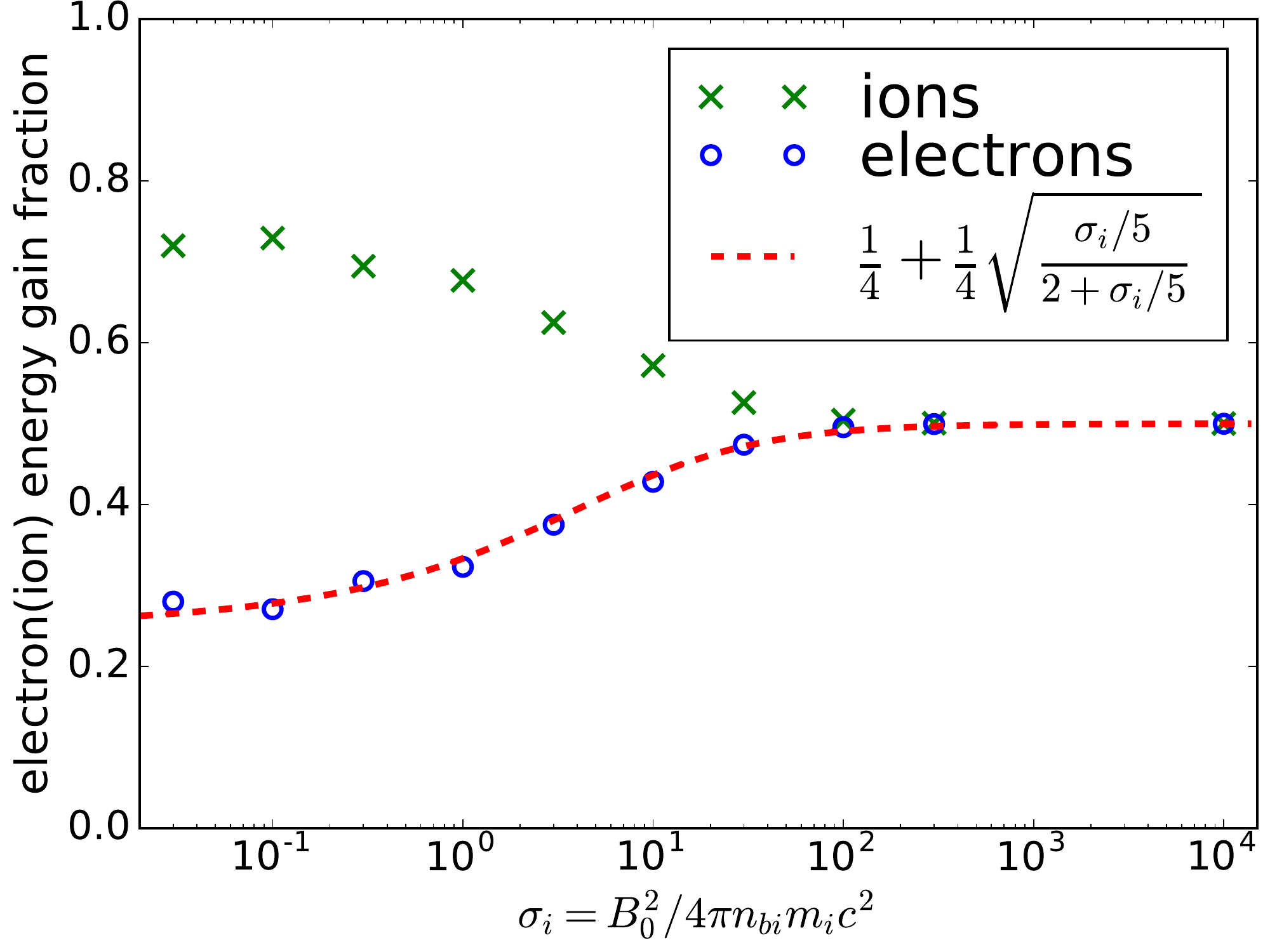}
\caption{
The final energy partition between background electrons and ions, versus~$\sigma_i$.
\label{fig:energyPartition}
}
\end{figure}

An important issue in reconnection research is the partitioning of released magnetic energy into various forms, including the division between electrons and ions.
In our simulation setup (the box aspect ratio $L_y/L_x=2$ and size $L_x=120\rho_c$) about 30\%--40\% of the initial magnetic energy is converted during reconnection into particle kinetic energy  (Fig.~\ref{fig:reconFluxVsSigma}, left); the precise amount of magnetic dissipation depends weakly on $\sigma_i$, with about 29\% of the magnetic energy being dissipated for $\sigma_i=0.03$, 35\% for $\sigma_i=1$, increasing to approximately 41\% in the ultrarelativistic limit.
Almost all the released in-plane ($xy$) magnetic energy goes to particles, with a comparatively small amount going to electric and out-of-plane magnetic field. The partitioning of released energy between electrons and ions is of great astrophysical importance because electrons are much more efficient radiators than ions, and so only the energy transferred to electrons may have direct observational consequences.  Knowing the partition is therefore an important step in estimating the energy budgets of astrophysical phenomena, based on intensity of observed radiation from, e.g., systems like black-hole-powered (blazar) jets and accreting black hole coronae \cite[e.g.,][]{Rees_etal-1982, Rees-1984}.

Our simulations are sufficiently large that over 90\% of the dissipated magnetic energy goes to background particles, with less than 10\% absorbed by initially-drifting particles in the Harris sheets.  
Ions (collectively) gain more energy than electrons, although the difference disappears as $\sigma_i$ increases, so that the scale separation between electrons and ions becomes negligible (Fig.~\ref{fig:energyPartition}).  
As $\sigma_i$ is lowered, transitioning into the semirelativistic regime, the ion-to-electron energy gain ratio increases (hence the electron heating fraction $q_e$ decreases). However, we find that this ratio does not continue to increase indefinitely as $\sigma_i$ drops well below~1, but levels out, with ions gaining about three times as much energy as electrons in our lowest-$\sigma_i$ runs ($\sigma_i = 0.03, \, 0.1$).
It is interesting to note that this asymptotic ratio is similar to what has been seen in nonrelativistic reconnection:
the ratio of ion-to-electron energy gain has been found to be $\simeq 2$ in MRX laboratory experiments \citep{Yamada_etal-2014}; PIC simulations have yielded ratios around 1.4--2.6 \citep{Yamada_etal-2015,Haggerty_etal-2015}; and spacecraft observations in Earth's magnetotail yielded values around 1.6--2.7 \citep{Eastwood_etal-2013}.  Observations of reconnection with asymmetric upstream plasma conditions in Earth's magnetopause reported higher ion-to-electron energy gain ratios from 3 to about 15, with an average around 8 \citep{Phan_etal-2013,Phan_etal-2014}; however, the ion/electron energy partition was shown to be sensitive to the upstream temperature ratio $T_e/T_i$ \citep{Haggerty_etal-2015}, and perhaps this sensitivity extends to differing upstream temperatures in asymmetric reconnection.
In addition, we note that in the very early stages of low-$\sigma_i$ simulation, background electrons (collectively) gain energy faster than background ions; as the simulation proceeds, the ions' energy gains overtake the electrons'.

Overall, we find that the electron fraction of the dissipated magnetic energy is well fitted across all our simulations by the simple, empirical formula
\begin{eqnarray} \label{eq:energyPartition}
    q_e &=& \frac{1}{4} \, 
    \left( 1+ \sqrt{{{\sigma_i/5}\over{2+ \sigma_i/5}}} \right) 
.\end{eqnarray}
This electron heating fraction, as a function of $\sigma_i$, may potentially serve as a physically motivated prescription for calculating radiative signatures of small-scale reconnection in global MHD simulations of black holes \citep[e.g.,][]{Ressler:2015}.

%----------------------------------------------------------------------------------

\subsection{Particle energy spectra}
\label{subsec-particle_spectra}

One of the main results of our simulations is the computation of energy spectra of particles accelerated during reconnection, which relate closely to observable (synchrotron and inverse Compton) radiation spectra.
As perhaps the most astrophysically important question in relativistic pair reconnection---due to its bearing on whether magnetic reconnection can explain observed high-energy radiation---nonthermal acceleration in relativistic pair-plasma reconnection has been much studied \citep{Zenitani_Hoshino-2001,Zenitani_Hoshino-2005,Zenitani_Hoshino-2007,Zenitani_Hoshino-2008, Larrabee_etal-2003, Jaroschek_etal-2004a, Lyubarsky_Liverts-2008,Liu_etal-2011, Cerutti_etal-2012a,Cerutti_etal-2012b,Cerutti_etal-2013,Cerutti_etal-2014b} with computational advances allowing clearer identification of the existence of nonthermal power-law energy spectra in more recent studies \citep{Sironi_Spitkovsky-2014,Guo:2014,Guo:2015,Werner_etal-2016}.
Nonthermal acceleration in relativistic electron-ion reconnection has thus far received less attention \citep{Werner_etal-2013,Melzani:2014b,Sironi_etal-2015,Guo:2016};
here we show that nonthermal particle acceleration occurs for semirelativistic electron-ion reconnection as well, and characterize the non-thermal power-law electron spectra.

\begin{figure}
(a)
\includegraphics*[width=8.cm,trim=0mm 0 0mm 0]{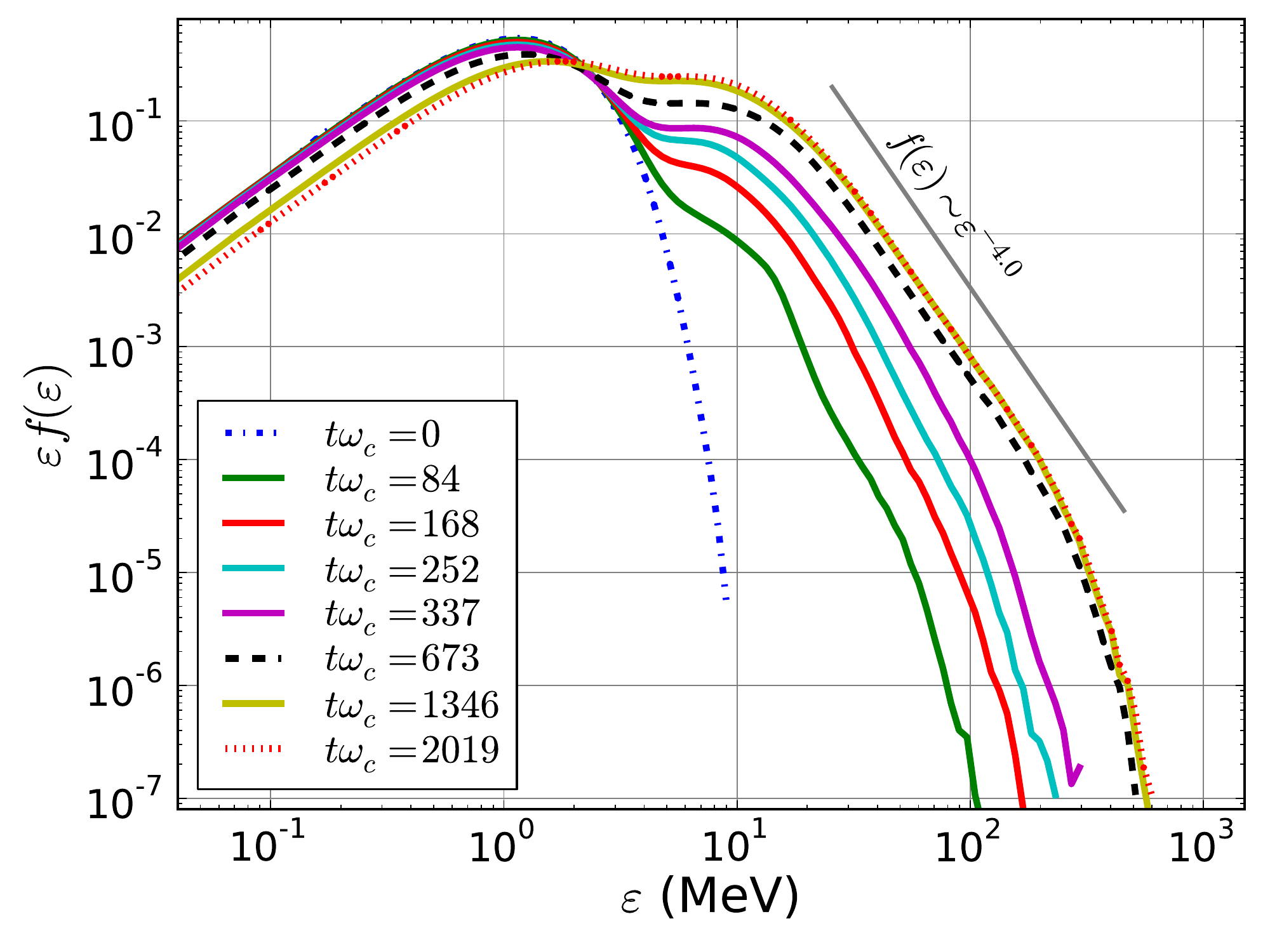}
(b)
\includegraphics*[width=8.cm,trim=0mm 0 0mm 0]{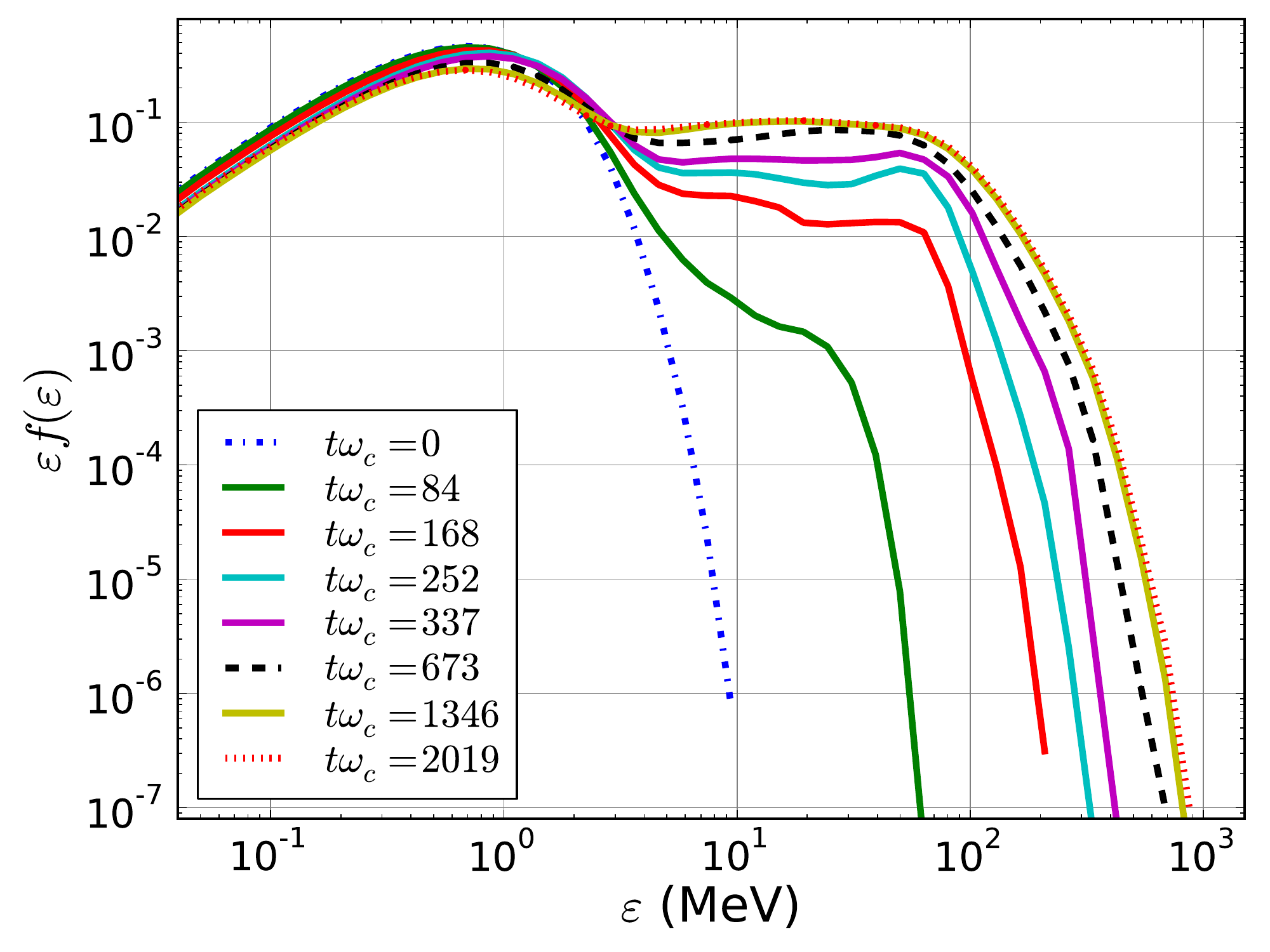}
\caption{
Time evolution of the (a) electron and (b) ion energy distributions,
$f(\varepsilon)$ (compensated by $\varepsilon$) for $\sigma_i=0.1$.
\label{fig:fvsTime}
}
\end{figure}

Figure~\ref{fig:fvsTime} shows electron and ion energy spectra $f(\varepsilon)$ (compensated by a factor $\varepsilon$) for the case of $\sigma_i=0.1$ at equally spaced times during the simulation; here $\varepsilon = (\gamma-1)mc^2$ is the particle kinetic energy (without the rest-mass energy).  
The spectra evolve from an initial Maxwellian to nonthermal/non-Maxwellian spectra (plus a remnant of the initial Maxwellian).

\begin{figure*}%[p]
\begin{tabular}{ll}
\includegraphics*[width=8.cm,trim=0mm 0 0mm 0]{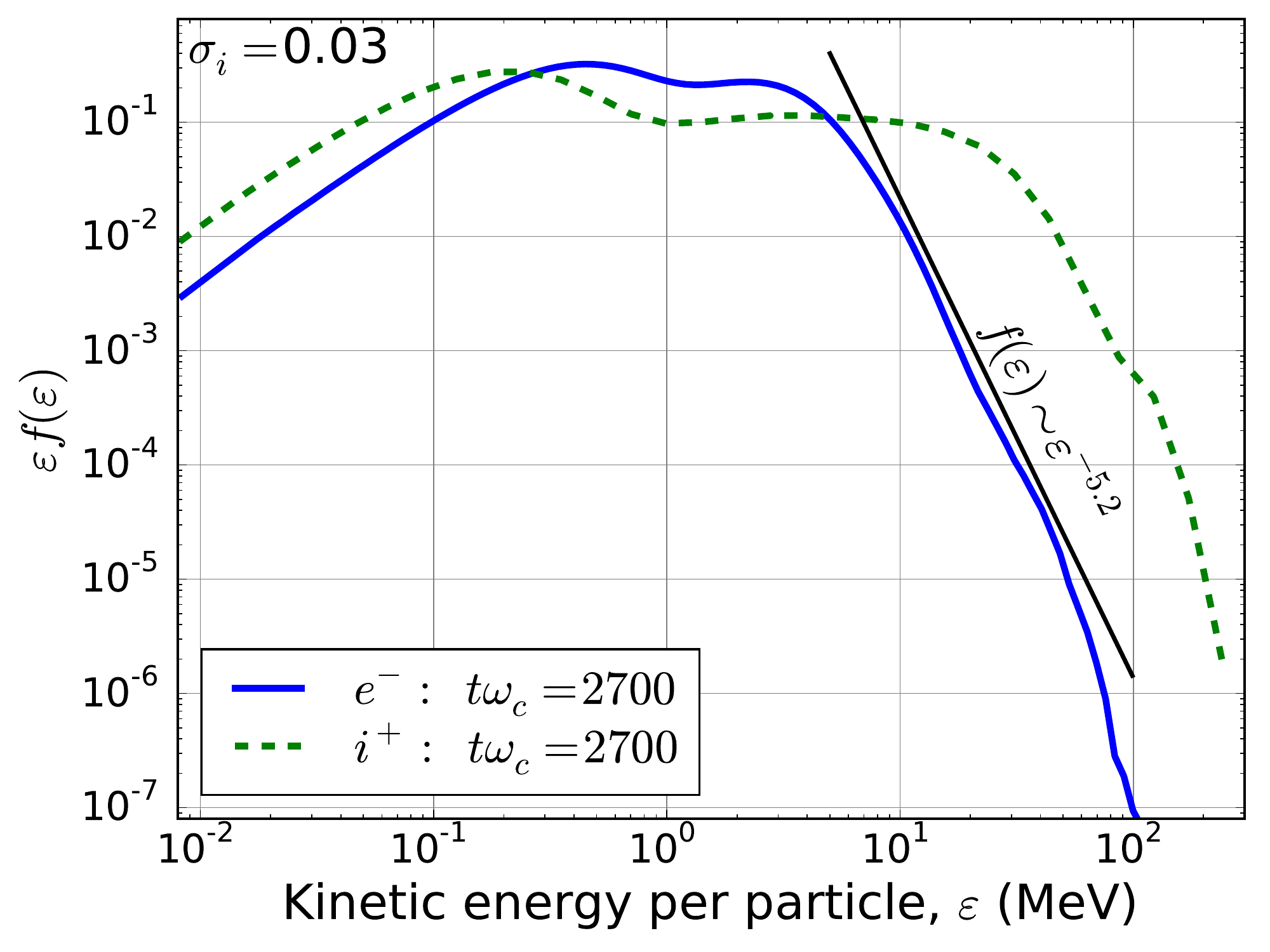}%
&\includegraphics*[width=8.cm,trim=0mm 0 0mm 0]{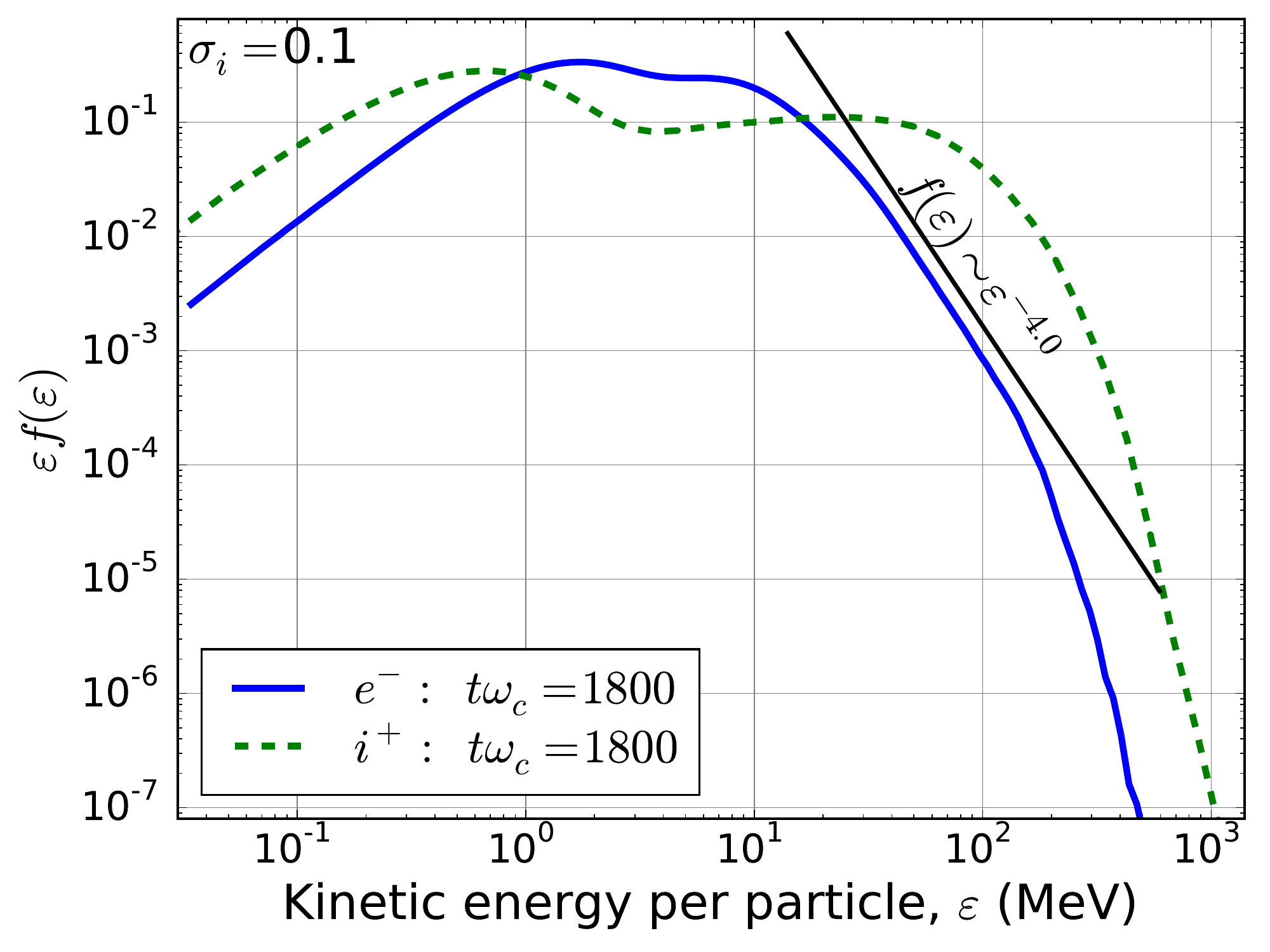}\\%
\includegraphics*[width=8.cm,trim=0mm 0 0mm 0]{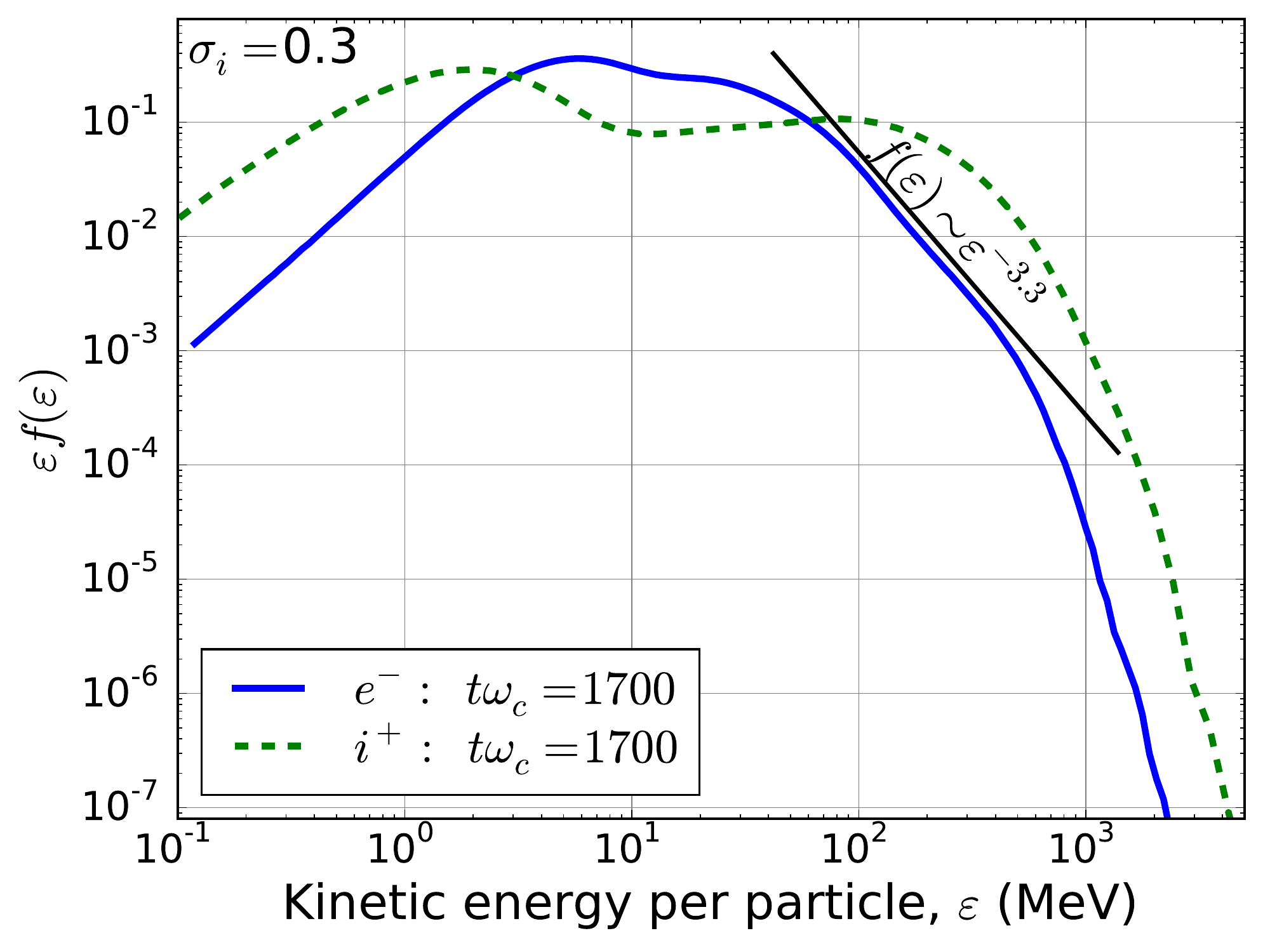}%
&\includegraphics*[width=8.cm,trim=0mm 0 0mm 0]{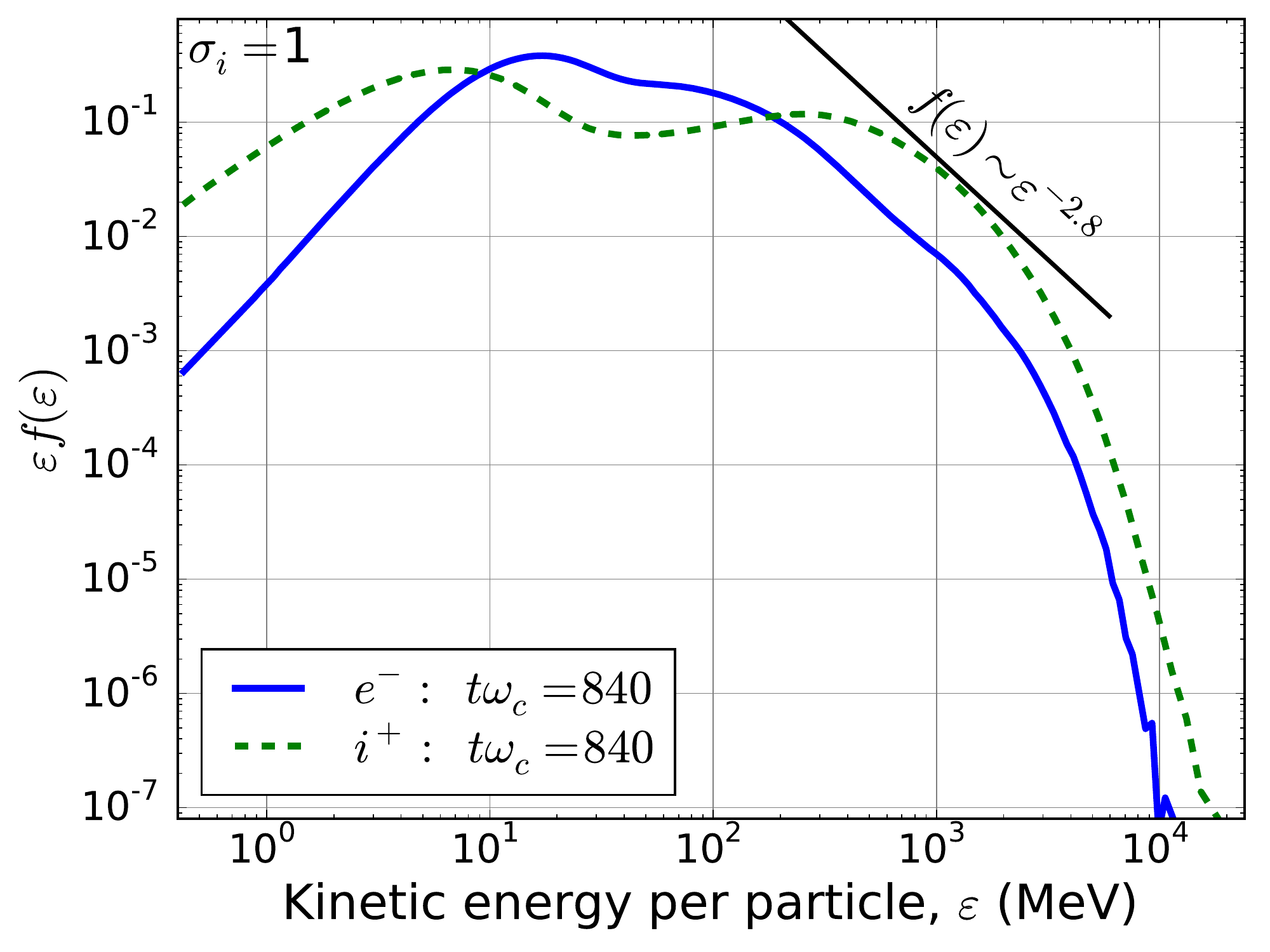}\\
\includegraphics*[width=8.cm,trim=0mm 0 0mm 0]{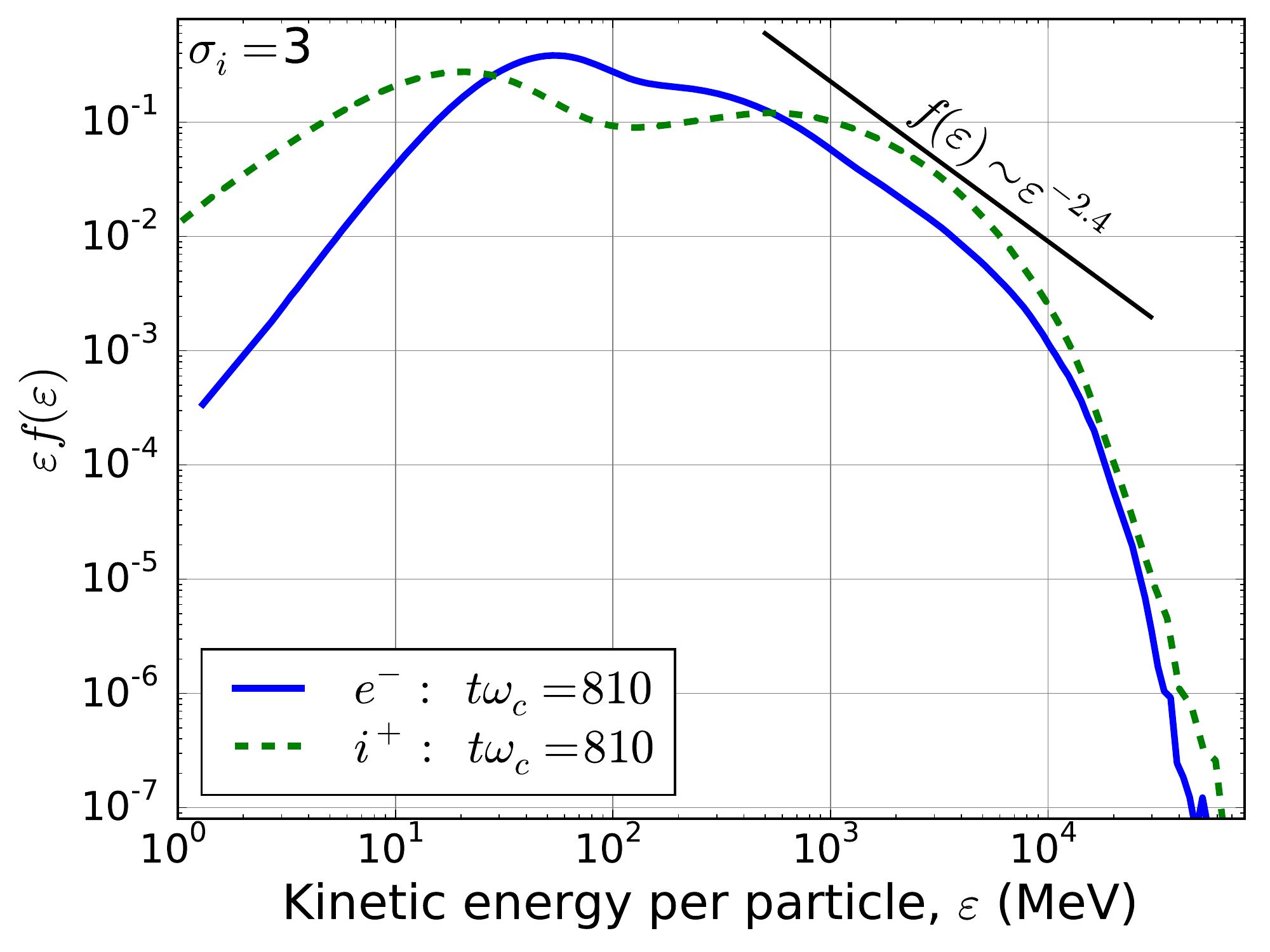}%
&\includegraphics*[width=8.cm,trim=0mm 0 0mm 0]{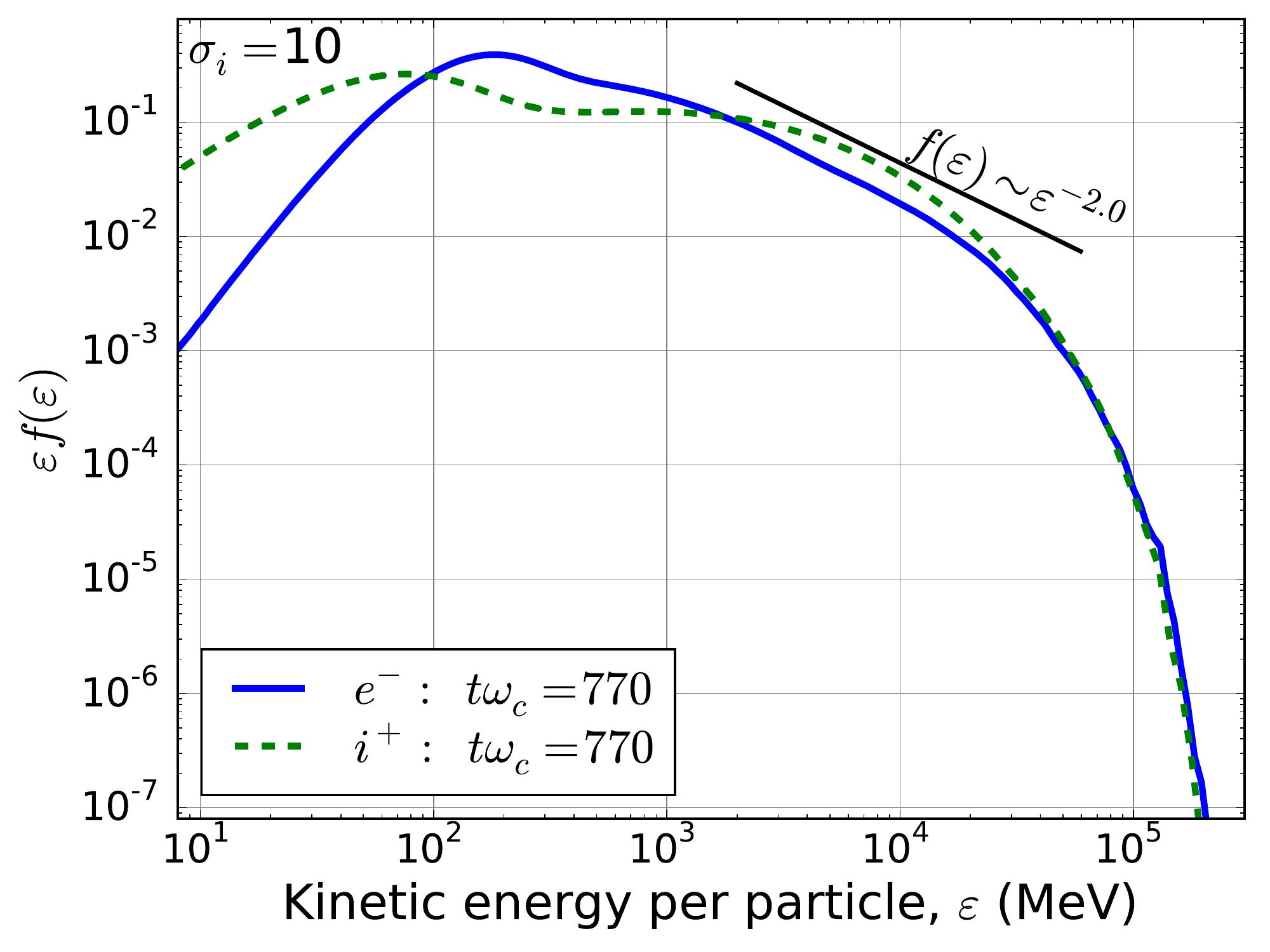}\\
 & \includegraphics*[width=8.cm,trim=0mm 0 0mm 0]{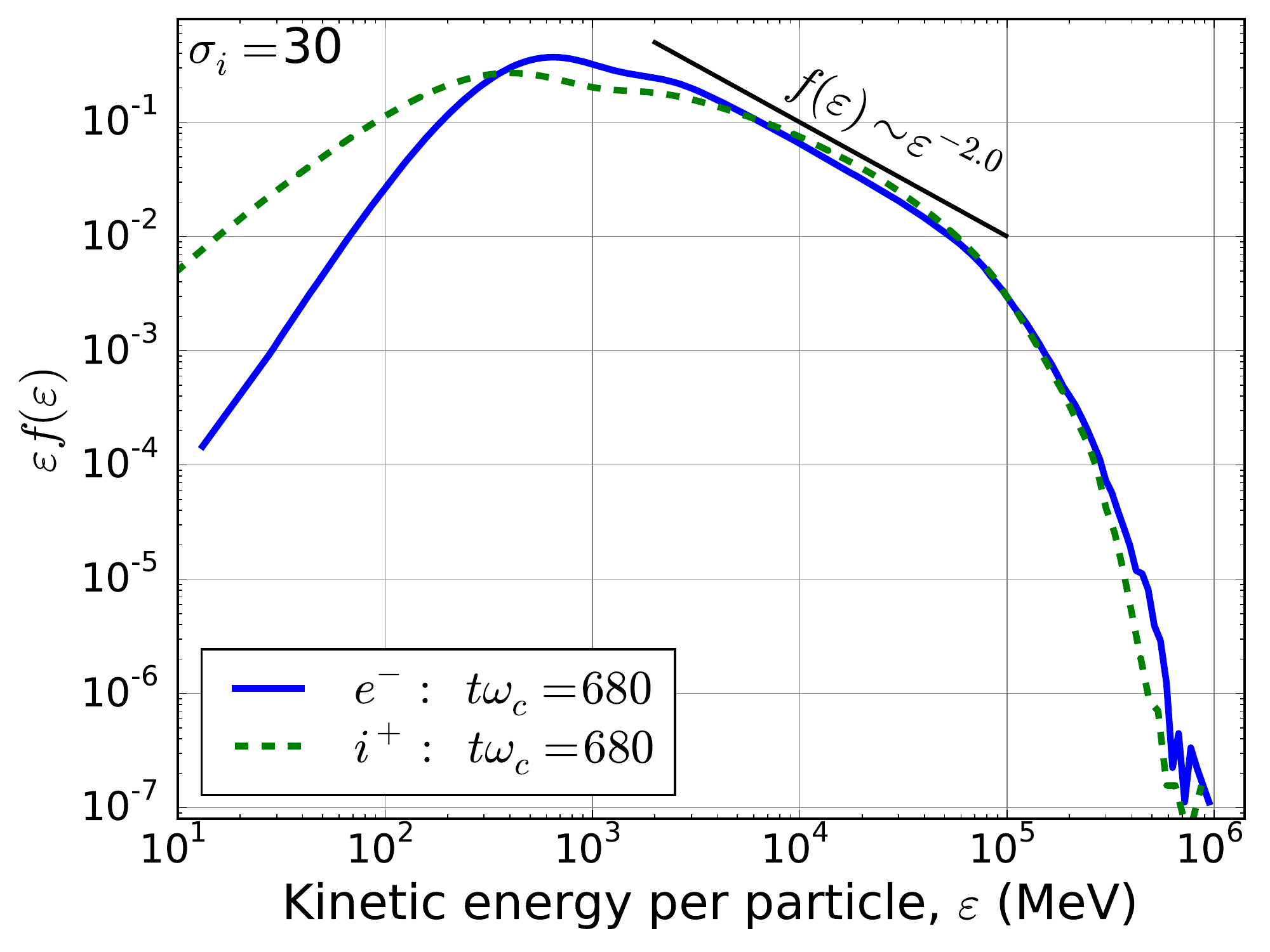}\\
\end{tabular}
\caption{
Electron and ion particle energy distributions (at the end of reconnection)
for a range of $\sigma_i$ from $0.03$ (upper left) to 30 (lower right) versus energy per particle; black lines indicate the slope corresponding to the fitted power-law indices of the electron distributions.
\label{fig:elecIonDist1}
}
\end{figure*}

The final spectra for many different~$\sigma_i$ can be compared in Fig.~\ref{fig:elecIonDist1}.
In each case (with the possible exception of $\sigma_i=0.03$) a power law can be identified in the high-energy part of the electron spectrum (well above the average electron energy).
We have not yet determined how to characterize the ion spectra for low $\sigma_i$, but focus instead on electrons as the primary emitters of observable radiation. Other interesting aspects of particle acceleration in relativistic reconnection, such as the asymptotic dependence on system size $L_x$ \citep{Werner_etal-2016,Guo:2015}, ion acceleration \citep{Melzani:2014b,Guo:2016}, angular distributions \citep{Cerutti_etal-2012b, Kagan_etal-2016}, acceleration mechanisms \citep{Guo:2014,Guo:2015,Nalewajko:2015}, etc., are left for future studies.

In contrast to our previous pair-reconnection work \citep{Werner_etal-2016}, where the high-energy parts of the electron energy spectra were well fitted by a power law times an exponential or super-exponential cutoff, the electron high-energy cutoffs in the present study are not well fitted by either exponential or super-exponential forms (nor a combination of both).  
This difference likely results from the somewhat different simulation parameters.  
Therefore we developed a new, more robust technique that finds a best-fitting power law $f(\varepsilon)\sim \varepsilon^{-p}$ by identifying the longest section of the spectrum with nearly-constant logarithmic slope $p(\varepsilon)\equiv -d \log f / d \log \varepsilon$; once a power law is fit over that section, we identify the high-energy cutoff $\varepsilon_c$ as the energy where the actual $f(\varepsilon)$ equals $e^{-1}$ times the pure power-law fit. 
The details of the fitting procedure are described in Appendix~\ref{sec:fitting}.
Because the fitting procedure depends on a few somewhat arbitrary parameters, such as the criterion for the slope to be `nearly-constant,' we performed multiple fits with different sets of such `under-the-hood' parameters for each run and display the median results; when all fits yield similar results, we judge the procedure to be robust, and characterize this robustness with `error' bars that encompass the middle 68\% of the fits.  
However (as will be shown below in Fig.~\ref{fig:alphaAndGcVsL}), the statistical simulation-to-simulation variation (for runs with the same $\sigma_i$ and~$L_x$) is generally larger than the fitting uncertainty; e.g., for $L_x=120\rho_c$, $p$ can vary over a range of roughly~$\pm 10$--20\%, and $\varepsilon_c$ over a factor of~2 or more.

\begin{figure}
\begin{tabular}{c}
\includegraphics*[width=8.cm,trim=0mm 0 0mm 0mm]{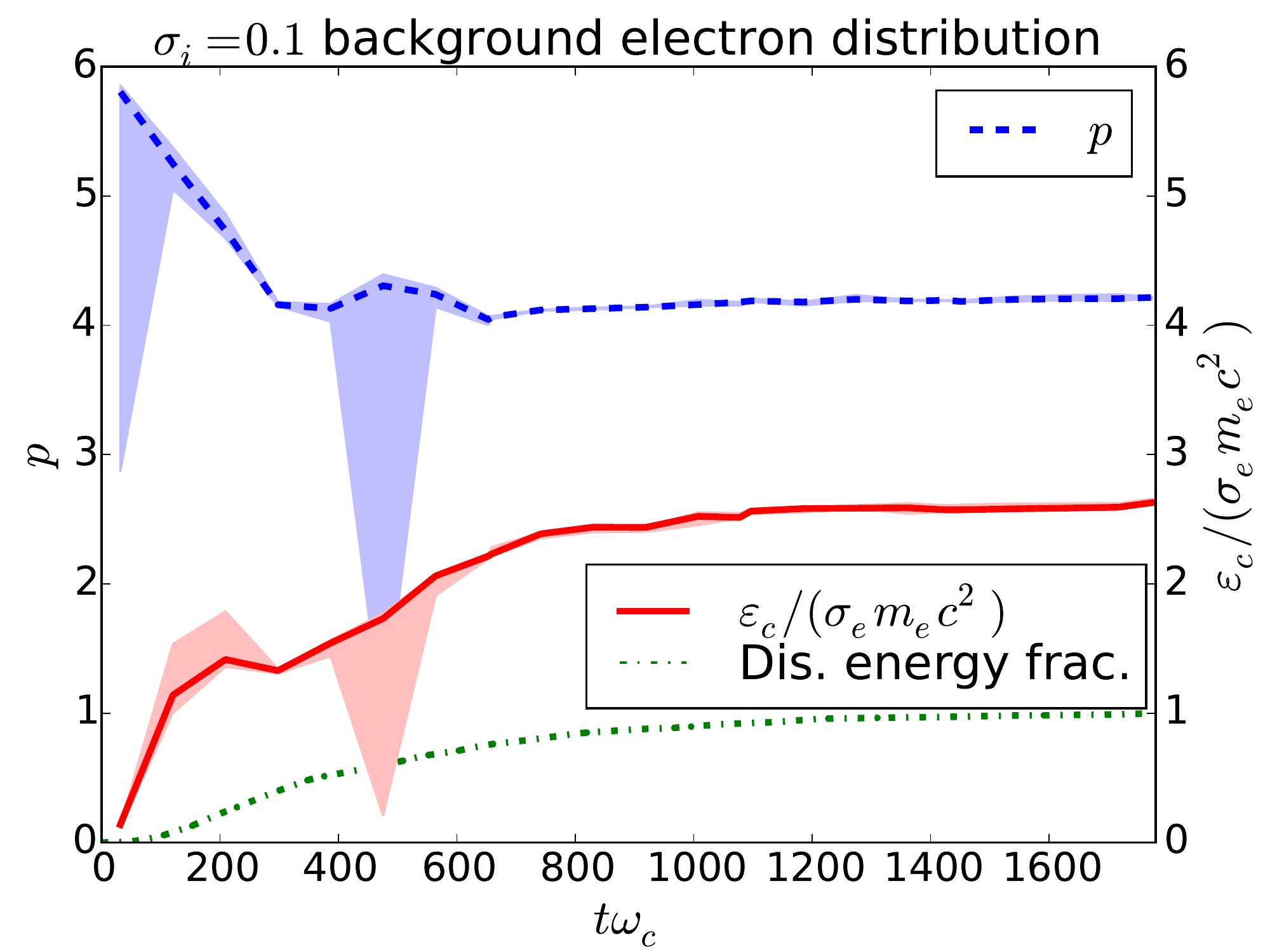}\\
\includegraphics*[width=8.cm,trim=0mm 0 0mm 0mm]{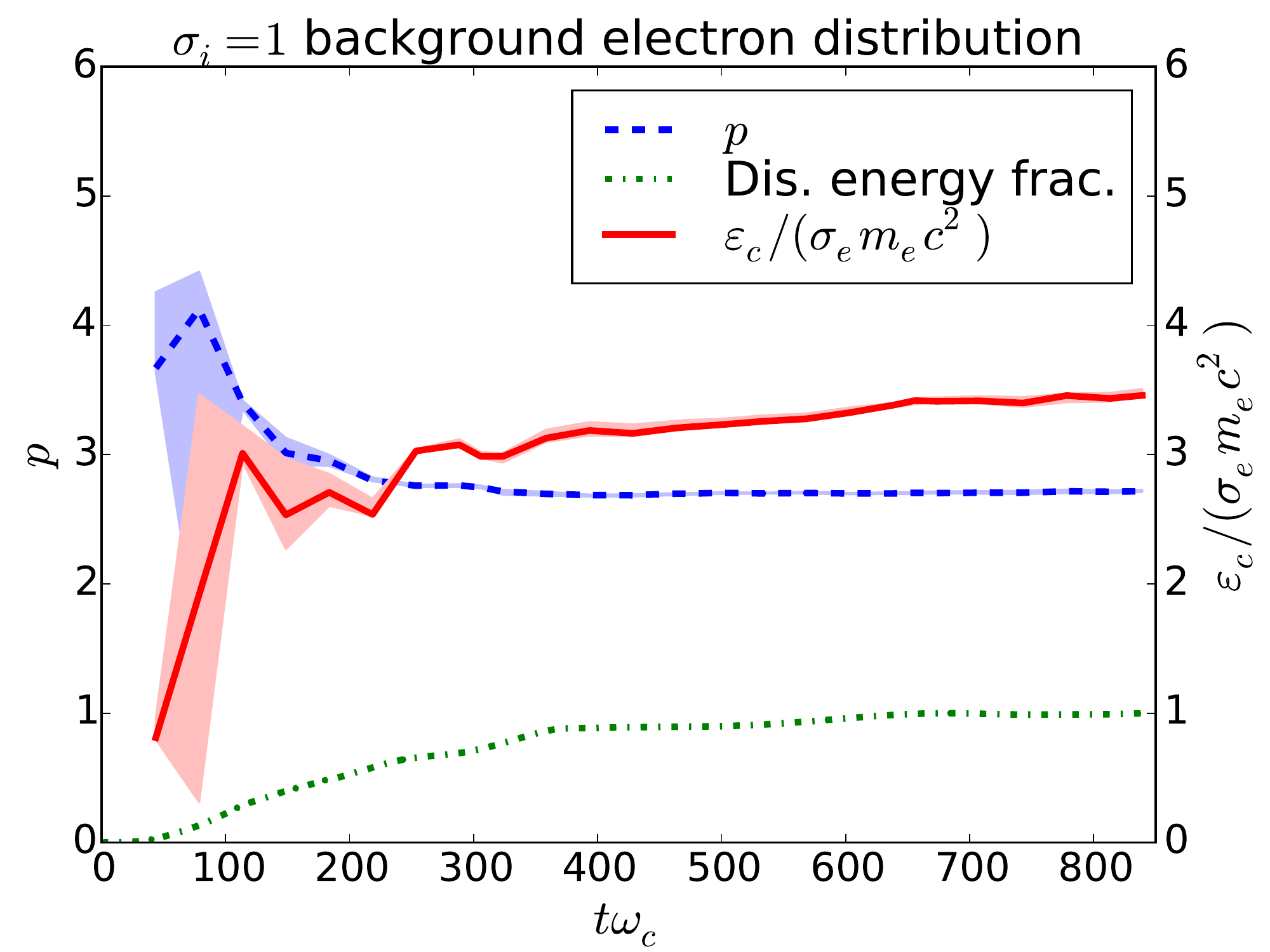}\vspace{-4mm}\\%
\end{tabular}
\caption{
The power-law index $p$ (blue) and the normalized high-energy cutoff 
$\varepsilon_c/(\sigma_e m_e c^2)$ (red)
of the electron distribution, graphed versus time, for $\sigma_i=0.1$ and $\sigma_i=1$, with the surrounding shaded regions indicating
uncertainty in the fitting procedure.  
For comparison, the green dashed line shows the
magnetic energy dissipated (arb. units).
\label{fig:pVsTime}
}
\end{figure}

At multiple snapshots in time, we analyse all background electrons in the simulation, keeping in mind that (especially for our aspect ratio, $L_y/L_x=2$) at any time, a substantial number remain in the upstream region without having been energized by reconnection.  The time dependence of power-law index and high-energy cutoff is shown in Fig.~\ref{fig:pVsTime} for two runs, $\sigma_i=0.1$ and~$1$. 
One can see that the index and cutoff reach levels close to their final values well before reconnection slows down due to the exhaustion of magnetic free energy; we take the final values at the end of reconnection to be the representative values of $p$ and $\varepsilon_c$ for each simulation.  

\begin{figure}
\includegraphics*[width=8.cm,trim=0mm 0 0mm 0]{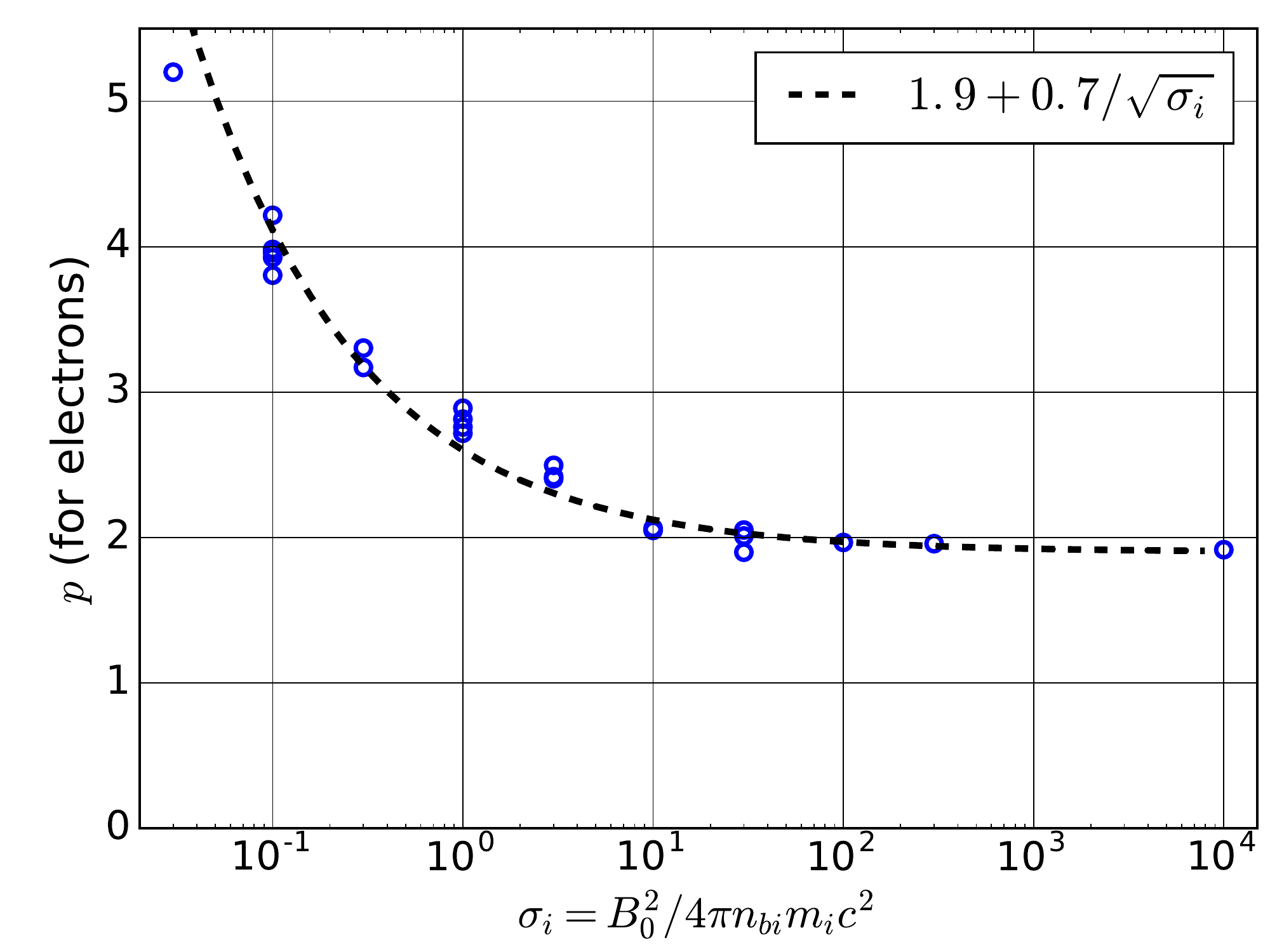}%
\caption{
The power-law index $p$ versus~$\sigma_i$, for electron energy spectra 
at the end of reconnection; error bars showing the robustness of fit are not shown because they are not significantly larger than the $\circ$-symbols.  (For $\sigma_i=0.03$, however, the automatic 
fitting procedure failed to identify a single convincing power-law slope, with error bars covering $1<p<8$;
we plot a value based on visual inspection (see Fig.~\ref{fig:elecIonDist1}), but consider it to be highly uncertain.)
\label{fig:alphaVsSigma}
}
\end{figure}

Figure~\ref{fig:alphaVsSigma} shows the variation in the electron power-law index $p$ with $\sigma_i$; for some values of $\sigma_i$, more than one simulation (identical except for random initial particle velocities) were
run to gauge statistical variations in the power-law index.
We find empirically that, for the range of $\sigma_i$ studied, $p$ can be approximated by the simple formula (which hopefully may find useful applications for astrophysical modelling, see \S\ref{sec-astro})
\begin{eqnarray} \label{eq:p_e}
  p(\sigma_i) & \approx & 1.9 + 0.7/\sqrt{\sigma_i} \,.
\end{eqnarray}

\begin{figure}
\includegraphics*[width=8.cm,trim=0mm 0 0mm 0]{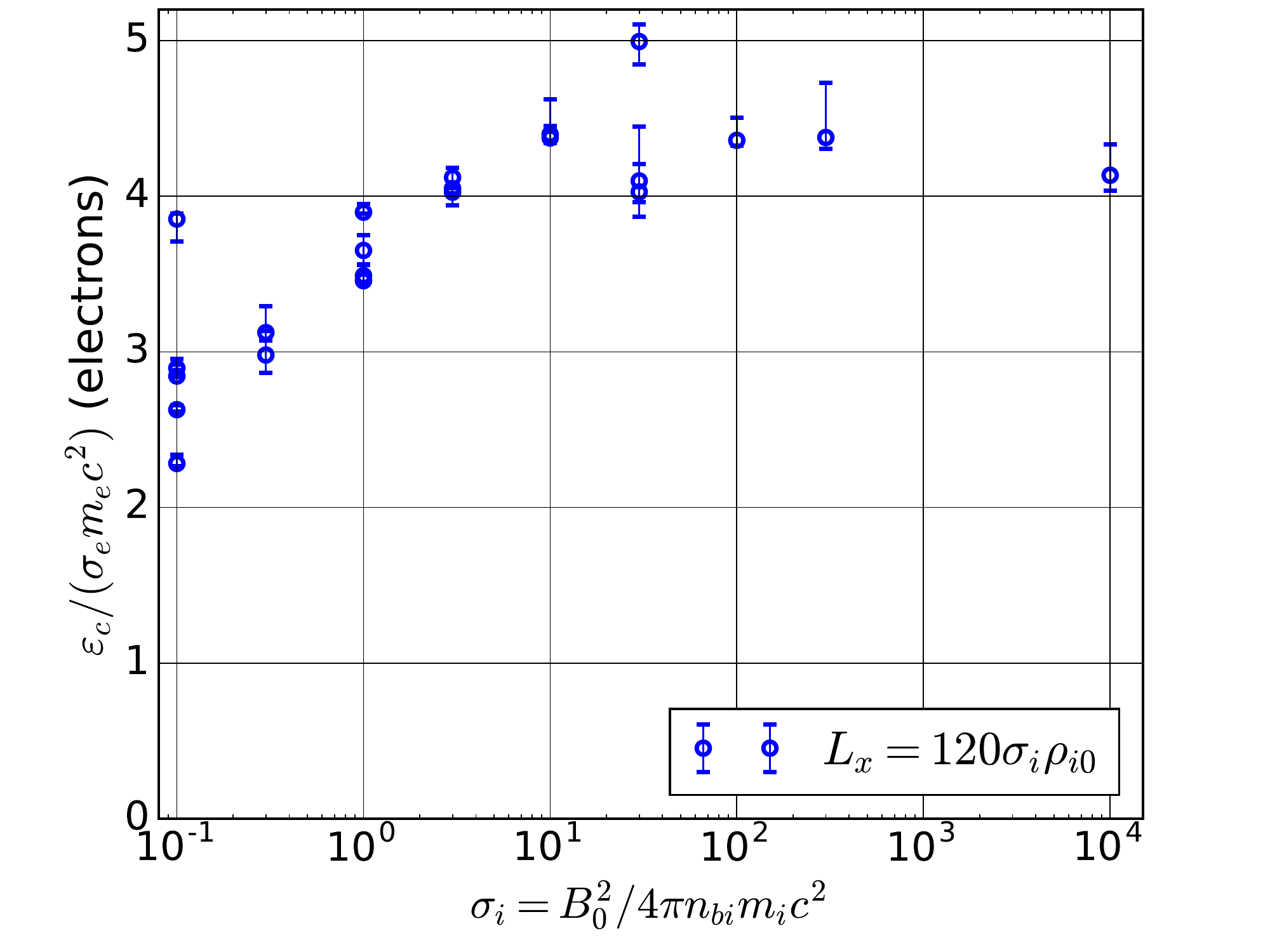}%
\caption{
$\varepsilon_c$ versus~$\sigma_i$.  The error bars show variation due to different choices of parameters used to fit the power law and cutoffs.  Results for
$\sigma_i=0.03$ are not shown because of the high uncertainty in the power-law index.
\label{fig:gcVsSigma}
}
\end{figure}

For large $\sigma_i$, where electron-ion reconnection should be identical to electron-positron reconnection, $p$ approaches a value around 1.9, consistent with previous studies of pair-plasma reconnection.  For these simulations, by virtue of our initial setup, $\sigma_{\rm hot} \rightarrow 25$ as $\sigma_i \rightarrow \infty$; a much larger $\sigma_{\rm hot}$ would yield $p$ closer to 1 than 2 \citep{Guo:2014,Guo:2015,Werner_etal-2016}.
In the semirelativistic regime we find $p$ varying between 2 and 4 as $\sigma_i$ decreases from about 10 to~$0.1$, reaching a value between $2.5$ and 3 at~$\sigma_i=1$.  
For $\sigma_i < 0.1$ we have only one simulation, $\sigma_i=0.03$, and its power law appears steeper than for $\sigma_i=0.1$; however, the steepness makes it hard to measure the index $p$ with much confidence.

We note that the $\sigma_i^{-1/2}$ scaling of the electron power-law index in the semirelativistic (small-$\sigma_i$) limit, expressed by Equation~\ref{eq:p_e}, can be understood within the framework of stochastic second-order Fermi acceleration of ultrarelativistic electrons bouncing back and forth between plasmoids  moving randomly along the reconnection layer. Denoting the typical separation between plasmoids in the plasmoid chain as $\lambda_{\rm pl}$ and the typical plasmoid speed as~$v_{\rm pl}$ (expected to be of order $V_A$ and hence $\ll c$ in the semirelativistic regime), we can estimate the average particle energy gain per bounce as $\Delta \varepsilon \sim (v_{\rm pl}/c)^2 \varepsilon$.  The typical time between bounces is simply $\Delta t_b \sim \lambda_{\rm pl}/c$, resulting in an effective acceleration time-scale of $t_{\rm acc} = \varepsilon \Delta t_b/{\Delta \varepsilon} = {\rm const} \cdot c \lambda_{\rm pl}/v_{\rm pl}^2$. Thus, a typical energetic electron, bouncing between two plasmoids large enough to contain it, undergoes random-walk diffusion in energy space until the two plasmoids come together and merge.  The particle then `escapes' from the interplasmoid acceleration region, e.g., by becoming trapped inside the merged plasmoid, with a typical `escape time' that can be estimated simply as the characteristic time for two plasmoids to approach each other, $t_{\rm esc} \sim \lambda_{\rm pl}/v_{\rm pl}$ (omitting factors of order unity). 
Thus, the second-order Fermi power-law index should scale as $ p = 1 +  t_{\rm acc}/t_{\rm esc} = 1 + {\rm const}\, c/v_{\rm pl}$. 
Since $v_{\rm pl} \simeq V_A \approx c \sigma_i^{1/2}$ in the $\sigma_i \ll 1$ regime, we see that the power-law index has a $\sigma_i$ dependence of the general form 
$p = 1 + C \sigma_i^{-1/2}$, 
where $C$ is a constant of order unity, consistent with our numerical findings.

The high-energy cutoff $\varepsilon_c$ depends mostly on $\sigma_i$ (Fig.~\ref{fig:gcVsSigma}); as $\sigma_i$ varies over more than 3 orders of magnitude, the cutoff energy scales as an $O(1)$ prefactor times $\mu \sigma_i = \sigma_e$, or roughly $\varepsilon_c \approx 4 \sigma_e m_e c^2$.
In the ultrarelativistic limit, the prefactor in this study is roughly twice that found in \cite{Werner_etal-2016}; the difference may be due to different simulations setups.
The normalized cutoff energy $\varepsilon_c/\sigma_e m_e c^2$ rises slowly with $\sigma_i$ in the semirelativistic regime, from around~2.5 to~4 or~4.5 as $\sigma_i$ goes from $0.1$ to~10.

Due to random fluctuations from simulation to simulation and the challenges of making precise measurements, as well as the difficulty of identifying convergence in the limit $L_x\rightarrow \infty$, we have not been able to conclude with confidence whether the computed power-law indices and cutoff energies are truly independent of $L_x$ for $L_x>120\rho_c$.
We have specifically examined the $L_x$ dependence of indices and cutoffs for the cases of $\sigma_i=0.1$ and $\sigma_i=1$ by extending our studies to $L_x=160\rho_c$ for these cases, as shown in Fig.~\ref{fig:alphaAndGcVsL}.
First, however, we note that our system size $L_x=120\rho_c$ is already squarely in the large-system regime described by \cite{Werner_etal-2016} (for electron-positron reconnection in simulations with aspect ratio $L_y/L_x=1$, the critical system size separating small- and large-system regimes, in the language of this paper, is reached when $L_x$ equals $L_c=20 \, \sigma_e \rho_{e0}=20\,\sigma_i \rho_{i0}$).  
Indeed, the $\sigma_i=1$ case yields an index and cutoff that look quite independent of $L_x$ for $L_x \gtrsim 40\sigma_i \rho_{i0}$.
It is much less clear whether the $\sigma_i=0.1$ simulation has converged with respect to $L_x$, although the trend suggests that $ p \gtrsim 4$ for larger $L_x$, and we suspect, based on the convergence of $p(\sigma_i=1)$ with $L_x$ and the apparent agreement between $L_x=120\, \rho_c$ and $L_x=160\,\rho_c$, that $p$ may near its asymptotic limit at $L_x \gtrsim 120 \sigma_i \rho_{i0}$.
That said, establishing an asymptotic limit is always difficult, especially since the computational resources needed to perform a (2D) simulation scale as $L_x^3$ (with our explicit algorithms), preventing us from exploring much larger~$L_x$.

\begin{figure*}
\includegraphics*[width=8.cm,trim=0mm 0 0mm 0]{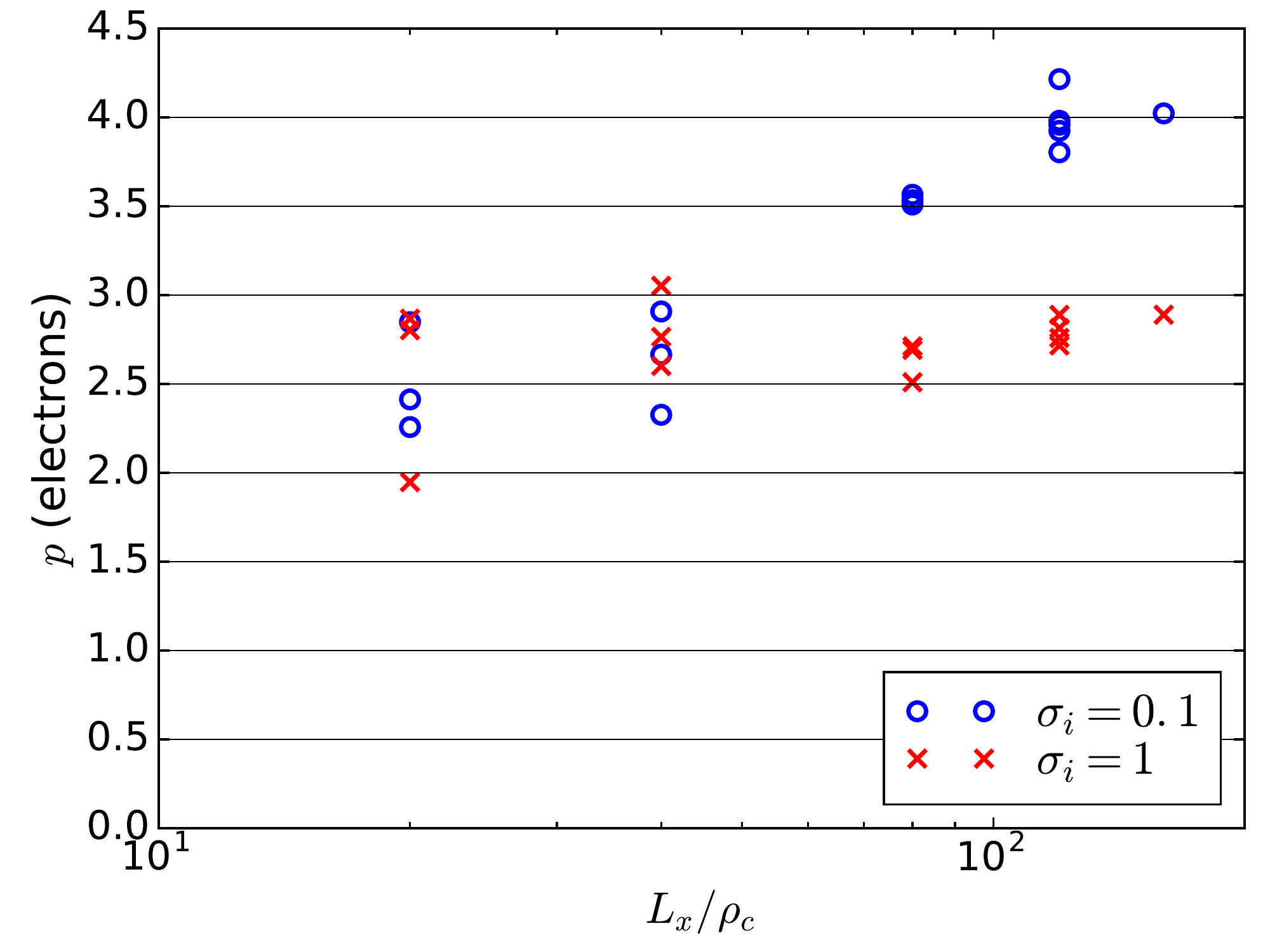}%
\hspace{5mm}%
\includegraphics*[width=8.cm,trim=0mm 0 0mm 0]{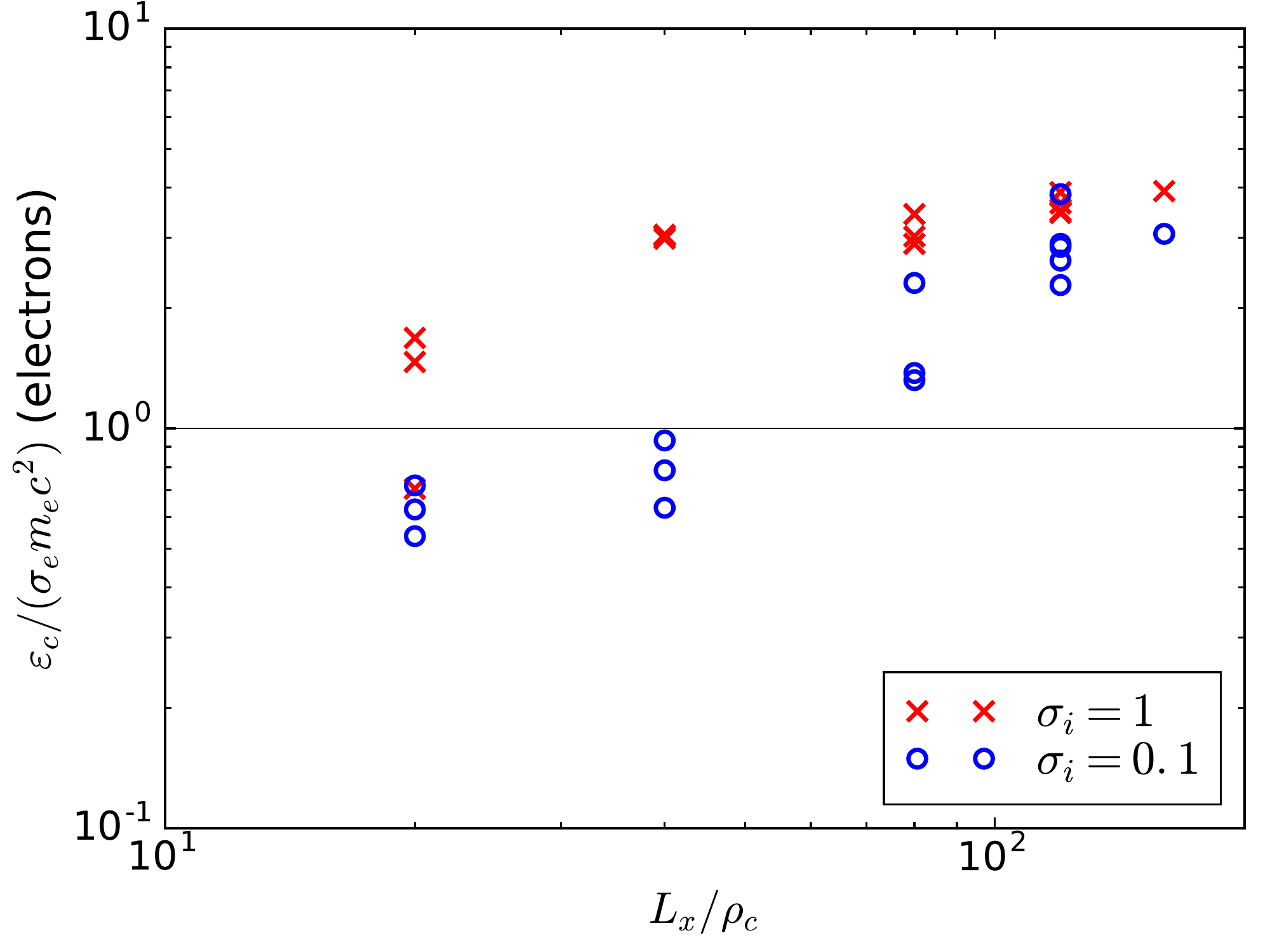}%
\caption{
The power-law index $p$ (left-hand panel) and cutoff energy $\varepsilon_c$ (right-hand panel) versus system size $L_x$ for electrons; each data point represents a separate simulation.
(The error bars representing uncertainty of fit are left off these plots because, except for the smallest system size, they are not much larger than the X and O symbols, hence smaller than the statistical variation between different simulations.) 
\label{fig:alphaAndGcVsL}
}
\end{figure*}

The ion spectra $f_i(\varepsilon)$ generally appear nonthermal, but exhibit obvious power laws only in the ultrarelativistic regime (Fig.~\ref{fig:elecIonDist1}).
For $\sigma_i > 10$, the ion spectra show nonthermal power laws that closely match the electron spectra at high energies; however,
for $\sigma_i \lesssim 10$ the ion spectra have a hard slope $p\approx 1$ (which appears horizontal in the compensated spectra of Fig.~\ref{fig:elecIonDist1}) that emerges immediately out of the cold (upstream) Maxwellian distribution and extends only to medium energies. 
While not a really convincing power law, $f_i(\varepsilon)$ in this region is flatter and broader than a Maxwellian (compare, e.g., the Maxwellian $t=0$ ion spectrum with late-time spectra in Fig.~\ref{fig:fvsTime}). 
At energies well below the \emph{electron} cutoff energy, the ion spectra turn significantly downward, although---at these high energies---there are always more ions than electrons at the same energy (consistent with the fact that ions collectively gain more energy than electrons).
We note that the ion spectra---even if described by power laws---might be expected to show a break around $\varepsilon\sim m_i c^2 \approx 10^3\:$MeV, where protons become transrelativistic.  Indeed, the hard $p\approx 1$ slope never extends beyond~$10^3\:$MeV (most notably for $\sigma_i=3$ in Fig.~\ref{fig:elecIonDist1}).

In summary, both electrons and ions are accelerated to nonthermal energy distributions, with the electron spectra forming pronounced power laws at high energies.  
The electron power-law index $p$ becomes steeper as $\sigma_i$ decreases, and extends to a cutoff energy $\varepsilon_c \sim \sigma_e m_e c^2$.  
In the ultrarelativistic limit, electron and ion spectra become similar, with a power-law slope $p \approx 1.9$ that may depend on $\sigma_{e,\rm hot} \approx 2\sigma_{\rm hot}$, which is~50 for this investigation. 
We know from electron-positron reconnection studies \citep{Sironi_Spitkovsky-2014,Guo:2015,Werner_etal-2016} that in the ultrarelativistic limit a smaller $\sigma_{e,\rm hot}$ results in a larger/steeper power-law index~$p$; however, in the pair case, small $\sigma_{e,\rm hot}$ means that electrons do not get accelerated very much (compared to their initial thermal energy).  
In contrast, electron-ion reconnection can (e.g., for $\sigma_i \approx 1$) yield steeper slopes $p>2$ while maintaining $\sigma_{e,\rm hot} \gg 1$ (allowing significant electron acceleration).
The ion spectra are harder to characterize, but at high energies there are more ions than electrons with any given energy.

%----------------------------------------------------------------------------------

\section{Astrophysical Implications}
\label{sec-astro}

The results obtained in this paper are relevant to a number of astrophysical systems, including accretion disc coronae (ADC) and blazar jets, which are composed of electron-ion plasma likely undergoing reconnection in the semirelativistic regime.  

The hard X-ray spectra of ADC in black-hole X-ray binaries in the high-soft (HS) and steep power law (SPL) states often exhibit relatively steep power-law tails that have spectral indices $\alpha \equiv - d\ln F_\nu/ d\ln\nu \approx 1.5$ and extend to well beyond 1~MeV \citep{Remillard_McClintock-2006,Done_etal-2007}. This high-energy radiation is generally believed to be produced by inverse Compton (IC) scattering of soft X-ray seed photons coming from the accretion disc by a nonthermal population of highly relativistic coronal electrons with an inferred power-law index $p=2\alpha+1 \approx 4$.

Blazars often feature double-humped broad-band spectral energy distributions (SEDs) attributed to synchrotron radiation (lower-energy component) and IC scattering (higher-energy component). 
There are two main spectral classes of blazars: flat-spectrum radio quasars (FSRQs) with synchrotron peaks in the far-infrared range and IC peaks in the MeV range; and high-frequency peaked BL Lacs (HBLs) with synchrotron peaks in ultraviolet (UV) and X-rays and IC peaks in the GeV--TeV range \citep[see][for a recent review]{Madejski_Sikora-2016}.
A systematic trend of increasing synchrotron luminosity and increasing Compton dominance with decreasing synchrotron peak frequency is known as the blazar sequence \citep{Fossati_etal-1998}. 
Typical spectral indices of FSRQs in the X-rays are $\alpha \sim 0.6$ \citep{Sikora_etal-2013}, indicating nonthermal distributions of ultrarelativistic electrons with $p \simeq 2.2$.
To explain the observed SEDs, these electron energy distributions should extend to $\gamma_{\rm max} \sim 10^3$ for FSRQs and to $\gamma_{\rm max} \gtrsim 10^4$ for HBLs. There are even a few ultrahigh-frequency-peaked BL Lacs (UHBLs) that demand $\gamma_{\rm max} \gtrsim 10^4$, possibly requiring hadronic models \citep{Cerruti_etal-2015}. 

The nonthermal electron power-law distributions inferred in these systems have often been attributed to first-order Fermi acceleration in strong shocks \citep{Spada_etal-2001,Sokolov_etal-2004}, but such a model is viable only if the plasma involved is weakly magnetized.  In contrast, ADCs are thought to be strongly magnetized; the buoyant rise of magnetic energy to form magnetically dominated coronae in accretion disc models with vertical stratification is well documented in local (shearing box) simulations \citep{Miller_Stone-2000,Bai_Stone-2013,Salvesen_etal-2016}, suggesting that reconnection is the dominant dissipation and particle acceleration mechanism \citep{Galeev_etal-1979}.  As for blazar jets, the likely importance of magnetic energy out to large distances in these systems is due to the nature of the hydromagnetic acceleration process, which is efficient as long as the jet is magnetically dominated but becomes extremely inefficient once the kinetic energy and Poynting fluxes become comparable \citep{Begelman_Li-1994,Vlahakis_Konigl-2003a,Vlahakis_Konigl-2003b,Komissarov_etal-2007}.  
Jet acceleration can be facilitated by dissipation of magnetic energy, e.g., through reconnection triggered by current-driven instabilities 
\citep{Begelman-1998,Giannios:2006,Tchekhovskoy_Bromberg-2016,Sironi_etal-2015}. 
Barring such dissipation, the conversion of Poynting flux to kinetic energy would get `stuck' at values of $\sigma_i \sim O(1)$.
 
It is thus interesting to observe that the nonthermal electron distributions with $2.2 < p < 4$, which emit radiation spectra with $0.6 \lesssim \alpha \lesssim 1.5$ seen in ADC and blazars, can be produced, according to our simulations, by magnetic reconnection with $4 \gtrsim \sigma_i \gtrsim 0.1$ --- just the range where substantial dissipation should go hand-in-hand with gradual acceleration of the jet beyond approximate equipartition.  We therefore suggest that reconnection in this semirelativistic, near-equipartition regime, is a natural way to produce the observed nonthermal spectra in blazar jets, and perhaps in ADC as well.  

Furthermore, adopting our estimated cutoff Lorentz factors for the accelerated electrons, $\gamma_c \sim 4\sigma_e = 4 \mu \sigma_i$, and assuming $\sigma_i = 4$ (corresponding to $p\simeq 2.2$ and $\alpha = 0.6$) for both FSRQs and HBLs, we obtain $\gamma_c \sim 3 \times 10^4$. This corresponds to a frequency of the observed synchrotron peak $\sim 10^{15} B_{\rm G} \Gamma_j\:$Hz, where $B_{\rm G}$ is the co-moving magnetic field in the emitting region, measured in Gauss, and $\Gamma_j$ is the Doppler boost due to the bulk Lorentz factor of the jet, which typically lies in the range $\Gamma_j \sim 10-40$ \citep{Savolainen_etal-2010}.  FSRQs, with synchrotron peak frequencies extending from as low as $\sim 10^{13}\:$Hz into the UV band, would therefore require low magnetic fields $B_{\rm G} \lesssim 1$, consistent with the deduction that gamma rays must be produced at rather large distances from the black hole, $r > 0.1\:{\rm pc} \sim 10^3 R_{\rm g}$, where $R_{\rm g} = GM/c^2$ is the gravitational radius, to avoid excessive pair-production opacity \citep{Ghisellini_Madau-1996,Nalewajko_etal-2014a}.
The gamma rays from HBLs, on the other hand, could be produced closer to the jet base (at $10-100 R_{\rm g}$) with corresponding magnetic field strengths $B_{\rm G} \sim 10^2-10^3$. Electron Lorentz factors $\sim O(10^4)$ are also consistent with the TeV photons being produced by the synchrotron-self-Compton mechanism in HBLs, in contrast to the MeV--GeV gamma rays in FSRQs, which most likely result from Comptonization of ambient photon sources \citep{Sikora_etal-1994,Ghisellini_etal-1998}.      
  
In addition to the distributed, and relatively steady, blazar emission discussed above, blazars often exhibit intense, rapid flares in gamma rays \citep{Abdo_etal-2010,Abdo_etal-2011b,Aharonian_etal-2007}. To avoid excessive pair production, which would prevent the gamma rays from escaping, the flaring regions must have local bulk Lorentz factors that exceed the mean for the jet by a factor of a few \citep{Begelman_etal-2008}.  Such `minijets' could represent outflow regions from relativistic reconnection sites \citep{Giannios:2009,Nalewajko:2011}, with local values of $\sigma_i$ several times larger than the mean.  We would then predict the spectral slopes to be correspondingly flatter and the cutoffs higher in proportion to $\sigma_i^2$ (assuming similar sources of seed photons for Comptonization).
An additional argument in support of the minijet model follows from our finding (see \S\ref{sec:PlasmaFlows})  that relativistic reconnection-driven electron fluid flows do not actually require high $\sigma_i$ but readily occur in the semirelativistic, small-$\sigma_i$ cases as well. Since gamma radiation is produced by the electrons, the minijet mechanism can thus operate successfully over a broad range of plasma magnetizations, including the $\sigma_i = O(1)$ regime favoured by the observed spectral indices as described above.

Finally, we would like to stress the crucial role of the electron-ion plasma composition, in the semirelativistic limit, for explaining the range of spectral indices typically seen in ADC and blazar jets.  Kinetic simulations of relativistic reconnection in electron-positron pair plasmas that produce highly relativistic energy cutoffs---necessary to explain the blazar observations---yield nonthermal electron power-law indices close to $p=1$ \citep{Sironi_Spitkovsky-2014, Guo:2014, Werner_etal-2016}. 
While such hard electron distributions may be necessary to explain the most extreme flaring events in blazars, characterized by very flat photon spectra with spectral indices near zero \citep{Hayashida_etal-2015}, they are  much too flat to be consistent with most of the regular blazar emission observations, even if one took cooling into account. In contrast, in electron-ion reconnection, electrons are accelerated to ultrarelativistic energies even for modest magnetizations $\sigma_i \sim 1$, which, according to our simulation results, are needed to explain the range of spectral indices observed in these systems. 

%----------------------------------------------------------------------------------

\section{Conclusions}
\label{sec-conclusions}

In this paper, we performed a comprehensive numerical investigation of antiparallel (no guide magnetic field) collisionless relativistic magnetic reconnection in an electron-ion plasma, relevant to many astrophysical systems, such as coronae of accreting BHs in XRBs and AGNs, and their relativistic jets, including blazars. We used a series of large 2D relativistic PIC simulations with real proton/electron mass ratio $\mu = m_i/m_e = 1836$ to probe various fundamental aspects of reconnection, such as dynamics (reconnection rate, electron and ion outflows, and Hall-effect signatures), energetics (e.g., the energy partitioning between electrons and ions), and nonthermal particle acceleration in different astrophysically-relevant physical regimes. The main focus of our study was on exploring how these aspects are affected as one transitions from the semirelativistic regime (ultrarelativistic electrons but nonrelativistic ions), which has not been systematically explored before, to the fully relativistic regime (both ions and electrons are ultrarelativistic). Since this transition is quantified by the initial `cold' upstream ion magnetization parameter $\sigma_i = B_0^2/4\pi n_{b,i} m_i c^2$ (basically, twice the average upstream magnetic energy per background ion normalized by the ion rest mass), our parameter-space study explored a broad range of~$\sigma_i$,  
from very small (semirelativistic regime, $\mu^{-1} < \sigma_i \lesssim 1$; down to $\sigma_i=0.03$ in our simulations) to very large (ultrarelativistic regime, $\sigma_i \gg 1$; up to $\sigma_i=10^4$). 
We also explored the dependence of various reconnection characteristics on the system size~$L_x$ (while keeping the aspect ratio of our computational box fixed, $L_y/L_x=2$), with our largest production runs (for all $\sigma_i$) reaching well into the large-system, plasmoid-dominated regime. 

Our main findings can be summarized as follows. 
First, the average dimensionless reconnection rate $\beta_{\rm rec} = E_{\rm rec}/B_0$ during the active reconnection phase approaches a constant value $\sim 0.1$ in the ultrarelativistic limit $\sigma_i \gg 1$. 
This value is consistent with previous studies of relativistic electron-positron pair-plasma reconnection \citep{Sironi_Spitkovsky-2014,Guo:2015}, which is in fact expected since any dynamical differences between ions and electrons disappear in the ultrarelativistic limit (and in the absence of radiative losses) and so electron-ion-plasma reconnection should become indistinguishable from the extensively studied pair case.  Our simulations also show that as $\sigma_i$ is lowered below $\sim 1$, $\beta_{\rm rec}$ decreases.  This dependence of the reconnection rate on $\sigma$ can be attributed almost entirely to the scaling of the upstream Alfv\'en speed $V_A$ with $\sigma_i$: the reconnection inflow velocity properly normalized to~$V_A$, $\beta_{{\rm rec},A} = cE_{\rm rec}/B_0 V_A$, remains essentially constant (about 0.1) across the physical regimes, varying very little as $\sigma_i$ is varied over 3 orders of magnitude. We also find that the reconnection rate is rather insensitive to the system size for large enough systems. 

Our simulations also show that, just as in the nonrelativistic case, reconnection diffusion regions around X-points develop a two-scale structure in the semirelativistic regime, with narrow electron current layers embedded in broader ion layers.
Interestingly, we find that electron fluid outflows in these electron layers can locally become relativistic, even in the small-$\sigma_i$ cases, when the overall (ion) reconnection outflows are subrelativistic; however, even for high-$\sigma_i$, these fluid outflows do not become more than mildly relativistic with Lorentz factors of a few.  Because the potentially observable radiative signatures are dominated by electrons, this finding may have important astrophysical implications, especially in situations where relativistic motions of the emitting, flaring plasma fireballs are inferred, as in the minijets model for ultrarapid TeV flares in blazar jets~\citep{Giannios:2009}. 
Another interesting consequence of the different electron and ion flow patterns in the reconnection region in the semirelativistic regime is the development of the classic Hall-effect signatures near magnetic X-points (again, similar to nonrelativistic electron-ion reconnection): the out-of-plane magnetic field $B_z$ with the characteristic quadrupole structure and the corresponding dipolar (directed into the layer and downstream) in-plane electric field, which can be described by an electrostatic potential that forms potential wells inside the plasmoids.  As one transitions into the ultrarelativistic regime, these signatures weaken, as expected. 

We also explored the question of the partitioning of the released magnetic energy between electrons and ions as a function of~$\sigma_i$.  This issue is important for interpreting astrophysical observations (e.g., for evaluating the energy budget of a high-energy flare) and for computing observational signatures of global MHD simulations of, e.g., coronae of accreting black holes, since only the electron fraction of the released energy can be promptly radiated away and hence directly observed as a flare. We found that while the electrons and ions get equal amounts of energy in the ultrarelativistic limit, as expected, the electron energy fraction decreases as $\sigma_i$ is decreased, but eventually seems to asymptote at a finite value of about 1/4 in the semirelativistic limit. We then proposed a simple  empirical formula [see Eq.~(\ref{eq:energyPartition})] that describes this dependence. 

Finally, we characterized the resulting energy spectra of electrons and ions accelerated by reconnection. 
We found that both electrons and ions attain nonthermal distributions by the end of the reconnection process. 
In particular, we were able to obtain clear evidence for nonthermal particle acceleration of electrons, manifested as robust power-law energy distributions. We mapped out the dependence of the electron power-law index $p$ and high-energy cutoff $\varepsilon_c$ on $\sigma_i$ and~$L_x$, and found that the index becomes fairly insensitive to the system size but exhibits a relatively strong dependence on $\sigma_i$ in the semirelativistic regime, well-fitted by the empirical relationship $p = 1.9 + 0.7 \sigma_e^{-1/2}$ [see~Eq.~(\ref{eq:p_e})]. The electron cutoff energy is found to scale approximately linearly with $\sigma_i$, namely, $\varepsilon_c \simeq (3-5) \, \sigma_i m_i c^2$, over several orders of magnitude in $\sigma_i$, although its scaling with $L_x$ is not completely clear; a conclusive detailed investigation of this issue is left for a future study.  As for ions, based on our present simulations we cannot claim a power-law ion energy distributions in the semirelativistic regime, although in the ultrarelativistic case the ion distribution of course becomes very similar to that of the electrons, with a clear power law.  

We believe that our results have important astrophysical implications, e.g., for high-energy emission from blazar jets. In particular, the electron power-law indices ($p \sim 2-3$) that correspond to the nonthermal radiation spectra typically observed in blazars ($\alpha \sim 0.6$) can be produced by magnetic reconnection with modest ambient magnetization $\sigma_i \sim 1$, in agreement with the values that are naturally expected in theoretical models of relativistic jet acceleration.

Our present study acts as a bridge connecting two previously extensively-investigated limits: nonrelativistic electron-ion reconnection  (usually done in the context of, e.g., solar flares and Earth's magnetosphere) and ultrarelativistic pair-plasma reconnection (with applications to, e.g., pulsar winds and PWN). 
This research provides baseline knowledge of magnetic reconnection in the astrophysically-important semirelativistic electron-ion regime, describing the basic dynamics of reconnection as well as resulting particle acceleration, and
offering a concrete way to connect theoretical models with astrophysical observations by providing specific fitting formulae for several key parameters of the accelerated particle populations (such as $q_e$, $p$, and~$\varepsilon_c$).
Our study also opens up a number of exciting new opportunities that should be explored in the near future; it will serve as a departure point for more detailed studies of the semirelativistic regime, including the examination of the effects of guide magnetic field, upstream plasma temperatures, initial configurations other than Harris sheets, 3D effects, and synchrotron and IC radiative cooling.

%----------------------------------------------------------------------------------

\section*{Acknowledgements}

This work was supported by DOE grants DE-SC0008409 and DE-SC0008655, NASA
grants NNX12AP17G and NNX16AB28G, and NSF grant AST-1411879.
D.~A.~Uzdensky gratefully acknowledges the hospitality of the Institute for Advanced Study and support from the Ambrose Monell Foundation; and K.~Nalewajko received support from the Polish National Science Centre grant 2015/18/E/ST9/00580.

This work required several supercomputer allocations.
This work used the Extreme Science and Engineering Discovery Environment (XSEDE), which is supported by National Science Foundation (NSF) grant number ACI-1053575 \citep{XSEDE2014}, and in particular we used the Stampede supercomputer at the Texas Advanced Computing Center (TACC) at The University of Texas at Austin.
This work also used the Pleiades supercomputer, a resource provided by the NASA High-End Computing (HEC) Program through the NASA Advanced Supercomputing (NAS) Division at Ames Research Center.  
In addition, this work made use of an award of computer time provided by the Innovative and Novel Computational Impact on Theory and Experiment (INCITE) program; in particular, we used the Mira supercomputer and associated resources of the Argonne Leadership Computing Facility, which is a DOE Office of Science User Facility supported under Contract DE-AC02-06CH11357.

We also gratefully acknowledge the developers responsible for the {\sc Vorpal}/{\sc VSim} simulation code.

%%%%%%%%%%%%%%%%%%%%%%%%%%%%%%%%%%%%%%%%%%%%%%%%%%



\bibliographystyle{mnras}
%\bibliography{accel}
%\bibliography{WernerRelElecIonRecon.bib}

\begin{thebibliography}{}
\makeatletter
\relax
\def\mn@urlcharsother{\let\do\@makeother \do\$\do\&\do\#\do\^\do\_\do\%\do\~}
\def\mn@doi{\begingroup\mn@urlcharsother \@ifnextchar [ {\mn@doi@}
  {\mn@doi@[]}}
\def\mn@doi@[#1]#2{\def\@tempa{#1}\ifx\@tempa\@empty \href
  {http://dx.doi.org/#2} {doi:#2}\else \href {http://dx.doi.org/#2} {#1}\fi
  \endgroup}
\def\mn@eprint#1#2{\mn@eprint@#1:#2::\@nil}
\def\mn@eprint@arXiv#1{\href {http://arxiv.org/abs/#1} {{\tt arXiv:#1}}}
\def\mn@eprint@dblp#1{\href {http://dblp.uni-trier.de/rec/bibtex/#1.xml}
  {dblp:#1}}
\def\mn@eprint@#1:#2:#3:#4\@nil{\def\@tempa {#1}\def\@tempb {#2}\def\@tempc
  {#3}\ifx \@tempc \@empty \let \@tempc \@tempb \let \@tempb \@tempa \fi \ifx
  \@tempb \@empty \def\@tempb {arXiv}\fi \@ifundefined
  {mn@eprint@\@tempb}{\@tempb:\@tempc}{\expandafter \expandafter \csname
  mn@eprint@\@tempb\endcsname \expandafter{\@tempc}}}

\bibitem[\protect\citeauthoryear{Abdo et~al.,}{Abdo
  et~al.}{2010}]{Abdo_etal-2010}
Abdo A.~A.,  et~al., 2010, ApJ, 722, 520

\bibitem[\protect\citeauthoryear{Abdo et~al.,}{Abdo
  et~al.}{2011}]{Abdo_etal-2011b}
Abdo A.,  et~al., 2011, ApJ Lett., 733, L26

\bibitem[\protect\citeauthoryear{Aharonian et~al.,}{Aharonian
  et~al.}{2007}]{Aharonian_etal-2007}
Aharonian F.,  et~al., 2007, ApJ Lett., 664, L71

\bibitem[\protect\citeauthoryear{Bai \& Stone}{Bai \&
  Stone}{2013}]{Bai_Stone-2013}
Bai X.-N.,  Stone J.~M.,  2013, ApJ, 767, 30

\bibitem[\protect\citeauthoryear{Begelman}{Begelman}{1998}]{Begelman-1998}
Begelman M.~C.,  1998, ApJ, 493, 291

\bibitem[\protect\citeauthoryear{Begelman \& Li}{Begelman \&
  Li}{1994}]{Begelman_Li-1994}
Begelman M.~C.,  Li Z.-Y.,  1994, ApJ, 426, 269

\bibitem[\protect\citeauthoryear{{Begelman}, {Fabian}  \& {Rees}}{{Begelman}
  et~al.}{2008}]{Begelman_etal-2008}
{Begelman} M.~C.,  {Fabian} A.~C.,   {Rees} M.~J.,  2008, \mn@doi [MNRAS]
  {10.1111/j.1745-3933.2007.00413.x}, \href
  {http://adsabs.harvard.edu/abs/2008MNRAS.384L..19B} {384, L19}

\bibitem[\protect\citeauthoryear{{Bessho} \& {Bhattacharjee}}{{Bessho} \&
  {Bhattacharjee}}{2012}]{Bessho_Bhattacharjee-2012}
{Bessho} N.,  {Bhattacharjee} A.,  2012, \mn@doi [\apj]
  {10.1088/0004-637X/750/2/129}, \href
  {http://adsabs.harvard.edu/abs/2012ApJ...750..129B} {750, 129}

\bibitem[\protect\citeauthoryear{Birn et~al.,}{Birn
  et~al.}{2001}]{Birn_etal-2001}
Birn J.,  et~al., 2001, J. of Geophys. Res.: Space Physics, 106, 3715

\bibitem[\protect\citeauthoryear{{Biskamp}}{{Biskamp}}{1994}]{Biskamp-1994}
{Biskamp} D.,  1994, \mn@doi [\physrep] {10.1016/0370-1573(94)90110-4}, \href
  {http://adsabs.harvard.edu/abs/1994PhR...237..179B} {237, 179}

\bibitem[\protect\citeauthoryear{Biskamp}{Biskamp}{2005}]{Biskamp:2005}
Biskamp D.,  2005, Magnetic reconnection in plasmas.
No.~3 in Cambridge monographs on plasma physics, Cambridge University Press

\bibitem[\protect\citeauthoryear{{Blackman} \& {Field}}{{Blackman} \&
  {Field}}{1994}]{Blackman_Field-1994}
{Blackman} E.~G.,  {Field} G.~B.,  1994, \mn@doi [Physical Review Letters]
  {10.1103/PhysRevLett.72.494}, \href
  {http://adsabs.harvard.edu/abs/1994PhRvL..72..494B} {72, 494}

\bibitem[\protect\citeauthoryear{Boris}{Boris}{1970}]{Boris:1970}
Boris J.~P.,  1970, in Proc. Fourth Conf. Num. Sim. Plasmas, Naval Res. Lab,
  Wash. DC. p.~3

\bibitem[\protect\citeauthoryear{{Cerruti}, {Zech}, {Boisson}  \&
  {Inoue}}{{Cerruti} et~al.}{2015}]{Cerruti_etal-2015}
{Cerruti} M.,  {Zech} A.,  {Boisson} C.,   {Inoue} S.,  2015, \mn@doi [MNRAS]
  {10.1093/mnras/stu2691}, \href
  {http://adsabs.harvard.edu/abs/2015MNRAS.448..910C} {448, 910}

\bibitem[\protect\citeauthoryear{{Cerutti}, {Uzdensky}  \&
  {Begelman}}{{Cerutti} et~al.}{2012a}]{Cerutti_etal-2012a}
{Cerutti} B.,  {Uzdensky} D.~A.,   {Begelman} M.~C.,  2012a, \mn@doi [\apj]
  {10.1088/0004-637X/746/2/148}, \href
  {http://adsabs.harvard.edu/abs/2012ApJ...746..148C} {746, 148}

\bibitem[\protect\citeauthoryear{{Cerutti}, {Werner}, {Uzdensky}  \&
  {Begelman}}{{Cerutti} et~al.}{2012b}]{Cerutti_etal-2012b}
{Cerutti} B.,  {Werner} G.~R.,  {Uzdensky} D.~A.,   {Begelman} M.~C.,  2012b,
  \mn@doi [\apjl] {10.1088/2041-8205/754/2/L33}, \href
  {http://adsabs.harvard.edu/abs/2012ApJ...754L..33C} {754, L33}

\bibitem[\protect\citeauthoryear{{Cerutti}, {Werner}, {Uzdensky}  \&
  {Begelman}}{{Cerutti} et~al.}{2013}]{Cerutti_etal-2013}
{Cerutti} B.,  {Werner} G.~R.,  {Uzdensky} D.~A.,   {Begelman} M.~C.,  2013,
  \mn@doi [\apj] {10.1088/0004-637X/770/2/147}, \href
  {http://adsabs.harvard.edu/abs/2013ApJ...770..147C} {770, 147}

\bibitem[\protect\citeauthoryear{{Cerutti}, {Werner}, {Uzdensky}  \&
  {Begelman}}{{Cerutti} et~al.}{2014a}]{Cerutti_etal-2014a}
{Cerutti} B.,  {Werner} G.~R.,  {Uzdensky} D.~A.,   {Begelman} M.~C.,  2014a,
  \mn@doi [Phys. Plasmas] {10.1063/1.4872024}, \href
  {http://adsabs.harvard.edu/abs/2014PhPl...21e6501C} {21, 056501}

\bibitem[\protect\citeauthoryear{{Cerutti}, {Werner}, {Uzdensky}  \&
  {Begelman}}{{Cerutti} et~al.}{2014b}]{Cerutti_etal-2014b}
{Cerutti} B.,  {Werner} G.~R.,  {Uzdensky} D.~A.,   {Begelman} M.~C.,  2014b,
  \mn@doi [\apj] {10.1088/0004-637X/782/2/104}, \href
  {http://adsabs.harvard.edu/abs/2014ApJ...782..104C} {782, 104}

\bibitem[\protect\citeauthoryear{Cerutti, Philippov, Parfrey  \&
  Spitkovsky}{Cerutti et~al.}{2015}]{Cerutti_etal-2015}
Cerutti B.,  Philippov A.,  Parfrey K.,   Spitkovsky A.,  2015, MNRAS, 448, 606

\bibitem[\protect\citeauthoryear{{Contopoulos}}{{Contopoulos}}{2007a}]{Contopoulos-2007a}
{Contopoulos} I.,  2007a, \mn@doi [\aap] {10.1051/0004-6361:20065973}, \href
  {http://adsabs.harvard.edu/abs/2007A%26A...466..301C} {466, 301}

\bibitem[\protect\citeauthoryear{{Contopoulos}}{{Contopoulos}}{2007b}]{Contopoulos-2007b}
{Contopoulos} I.,  2007b, \mn@doi [\aap] {10.1051/0004-6361:20077167}, \href
  {http://adsabs.harvard.edu/abs/2007A%26A...472..219C} {472, 219}

\bibitem[\protect\citeauthoryear{{Coroniti}}{{Coroniti}}{1990}]{Coroniti:1990}
{Coroniti} F.~V.,  1990, \mn@doi [ApJ] {10.1086/168340}, \href
  {http://adsabs.harvard.edu/abs/1990ApJ...349..538C} {349, 538}

\bibitem[\protect\citeauthoryear{Di~Matteo}{Di~Matteo}{1998}]{DiMatteo-1998}
Di~Matteo T.,  1998, MNRAS, 299, L15

\bibitem[\protect\citeauthoryear{Ding, Yuan  \& Liang}{Ding
  et~al.}{2010}]{Ding:2010}
Ding J.,  Yuan F.,   Liang E.,  2010, ApJ, 708, 1545

\bibitem[\protect\citeauthoryear{Done, Gierli{\'n}ski  \& Kubota}{Done
  et~al.}{2007}]{Done_etal-2007}
Done C.,  Gierli{\'n}ski M.,   Kubota A.,  2007, A\& A Rev., 15, 1

\bibitem[\protect\citeauthoryear{Drake, Cassak, Shay, Swisdak  \&
  Quataert}{Drake et~al.}{2009}]{Drake:2009}
Drake J.,  Cassak P.,  Shay M.,  Swisdak M.,   Quataert E.,  2009, ApJ Lett.,
  700, L16

\bibitem[\protect\citeauthoryear{{Drenkhahn} \& {Spruit}}{{Drenkhahn} \&
  {Spruit}}{2002}]{Drenkhahn:2002}
{Drenkhahn} G.,  {Spruit} H.~C.,  2002, \mn@doi [A\& A]
  {10.1051/0004-6361:20020839}, \href
  {http://adsabs.harvard.edu/abs/2002A%26A...391.1141D} {391, 1141}

\bibitem[\protect\citeauthoryear{Eastwood, Phan, Drake, Shay, Borg, Lavraud  \&
  Taylor}{Eastwood et~al.}{2013}]{Eastwood_etal-2013}
Eastwood J.,  Phan T.,  Drake J.,  Shay M.,  Borg A.,  Lavraud B.,   Taylor M.,
   2013, Phys. Rev. Lett., 110, 225001

\bibitem[\protect\citeauthoryear{Eckart}{Eckart}{1940}]{Eckart-1940}
Eckart C.,  1940, Phys. Rev., 58, 919

\bibitem[\protect\citeauthoryear{Esirkepov}{Esirkepov}{2001}]{Esirkepov:2001}
Esirkepov T.~Z.,  2001, Comput. Phys. Commun., 135, 144

\bibitem[\protect\citeauthoryear{Fossati, Maraschi, Celotti, Comastri  \&
  Ghisellini}{Fossati et~al.}{1998}]{Fossati_etal-1998}
Fossati G.~a.,  Maraschi L.,  Celotti A.,  Comastri A.,   Ghisellini G.,  1998,
  MNRAS, 299, 433

\bibitem[\protect\citeauthoryear{Galeev, Rosner  \& Vaiana}{Galeev
  et~al.}{1979}]{Galeev_etal-1979}
Galeev A.,  Rosner R.,   Vaiana G.,  1979, ApJ, 229, 318

\bibitem[\protect\citeauthoryear{Ghisellini \& Madau}{Ghisellini \&
  Madau}{1996}]{Ghisellini_Madau-1996}
Ghisellini G.,  Madau P.,  1996, MNRAS, 280, 67

\bibitem[\protect\citeauthoryear{Ghisellini, Celotti, Fossati, Maraschi  \&
  Comastri}{Ghisellini et~al.}{1998}]{Ghisellini_etal-1998}
Ghisellini G.,  Celotti A.,  Fossati G.,  Maraschi L.,   Comastri A.,  1998,
  MNRAS, 301, 451

\bibitem[\protect\citeauthoryear{{Giannios}}{{Giannios}}{2013}]{Giannios-2013}
{Giannios} D.,  2013, \mn@doi [\mnras] {10.1093/mnras/stt167}, \href
  {http://adsabs.harvard.edu/abs/2013MNRAS.431..355G} {431, 355}

\bibitem[\protect\citeauthoryear{Giannios \& Spruit}{Giannios \&
  Spruit}{2006}]{Giannios:2006}
Giannios D.,  Spruit H.~C.,  2006, A\& A, 450, 887

\bibitem[\protect\citeauthoryear{{Giannios}, {Uzdensky}  \&
  {Begelman}}{{Giannios} et~al.}{2009}]{Giannios:2009}
{Giannios} D.,  {Uzdensky} D.~A.,   {Begelman} M.~C.,  2009, \mn@doi [MNRAS]
  {10.1111/j.1745-3933.2009.00635.x}, \href
  {http://adsabs.harvard.edu/abs/2009MNRAS.395L..29G} {395, L29}

\bibitem[\protect\citeauthoryear{Giannios, Uzdensky  \& Begelman}{Giannios
  et~al.}{2010}]{Giannios:2010}
Giannios D.,  Uzdensky D.~A.,   Begelman M.~C.,  2010, MNRAS, 402, 1649

\bibitem[\protect\citeauthoryear{Goodman \& Uzdensky}{Goodman \&
  Uzdensky}{2008}]{Goodman_Uzdensky-2008}
Goodman J.,  Uzdensky D.,  2008, ApJ, 688, 555

\bibitem[\protect\citeauthoryear{Gordovskyy, Browning  \& Vekstein}{Gordovskyy
  et~al.}{2010}]{Gordovskyy:2010b}
Gordovskyy M.,  Browning P.,   Vekstein G.,  2010, ApJ, 720, 1603

\bibitem[\protect\citeauthoryear{Guo, Li, Daughton  \& Liu}{Guo
  et~al.}{2014}]{Guo:2014}
Guo F.,  Li H.,  Daughton W.,   Liu Y.-H.,  2014, \mn@doi [Phys. Rev. Lett.]
  {10.1103/PhysRevLett.113.155005}, 113, 155005

\bibitem[\protect\citeauthoryear{Guo, Liu, Daughton  \& Li}{Guo
  et~al.}{2015}]{Guo:2015}
Guo F.,  Liu Y.-H.,  Daughton W.,   Li H.,  2015, ApJ, 806, 167

\bibitem[\protect\citeauthoryear{Guo et~al.,}{Guo et~al.}{2016}]{Guo:2016}
Guo F.,  et~al., 2016, ApJ Lett., 818, L9

\bibitem[\protect\citeauthoryear{Haggerty, Shay, Drake, Phan  \&
  McHugh}{Haggerty et~al.}{2015}]{Haggerty_etal-2015}
Haggerty C.,  Shay M.,  Drake J.,  Phan T.,   McHugh C.,  2015, Geophys. Res.
  Lett., 42, 9657

\bibitem[\protect\citeauthoryear{Hayashida et~al.,}{Hayashida
  et~al.}{2015}]{Hayashida_etal-2015}
Hayashida M.,  et~al., 2015, ApJ, 807, 79

\bibitem[\protect\citeauthoryear{{Hoh}}{{Hoh}}{1966}]{Hoh-1966}
{Hoh} F.~C.,  1966, \mn@doi [Physics of Fluids] {10.1063/1.1761670}, \href
  {http://adsabs.harvard.edu/abs/1966PhFl....9..277H} {9, 277}

\bibitem[\protect\citeauthoryear{{Hoshino} \& {Lyubarsky}}{{Hoshino} \&
  {Lyubarsky}}{2012}]{Hoshino_Lyubarsky-2012}
{Hoshino} M.,  {Lyubarsky} Y.,  2012, \mn@doi [\ssr]
  {10.1007/s11214-012-9931-z}, \href
  {http://adsabs.harvard.edu/abs/2012SSRv..173..521H} {173, 521}

\bibitem[\protect\citeauthoryear{{Jaroschek} \& {Hoshino}}{{Jaroschek} \&
  {Hoshino}}{2009}]{Jaroschek_Hoshino-2009}
{Jaroschek} C.~H.,  {Hoshino} M.,  2009, \mn@doi [Physical Review Letters]
  {10.1103/PhysRevLett.103.075002}, \href
  {http://adsabs.harvard.edu/abs/2009PhRvL.103g5002J} {103, 075002}

\bibitem[\protect\citeauthoryear{{Jaroschek}, {Treumann}, {Lesch}  \&
  {Scholer}}{{Jaroschek} et~al.}{2004}]{Jaroschek_etal-2004a}
{Jaroschek} C.~H.,  {Treumann} R.~A.,  {Lesch} H.,   {Scholer} M.,  2004,
  \mn@doi [Physics of Plasmas] {10.1063/1.1644814}, \href
  {http://adsabs.harvard.edu/abs/2004PhPl...11.1151J} {11, 1151}

\bibitem[\protect\citeauthoryear{{Kagan}, {Milosavljevi{\'c}}  \&
  {Spitkovsky}}{{Kagan} et~al.}{2013}]{Kagan_etal-2013}
{Kagan} D.,  {Milosavljevi{\'c}} M.,   {Spitkovsky} A.,  2013, \mn@doi [\apj]
  {10.1088/0004-637X/774/1/41}, \href
  {http://adsabs.harvard.edu/abs/2013ApJ...774...41K} {774, 41}

\bibitem[\protect\citeauthoryear{{Kagan}, {Nakar}  \& {Piran}}{{Kagan}
  et~al.}{2016}]{Kagan_etal-2016}
{Kagan} D.,  {Nakar} E.,   {Piran} T.,  2016, \mn@doi [\apj]
  {10.3847/0004-637X/826/2/221}, \href
  {http://adsabs.harvard.edu/abs/2016ApJ...826..221K} {826, 221}

\bibitem[\protect\citeauthoryear{{Kirk}}{{Kirk}}{2004}]{Kirk-2004}
{Kirk} J.~G.,  2004, \mn@doi [Physical Review Letters]
  {10.1103/PhysRevLett.92.181101}, \href
  {http://adsabs.harvard.edu/abs/2004PhRvL..92r1101K} {92, 181101}

\bibitem[\protect\citeauthoryear{{Kirk} \& {Skj{\ae}raasen}}{{Kirk} \&
  {Skj{\ae}raasen}}{2003}]{Kirk_Skjaeraasen-2003}
{Kirk} J.~G.,  {Skj{\ae}raasen} O.,  2003, \mn@doi [\apj] {10.1086/375215},
  \href {http://adsabs.harvard.edu/abs/2003ApJ...591..366K} {591, 366}

\bibitem[\protect\citeauthoryear{Komissarov, Barkov, Vlahakis  \&
  K{\"o}nigl}{Komissarov et~al.}{2007}]{Komissarov_etal-2007}
Komissarov S.~S.,  Barkov M.~V.,  Vlahakis N.,   K{\"o}nigl A.,  2007, MNRAS,
  380, 51

\bibitem[\protect\citeauthoryear{Kowal, de Gouveia Dal~Pino  \& Lazarian}{Kowal
  et~al.}{2011}]{Kowal:2011}
Kowal G.,  de Gouveia Dal~Pino E.~M.,   Lazarian A.,  2011, ApJ, 735, 102

\bibitem[\protect\citeauthoryear{Landau \& Lifshitz}{Landau \&
  Lifshitz}{1959}]{Landau_Lifshitz_Fluid_Sec127-1959}
Landau L.~D.,  Lifshitz E.~M.,  1959, Fluid Mechanics, \S{}127.
Addison-Wesley

\bibitem[\protect\citeauthoryear{{Larrabee}, {Lovelace}  \&
  {Romanova}}{{Larrabee} et~al.}{2003}]{Larrabee_etal-2003}
{Larrabee} D.~A.,  {Lovelace} R.~V.~E.,   {Romanova} M.~M.,  2003, \mn@doi
  [\apj] {10.1086/367640}, \href
  {http://adsabs.harvard.edu/abs/2003ApJ...586...72L} {586, 72}

\bibitem[\protect\citeauthoryear{{Liu}, {Li}, {Yin}, {Albright}, {Bowers}  \&
  {Liang}}{{Liu} et~al.}{2011}]{Liu_etal-2011}
{Liu} W.,  {Li} H.,  {Yin} L.,  {Albright} B.~J.,  {Bowers} K.~J.,   {Liang}
  E.~P.,  2011, \mn@doi [Physics of Plasmas] {10.1063/1.3589304}, \href
  {http://adsabs.harvard.edu/abs/2011PhPl...18e2105L} {18, 052105}

\bibitem[\protect\citeauthoryear{Liu, Guo, Daughton, Li  \& Hesse}{Liu
  et~al.}{2015}]{Liu_etal-2015}
Liu Y.-H.,  Guo F.,  Daughton W.,  Li H.,   Hesse M.,  2015, Phys. Rev. Lett.,
  114, 095002

\bibitem[\protect\citeauthoryear{{Lyubarsky}}{{Lyubarsky}}{2005}]{Lyubarsky-2005}
{Lyubarsky} Y.~E.,  2005, \mn@doi [\mnras] {10.1111/j.1365-2966.2005.08767.x},
  \href {http://adsabs.harvard.edu/abs/2005MNRAS.358..113L} {358, 113}

\bibitem[\protect\citeauthoryear{{Lyubarsky} \& {Kirk}}{{Lyubarsky} \&
  {Kirk}}{2001}]{Lyubarsky:2001}
{Lyubarsky} Y.,  {Kirk} J.~G.,  2001, \mn@doi [ApJ] {10.1086/318354}, \href
  {http://adsabs.harvard.edu/abs/2001ApJ...547..437L} {547, 437}

\bibitem[\protect\citeauthoryear{{Lyubarsky} \& {Liverts}}{{Lyubarsky} \&
  {Liverts}}{2008}]{Lyubarsky_Liverts-2008}
{Lyubarsky} Y.,  {Liverts} M.,  2008, \mn@doi [\apj] {10.1086/589640}, \href
  {http://adsabs.harvard.edu/abs/2008ApJ...682.1436L} {682, 1436}

\bibitem[\protect\citeauthoryear{Lyutikov}{Lyutikov}{2003}]{Lyutikov:2003}
Lyutikov M.,  2003, MNRAS, 346, 540

\bibitem[\protect\citeauthoryear{Lyutikov}{Lyutikov}{2006}]{Lyutikov:2006}
Lyutikov M.,  2006, MNRAS, 367, 1594

\bibitem[\protect\citeauthoryear{{Lyutikov} \& {Uzdensky}}{{Lyutikov} \&
  {Uzdensky}}{2003}]{Lyutikov_Uzdensky-2003}
{Lyutikov} M.,  {Uzdensky} D.,  2003, \mn@doi [\apj] {10.1086/374808}, \href
  {http://adsabs.harvard.edu/abs/2003ApJ...589..893L} {589, 893}

\bibitem[\protect\citeauthoryear{Madejski \& Sikora}{Madejski \&
  Sikora}{2016}]{Madejski_Sikora-2016}
Madejski G.,  Sikora M.,  2016, ARA\& A, 54, 725

\bibitem[\protect\citeauthoryear{Mair, Hornik  \& de Leeuw}{Mair
  et~al.}{2009}]{Mair_etal-2009}
Mair P.,  Hornik K.,   de Leeuw J.,  2009, Journal of statistical software, 32,
  1

\bibitem[\protect\citeauthoryear{{McKinney} \& {Uzdensky}}{{McKinney} \&
  {Uzdensky}}{2012}]{McKinney:2012}
{McKinney} J.~C.,  {Uzdensky} D.~A.,  2012, \mn@doi [MNRAS]
  {10.1111/j.1365-2966.2011.19721.x}, \href
  {http://adsabs.harvard.edu/abs/2012MNRAS.419..573M} {419, 573}

\bibitem[\protect\citeauthoryear{Melzani, Walder, Folini, Winisdoerffer  \&
  Favre}{Melzani et~al.}{2014a}]{Melzani:2014a}
Melzani M.,  Walder R.,  Folini D.,  Winisdoerffer C.,   Favre J.~M.,  2014a,
  A\& A, 570, A111

\bibitem[\protect\citeauthoryear{Melzani, Walder, Folini, Winisdoerffer  \&
  Favre}{Melzani et~al.}{2014b}]{Melzani:2014b}
Melzani M.,  Walder R.,  Folini D.,  Winisdoerffer C.,   Favre J.~M.,  2014b,
  A\& A, 570, A112

\bibitem[\protect\citeauthoryear{Miller \& Stone}{Miller \&
  Stone}{2000}]{Miller_Stone-2000}
Miller K.~A.,  Stone J.~M.,  2000, ApJ, 534, 398

\bibitem[\protect\citeauthoryear{{Nalewajko}, {Giannios}, {Begelman},
  {Uzdensky}  \& {Sikora}}{{Nalewajko} et~al.}{2011}]{Nalewajko:2011}
{Nalewajko} K.,  {Giannios} D.,  {Begelman} M.~C.,  {Uzdensky} D.~A.,
  {Sikora} M.,  2011, \mn@doi [MNRAS] {10.1111/j.1365-2966.2010.18140.x}, \href
  {http://adsabs.harvard.edu/abs/2011MNRAS.413..333N} {413, 333}

\bibitem[\protect\citeauthoryear{Nalewajko, Begelman, Cerutti, Uzdensky  \&
  Sikora}{Nalewajko et~al.}{2012}]{Nalewajko:2012}
Nalewajko K.,  Begelman M.~C.,  Cerutti B.,  Uzdensky D.~A.,   Sikora M.,
  2012, MNRAS, 425, 2519

\bibitem[\protect\citeauthoryear{Nalewajko, Begelman  \& Sikora}{Nalewajko
  et~al.}{2014}]{Nalewajko_etal-2014a}
Nalewajko K.,  Begelman M.~C.,   Sikora M.,  2014, ApJ, 789, 161

\bibitem[\protect\citeauthoryear{Nalewajko, Uzdensky, Cerutti, Werner  \&
  Begelman}{Nalewajko et~al.}{2015}]{Nalewajko:2015}
Nalewajko K.,  Uzdensky D.~A.,  Cerutti B.,  Werner G.~R.,   Begelman M.~C.,
  2015, ApJ, 815, 101

\bibitem[\protect\citeauthoryear{Nieter \& Cary}{Nieter \&
  Cary}{2004}]{Nieter:2004}
Nieter C.,  Cary J.~R.,  2004, J. Comput. Phys., 196, 448

\bibitem[\protect\citeauthoryear{Nodes, Birk, Lesch  \& Schopper}{Nodes
  et~al.}{2003}]{Nodes:2003}
Nodes C.,  Birk G.~T.,  Lesch H.,   Schopper R.,  2003, Phys. Plasmas, 10, 835

\bibitem[\protect\citeauthoryear{Onofri, Isliker  \& Vlahos}{Onofri
  et~al.}{2006}]{Onofri:2006}
Onofri M.,  Isliker H.,   Vlahos L.,  2006, \mn@doi [Phys. Rev. Lett.]
  {10.1103/PhysRevLett.96.151102}, 96, 151102

\bibitem[\protect\citeauthoryear{{Petropoulou}, {Giannios}  \&
  {Sironi}}{{Petropoulou} et~al.}{2016}]{Petropoulou_etal-2016}
{Petropoulou} M.,  {Giannios} D.,   {Sironi} L.,  2016, \mn@doi [\mnras]
  {10.1093/mnras/stw1832}, \href
  {http://adsabs.harvard.edu/abs/2016MNRAS.462.3325P} {462, 3325}

\bibitem[\protect\citeauthoryear{Phan et~al.,}{Phan
  et~al.}{2013}]{Phan_etal-2013}
Phan T.,  et~al., 2013, GRL, 40, 4475

\bibitem[\protect\citeauthoryear{Phan et~al.,}{Phan
  et~al.}{2014}]{Phan_etal-2014}
Phan T.,  et~al., 2014, Geophys. Res. Lett., 41, 7002

\bibitem[\protect\citeauthoryear{Philippov \& Spitkovsky}{Philippov \&
  Spitkovsky}{2014}]{Philippov_etal-2014a}
Philippov A.~A.,  Spitkovsky A.,  2014, ApJ Lett., 785, L33

\bibitem[\protect\citeauthoryear{Philippov, Tchekhovskoy  \& Li}{Philippov
  et~al.}{2014}]{Philippov_etal-2014b}
Philippov A.,  Tchekhovskoy A.,   Li J.~G.,  2014, MNRAS, 441, 1879

\bibitem[\protect\citeauthoryear{{Rees}}{{Rees}}{1984}]{Rees-1984}
{Rees} M.~J.,  1984, \mn@doi [\araa] {10.1146/annurev.aa.22.090184.002351},
  \href {http://adsabs.harvard.edu/abs/1984ARA%26A..22..471R} {22, 471}

\bibitem[\protect\citeauthoryear{{Rees}, {Begelman}, {Blandford}  \&
  {Phinney}}{{Rees} et~al.}{1982}]{Rees_etal-1982}
{Rees} M.~J.,  {Begelman} M.~C.,  {Blandford} R.~D.,   {Phinney} E.~S.,  1982,
  \mn@doi [\nat] {10.1038/295017a0}, \href
  {http://adsabs.harvard.edu/abs/1982Natur.295...17R} {295, 17}

\bibitem[\protect\citeauthoryear{{Remillard} \& {McClintock}}{{Remillard} \&
  {McClintock}}{2006}]{Remillard_McClintock-2006}
{Remillard} R.~A.,  {McClintock} J.~E.,  2006, \mn@doi [ARA\& A]
  {10.1146/annurev.astro.44.051905.092532}, \href
  {http://adsabs.harvard.edu/abs/2006ARA%26A..44...49R} {44, 49}

\bibitem[\protect\citeauthoryear{Ren, Yamada, Gerhardt, Ji, Kulsrud  \&
  Kuritsyn}{Ren et~al.}{2005}]{Ren_etal-2005}
Ren Y.,  Yamada M.,  Gerhardt S.,  Ji H.,  Kulsrud R.,   Kuritsyn A.,  2005,
  Phys. Rev. Lett., 95, 055003

\bibitem[\protect\citeauthoryear{Ressler, Tchekhovskoy, Quataert, Chandra  \&
  Gammie}{Ressler et~al.}{2015}]{Ressler:2015}
Ressler S.~M.,  Tchekhovskoy A.,  Quataert E.,  Chandra M.,   Gammie C.~F.,
  2015, \mn@doi [MNRAS] {10.1093/mnras/stv2084}, 454, 1848

\bibitem[\protect\citeauthoryear{Rogers, Denton, Drake  \& Shay}{Rogers
  et~al.}{2001}]{Rogers_etal-2001}
Rogers B.,  Denton R.,  Drake J.,   Shay M.,  2001, Phys. Rev. Lett., 87,
  195004

\bibitem[\protect\citeauthoryear{Salvesen, Simon, Armitage  \&
  Begelman}{Salvesen et~al.}{2016}]{Salvesen_etal-2016}
Salvesen G.,  Simon J.~B.,  Armitage P.~J.,   Begelman M.~C.,  2016, MNRAS,
  457, 857

\bibitem[\protect\citeauthoryear{Savolainen, Homan, Hovatta, Kadler, Kovalev,
  Lister, Ros  \& Zensus}{Savolainen et~al.}{2010}]{Savolainen_etal-2010}
Savolainen T.,  Homan D.,  Hovatta T.,  Kadler M.,  Kovalev Y.,  Lister M.,
  Ros E.,   Zensus J.~A.,  2010, A\& A, 512, A24

\bibitem[\protect\citeauthoryear{Shay, Drake, Denton  \& Biskamp}{Shay
  et~al.}{1998}]{Shay_etal-1998}
Shay M.~A.,  Drake J.~F.,  Denton R.~E.,   Biskamp D.,  1998, J. Geophys. Res:
  Space Physics, 103, 9165

\bibitem[\protect\citeauthoryear{Sikora, Begelman  \& Rees}{Sikora
  et~al.}{1994}]{Sikora_etal-1994}
Sikora M.,  Begelman M.~C.,   Rees M.~J.,  1994, ApJ, 421, 153

\bibitem[\protect\citeauthoryear{Sikora, Janiak, Nalewajko, Madejski  \&
  Moderski}{Sikora et~al.}{2013}]{Sikora_etal-2013}
Sikora M.,  Janiak M.,  Nalewajko K.,  Madejski G.~M.,   Moderski R.,  2013,
  ApJ, 779, 68

\bibitem[\protect\citeauthoryear{{Sironi} \& {Spitkovsky}}{{Sironi} \&
  {Spitkovsky}}{2011}]{Sironi_Spitkovsky-2011}
{Sironi} L.,  {Spitkovsky} A.,  2011, \mn@doi [\apj]
  {10.1088/0004-637X/741/1/39}, \href
  {http://adsabs.harvard.edu/abs/2011ApJ...741...39S} {741, 39}

\bibitem[\protect\citeauthoryear{{Sironi} \& {Spitkovsky}}{{Sironi} \&
  {Spitkovsky}}{2014}]{Sironi_Spitkovsky-2014}
{Sironi} L.,  {Spitkovsky} A.,  2014, \mn@doi [\apjl]
  {10.1088/2041-8205/783/1/L21}, \href
  {http://adsabs.harvard.edu/abs/2014ApJ...783L..21S} {783, L21}

\bibitem[\protect\citeauthoryear{{Sironi}, {Petropoulou}  \&
  {Giannios}}{{Sironi} et~al.}{2015}]{Sironi_etal-2015}
{Sironi} L.,  {Petropoulou} M.,   {Giannios} D.,  2015, \mn@doi [\mnras]
  {10.1093/mnras/stv641}, \href
  {http://adsabs.harvard.edu/abs/2015MNRAS.450..183S} {450, 183}

\bibitem[\protect\citeauthoryear{{Sironi}, {Giannios}  \&
  {Petropoulou}}{{Sironi} et~al.}{2016}]{Sironi_etal-2016}
{Sironi} L.,  {Giannios} D.,   {Petropoulou} M.,  2016, \mn@doi [\mnras]
  {10.1093/mnras/stw1620}, \href
  {http://adsabs.harvard.edu/abs/2016MNRAS.462...48S} {462, 48}

\bibitem[\protect\citeauthoryear{Sokolov, Marscher  \& McHardy}{Sokolov
  et~al.}{2004}]{Sokolov_etal-2004}
Sokolov A.,  Marscher A.~P.,   McHardy I.~M.,  2004, ApJ, 613, 725

\bibitem[\protect\citeauthoryear{Sonnerup}{Sonnerup}{1979}]{Sonnerup-1979}
Sonnerup B.,  1979, Solar system plasma physics, 1, 45

\bibitem[\protect\citeauthoryear{Spada, Ghisellini, Lazzati  \& Celotti}{Spada
  et~al.}{2001}]{Spada_etal-2001}
Spada M.,  Ghisellini G.,  Lazzati D.,   Celotti A.,  2001, MNRAS, 325, 1559

\bibitem[\protect\citeauthoryear{{Tchekhovskoy} \& {Bromberg}}{{Tchekhovskoy}
  \& {Bromberg}}{2016}]{Tchekhovskoy_Bromberg-2016}
{Tchekhovskoy} A.,  {Bromberg} O.,  2016, \mn@doi [MNRAS]
  {10.1093/mnrasl/slw064}, \href
  {http://adsabs.harvard.edu/abs/2016MNRAS.461L..46T} {461, L46}

\bibitem[\protect\citeauthoryear{Terasawa}{Terasawa}{1983}]{Terasawa-1983}
Terasawa T.,  1983, Geophys. Res. Lett., 10, 475

\bibitem[\protect\citeauthoryear{Towns et~al.,}{Towns et~al.}{2014}]{XSEDE2014}
Towns J.,  et~al., 2014, Computing in Science \& Engineering, 16, 62

\bibitem[\protect\citeauthoryear{Uzdensky}{Uzdensky}{2011}]{Uzdensky:2011}
Uzdensky D.~A.,  2011, Space science reviews, 160, 45

\bibitem[\protect\citeauthoryear{Uzdensky \& Goodman}{Uzdensky \&
  Goodman}{2008}]{Uzdensky_Goodman-2008}
Uzdensky D.~A.,  Goodman J.,  2008, ApJ, 682, 608

\bibitem[\protect\citeauthoryear{Uzdensky \& Kulsrud}{Uzdensky \&
  Kulsrud}{2006}]{UzdenskyKulsrud:2006}
Uzdensky D.~A.,  Kulsrud R.~M.,  2006, Phys. Plasmas, 13, 062305

\bibitem[\protect\citeauthoryear{{Uzdensky} \& {Spitkovsky}}{{Uzdensky} \&
  {Spitkovsky}}{2014}]{Uzdensky:2014}
{Uzdensky} D.~A.,  {Spitkovsky} A.,  2014, \mn@doi [ApJ]
  {10.1088/0004-637X/780/1/3}, \href
  {http://adsabs.harvard.edu/abs/2014ApJ...780....3U} {780, 3}

\bibitem[\protect\citeauthoryear{{Uzdensky}, {Cerutti}  \&
  {Begelman}}{{Uzdensky} et~al.}{2011}]{Uzdensky_etal-2011}
{Uzdensky} D.~A.,  {Cerutti} B.,   {Begelman} M.~C.,  2011, \mn@doi [\apjl]
  {10.1088/2041-8205/737/2/L40}, \href
  {http://adsabs.harvard.edu/abs/2011ApJ...737L..40U} {737, L40}

\bibitem[\protect\citeauthoryear{Villasenor \& Buneman}{Villasenor \&
  Buneman}{1992}]{Villasenor:1992}
Villasenor J.,  Buneman O.,  1992, Comput. Phys. Commun., 69, 306

\bibitem[\protect\citeauthoryear{Vlahakis \& K{\"o}nigl}{Vlahakis \&
  K{\"o}nigl}{2003a}]{Vlahakis_Konigl-2003a}
Vlahakis N.,  K{\"o}nigl A.,  2003a, ApJ, 596, 1080

\bibitem[\protect\citeauthoryear{Vlahakis \& K{\"o}nigl}{Vlahakis \&
  K{\"o}nigl}{2003b}]{Vlahakis_Konigl-2003b}
Vlahakis N.,  K{\"o}nigl A.,  2003b, ApJ, 596, 1104

\bibitem[\protect\citeauthoryear{Wang, Bhattacharjee  \& Ma}{Wang
  et~al.}{2000}]{Wang_etal-2000}
Wang X.,  Bhattacharjee A.,   Ma Z.,  2000, J. of Geophys. Res.: Space Physics,
  105, 27633

\bibitem[\protect\citeauthoryear{{Werner}}{{Werner}}{2015}]{WernerDPP:2015}
{Werner} G.,  2015, in APS Division of Plasma Physics Meeting Abstracts.

\bibitem[\protect\citeauthoryear{{Werner}, {Begelman}, {Cerutti}, {Nalewajko}
  \& {Uzdensky}}{{Werner} et~al.}{2013}]{Werner_etal-2013}
{Werner} G.,  {Begelman} M.,  {Cerutti} B.,  {Nalewajko} K.,   {Uzdensky} D.,
  2013, in APS Division of Plasma Physics Meeting Abstracts.

\bibitem[\protect\citeauthoryear{{Werner}, {Uzdensky}, {Cerutti}, {Nalewajko}
  \& {Begelman}}{{Werner} et~al.}{2016}]{Werner_etal-2016}
{Werner} G.~R.,  {Uzdensky} D.~A.,  {Cerutti} B.,  {Nalewajko} K.,   {Begelman}
  M.~C.,  2016, \mn@doi [ApJ Lett.] {10.3847/2041-8205/816/1/L8}, \href
  {http://adsabs.harvard.edu/abs/2016ApJ...816L...8W} {816, L8}

\bibitem[\protect\citeauthoryear{Yamada, Kulsrud  \& Ji}{Yamada
  et~al.}{2010}]{Yamada_etal-2010}
Yamada M.,  Kulsrud R.,   Ji H.,  2010, Rev. Mod. Phys., 82, 603

\bibitem[\protect\citeauthoryear{Yamada, Yoo, Jara-Almonte, Ji, Kulsrud  \&
  Myers}{Yamada et~al.}{2014}]{Yamada_etal-2014}
Yamada M.,  Yoo J.,  Jara-Almonte J.,  Ji H.,  Kulsrud R.~M.,   Myers C.~E.,
  2014, Nature Commun., 5, 4774

\bibitem[\protect\citeauthoryear{Yamada, Yoo, Jara-Almonte, Daughton, Ji,
  Kulsrud  \& Myers}{Yamada et~al.}{2015}]{Yamada_etal-2015}
Yamada M.,  Yoo J.,  Jara-Almonte J.,  Daughton W.,  Ji H.,  Kulsrud R.~M.,
  Myers C.~E.,  2015, Phys. Plasmas, 22, 056501

\bibitem[\protect\citeauthoryear{Yee}{Yee}{1966}]{Yee:1966}
Yee K.~S.,  1966, IEEE Trans. Antennas Propag., 14, 302

\bibitem[\protect\citeauthoryear{Yoo, Yamada, Ji  \& Myers}{Yoo
  et~al.}{2013}]{Yoo_etal-2013}
Yoo J.,  Yamada M.,  Ji H.,   Myers C.~E.,  2013, Phys. Rev. Lett., 110, 215007

\bibitem[\protect\citeauthoryear{{Zenitani} \& {Hoshino}}{{Zenitani} \&
  {Hoshino}}{2001}]{Zenitani_Hoshino-2001}
{Zenitani} S.,  {Hoshino} M.,  2001, \mn@doi [\apjl] {10.1086/337972}, \href
  {http://adsabs.harvard.edu/abs/2001ApJ...562L..63Z} {562, L63}

\bibitem[\protect\citeauthoryear{{Zenitani} \& {Hoshino}}{{Zenitani} \&
  {Hoshino}}{2005}]{Zenitani_Hoshino-2005}
{Zenitani} S.,  {Hoshino} M.,  2005, \mn@doi [\apjl] {10.1086/427873}, \href
  {http://adsabs.harvard.edu/abs/2005ApJ...618L.111Z} {618, L111}

\bibitem[\protect\citeauthoryear{{Zenitani} \& {Hoshino}}{{Zenitani} \&
  {Hoshino}}{2007}]{Zenitani_Hoshino-2007}
{Zenitani} S.,  {Hoshino} M.,  2007, \mn@doi [\apj] {10.1086/522226}, \href
  {http://adsabs.harvard.edu/abs/2007ApJ...670..702Z} {670, 702}

\bibitem[\protect\citeauthoryear{{Zenitani} \& {Hoshino}}{{Zenitani} \&
  {Hoshino}}{2008}]{Zenitani_Hoshino-2008}
{Zenitani} S.,  {Hoshino} M.,  2008, \mn@doi [\apj] {10.1086/528708}, \href
  {http://adsabs.harvard.edu/abs/2008ApJ...677..530Z} {677, 530}

\bibitem[\protect\citeauthoryear{{Zweibel} \& {Yamada}}{{Zweibel} \&
  {Yamada}}{2009}]{Zweibel_Yamada-2009}
{Zweibel} E.~G.,  {Yamada} M.,  2009, \mn@doi [ARA\& A]
  {10.1146/annurev-astro-082708-101726}, \href
  {http://adsabs.harvard.edu/abs/2009ARA%26A..47..291Z} {47, 291}

\makeatother
\end{thebibliography}

%\begin{thebibliography}{99}
%\end{thebibliography}

%%%%%%%%%%%%%%%%%%%%%%%%%%%%%%%%%%%%%%%%%%%%%%%%%%

%%%%%%%%%%%%%%%%% APPENDICES %%%%%%%%%%%%%%%%%%%%%

\appendix

\section{Eckart and Landau fluid flow}
\label{sec:fluidFrames}

In this appendix, we describe how to calculate relativistic fluid flow velocities from particles in our kinetic simulations.
For nonrelativistic motion, bulk fluid flow is unambiguously defined
as the average velocity; in a local reference frame travelling at this velocity,
the net particle current density and momentum density both vanish.
For relativistic motion, however, these quantities may vanish in 
different frames and so the task/problem of defining the bulk flow velocity in terms of the moments of particle distribution function becomes less trivial.

We first describe the Eckart flow velocity, defined to be the boost
velocity that transforms to a reference frame in which the current
density vanishes \cite{Eckart-1940}.
Given a particle distribution $f(p)$, where $p$ is the four-momentum,
we can define a density-current four-vector
\begin{eqnarray}
  N^\mu &=& 
    \int v^\mu f(p) d^3 p 
    \; = \; \int p^\mu f(p) \frac{d^3p}{p^0}
\end{eqnarray}
where $p^0 = \gamma m c$ and $v^0=1$; $f(p)$ and
$d^3p/p^0$ are invariant scalars, showing that $N^\mu$ transforms like the
four-vector $p^\mu$.  Straightforward calculation shows that
any (time-like) four-vector $N^\mu$ can be boosted to a reference frame 
in which ${N'}^{i} = 0$ for $i=1,2,3$ by a velocity
\begin{eqnarray}
  v_{\rm Eckart}^i &=& \frac{N^i}{N^0} = 
    \frac{ \int v^i f(p) d^3p }{ \int f(p) d^3 p }
    = \langle v^i \rangle
.\end{eqnarray}
Thus a boost by the average velocity transforms to a frame---the Eckart local rest frame---in which the net current density is zero.

On the other hand, it is equally valid to consider the Landau fluid rest frame \citep{Landau_Lifshitz_Fluid_Sec127-1959}, in which
the net momentum density vanishes; in this frame, the components 
${T'}^{i0}={T'}^{0i}$ of the stress-energy tensor vanish.
In a general frame, the stress-energy tensor is
\begin{eqnarray}
  T^{\mu \nu} &=& \int v^\mu p^\nu f(p) d^3 p
               \; = \; \int p^\mu p^\nu f(p) \frac{d^3p}{p^0}
.\end{eqnarray}
The components $T^{i0}$ are proportional to the local average momentum
$\langle p^i \rangle$.
Given $T^{\mu \nu}$, one can determine the boost that yields 
${T'}^{i0}=0$ by first finding the boost along $x$ that reduces
${T'}^{10}$ to zero, then boosting from that frame along $y$ to
reduce ${T'}^{20}$ to zero, and finally along $z$.

However, we can also determine the Landau four-velocity $u$ by considering the
four-vector $w^\mu = {T^{\mu}}_{\nu} u^\nu$.
Since boosting $u$ by itself yields $u' = (1,0,0,0)$, and
${T'}^{i0} = 0$ in the boosted frame, 
\begin{eqnarray}
  {w'}^{\mu} &=& {{T'}^{\mu}}_{\nu} {u'}^{\nu} = {{T'}^{\mu}}_{0}
  = \left({T'}^{00}, \, 0,\, 0,\, 0 \right)
  = {T'}^{00} u'
.\end{eqnarray}
Thus $u'$ is an eigenvector of the matrix ${{T'}^{\mu}}_{\nu}$ with eigenvalue
${{T'}^{0}}_{0}={T'}^{00}$; transforming this eigenvalue equation to the original frame, we see that $u$ is an eigenvector of ${T^{\mu}}_{\nu}$ with the same eigenvalue:
\begin{eqnarray}
  {T^{\mu}}_{\nu} u^{\nu} &=& {T'}^{00} u^{\mu}
.\end{eqnarray}
In the boosted frame, the matrix ${{T'}^{\mu}}_{\nu}$ is block diagonal, with the $1\times 1$
block consisting of ${T'}^{00}$, corresponding to the eigenvector
$u'=(1,0,0,0)$, and the $3\times 3$ block, which is symmetric and therefore
has three real eigenvalues and orthogonal eigenvectors, which are space-like because they are orthogonal to $(1,0,0,0)$.
Therefore, the matrix ${T^{\mu}}_{\nu}$ also has four real eigenvalues, along with one time-like and three space-like eigenvectors.  The time-like eigenvector,
if normalized to $u^\mu u_\mu = 1$ (and, if necessary, multiplied by
$-1$ so that $u^0>0$), yields the Landau four-velocity $u$.

To be clear, we note that ${T^{\mu}}_{\nu} = T^{\mu \kappa} \eta_{\kappa\nu }$, where $\eta$ is the Minkowski metric with diagonal $(1,-1,-1,-1)$.  It is the non-symmetric matrix $T\eta$, and not the symmetric stress-energy tensor $T$, that must be diagonalized.
In the boosted frame, however, $T'\eta$ is symmetric because
$T'$ is symmetric and ${T'}^{i0} = 0$.

Thus, given a local stress-energy tensor $T^{\mu\nu}$, the Landau four-velocity $u$ can be found as follows:
\begin{enumerate}
  \item Diagonalize the matrix $T\eta$ and find its eigenvectors.
  \item Only one eigenvector $u$ is time-like; normalize it so 
   that $u^{\mu} u_{\mu} = c^2$ and $u^{0} > 0$; $u$ is the 
   Landau fluid  velocity.
\end{enumerate}

\section{Power-law fitting}
\label{sec:fitting}

We have developed a robust procedure to fit the high-energy portion of particle energy spectra to a power law; subsequently we can determine the high-energy cutoff of the power law.  
Essentially, the procedure attempts to identify the longest section of the spectrum $f(\varepsilon)$ that resembles a power law $\propto \varepsilon^{-p}$, and then fits a power law to that section only.
[Here, $\varepsilon = (\gamma-1)mc^2$ is the particle kinetic energy.]

Because $f(\varepsilon)$ is generally an excellent power law at very low energies (the low-energy part of the initial Maxwellian)---and because we are mainly interested in the nonthermal spectrum---we look only at $\varepsilon > \varepsilon_{\rm min}$ where $\varepsilon_{\rm min}$ is around the average particle energy (for this study, we used 1, 2, and 4 times the average energy).

We begin by smoothing $f(\varepsilon)$.  
Empirically, we expect $f(\varepsilon)$ to decrease monotonically above $\varepsilon>\varepsilon_{\rm min}$, and so any non-monotonicity indicates noise.  Therefore, we find the monotonic function closest to $f(\varepsilon)$ using the Pool-Adjacent-Violators algorithm \citep{Mair_etal-2009}, and use this to estimate noise; that is, we subsequently approximate $f(\varepsilon)$ with the minimal number of cubic splines necessary so that the best spline fit is, on average, no further from $f(\varepsilon)$ than the monotonic fit.  The spline fit yields a smoother $f(\varepsilon)$.

Working with the smoothed $f(\varepsilon)$, we calculate the local slope $p(\varepsilon) = -d \ln f(\varepsilon) / d \ln \varepsilon$, and then, with brute force, search for the longest 
interval [in terms of $\ln(\varepsilon)$] over which $p(\varepsilon)$  remains approximately constant (for example, $p$ varies at most within $\pm 0.1$ or $\pm 0.2$ of a central value).  We estimate `the' power-law index~$p$ as the median value of $p(\varepsilon)$ over that interval.
After fitting a pure power law over the same interval, we calculate the cutoff energy $\varepsilon_c$ where the measured $f(\varepsilon)$ drops to a value that is $e^{-1}$ times the pure power-law fit.

This procedure for determining a power law has the advantage that, within the identified power-law section, the local slope $p(\varepsilon)$ does not vary much, and so guarantees that the identified section will really resemble a power law (although the procedure might fail to find such a section of reasonable length).  
However, this procedure depends on several (somewhat arbitrary though reasonably chosen) parameters, such as $\varepsilon_{\rm min}$, the amount of smoothing, and the allowed maximum variation in $p(\varepsilon)$.
Therefore, for each particle spectrum obtained from simulation, 
we perform many fits with different sets of these parameters, taking the median $p$ (over all fits) as the final value.
We estimate the robustness by determining the range of $p$ that encompasses the middle 68\% of the fits, and express this range as `error' bars around the median.

%%%%%%%%%%%%%%%%%%%%%%%%%%%%%%%%%%%%%%%%%%%%%%%%%%

% Don't change these lines
\bsp	% typesetting comment
\label{lastpage}
\end{document}